\documentclass[prb,aps,twocolumn,eqsecnum,showpacs]{revtex4}
\usepackage{graphicx}
\usepackage{dcolumn}
\begin{document}

\title{
Fingerprints of spin-orbital physics in cubic Mott insulators:\\
Magnetic exchange interactions and optical spectral weights
}

\author {     Andrzej M. Ole\'{s} }
\affiliation{ Max-Planck-Institut f\"ur Festk\"orperforschung,
              Heisenbergstrasse 1, D-70569 Stuttgart, Germany, and  \\
              Marian Smoluchowski Institute of Physics, Jagellonian
              University, Reymonta 4, PL-30059 Krak\'ow, Poland }
\author {     Giniyat Khaliullin }
\affiliation{ Max-Planck-Institut f\"ur Festk\"orperforschung,
              Heisenbergstrasse 1, D-70569 Stuttgart, Germany }
\author {     Peter Horsch }
\affiliation{ Max-Planck-Institut f\"ur Festk\"orperforschung,
              Heisenbergstrasse 1, D-70569 Stuttgart, Germany }
\author {     Louis Felix Feiner }
\affiliation{ Institute for Theoretical Physics, Utrecht University,
              Leuvenlaan 4, NL-3584 CC Utrecht, and \\
              Philips Research Laboratories, Prof. Holstlaan 4,
              NL-5656 AA Eindhoven, The Netherlands }

\date{31 August, 2005}

\begin{abstract}
The temperature dependence and anisotropy of optical spectral weights
associated with different multiplet transitions is determined by the
spin and orbital correlations. To provide a systematic basis to exploit
this close relationship between magnetism and optical spectra,
we present and analyze the spin-orbital superexchange models for a
series of representative orbital-degenerate transition metal oxides with
different multiplet structure. For each case we derive the magnetic
exchange constants, which determine the spin wave dispersions, as well
as the  partial optical sum rules. The magnetic and optical properties
of early transition metal oxides with degenerate $t_{2g}$ orbitals
(titanates and vanadates with  perovskite structure) are shown to depend
only on two parameters, viz. the superexchange energy $J$ and the ratio
$\eta$ of Hund's exchange to the intraorbital Coulomb interaction,
and on the actual orbital state. In $e_g$ systems important corrections
follow from charge transfer excitations, and we show that KCuF$_3$ can
be classified as a charge transfer insulator, while LaMnO$_3$ is a
Mott insulator with moderate charge transfer contributions. In some 
cases orbital fluctuations are quenched and decoupling of spin and 
orbital degrees of freedom with static orbital order gives satisfactory 
results for the optical weights. On the example of cubic vanadates we 
describe a case where the full quantum spin-orbital physics must be 
considered. Thus information on optical excitations, their energies, 
temperature dependence and anisotropy, combined with the results of 
magnetic neutron scattering experiments, provides an important 
consistency test of the spin-orbital models, and indicates whether 
orbital and/or spin fluctuations are important in a given compound.
[{\it Published in: Phys. Rev. B {\bf 72}, 214431 (2005).}]

\end{abstract}


\maketitle

\section{Superexchange and optical excitations at orbital degeneracy}
\label{sec:motiv}

The physical properties of Mott (or charge transfer) insulators are
dominated by large on-site Coulomb interactions $\propto U$ which
suppress charge fluctuations. Quite generally, the Coulomb interactions
lead then to strong electron correlations which frequently involve
orbitally degenerate states, such as $3d$ (or $4d$) states in transition
metal compounds, and are responsible for quite complex behavior with
often puzzling transport and magnetic properties.\cite{Ima98} The
theoretical understanding of this class of compounds, with the colossal
magnetoresistance (CMR) manganites as a prominent example,
\cite{Dag02,Dag03} has substantially advanced over the last decade,
\cite{Mae04} after it became clear that orbital degrees of freedom play
a crucial role in these materials and have to be treated on equal
footing with the electron spins, which has led to a rapidly developing
field --- {\it orbital physics\/}.\cite{Tok00}
Due to the strong onsite Coulomb repulsion, charge fluctuations in the
undoped parent compounds are almost entirely suppressed, and an
adequate description of these strongly correlated insulators appears
possible in terms of superexchange.\cite{And59} At orbital degeneracy
the superexchange interactions have a rather rich
structure, represented by the so-called spin-orbital models,
discovered three decades ago,\cite{Kug82,Cyr75} and extensively
studied in recent years.
\cite{Fei97,Kha98,Ole00,Ish97,Fei99,Feh04,Kha00,Kha01,Dima2,Ole01}

Although this field is already quite mature, and the first textbooks
have already appeared,\cite{Dag03,Mae04,Faz99} it has been realized only
recently that the {\it magnetic\/} and the {\it optical\/} properties
of such correlated insulators with partly filled $d$ orbitals are
intimately related to each other, being just different experimental
manifestations of the same underlying spin-orbital physics.
\cite{Kha04,Lee05} While it is clear that the {\it low-energy\/}
effective superexchange Hamiltonian decides about the magnetic
interactions, it is not immediately obvious that the {\it high-energy\/}
optical excitations and their partial sum rules have the same roots and
may be described by the superexchange as well.
In fact, this interrelation between
the magnetic and the optical properties makes it necessary to reanalyze
the spin-orbital superexchange models, and to extract from them
important constraints imposed by the theory on the system parameters. We
will show that also the opposite holds --- some general rules apply for
the magnetic interactions in the correlated insulators with degenerate
(or almost degenerate) orbitals, and therefore the magnetic measurements
impose constraints on any realistic theory. At the same time, we shall
argue that such experiments provide very useful information concerning
the orbital order (OO) and the strength of quantum fluctuations
in a given compound, which can next be employed to
interpret other experiments, including the optical spectroscopy.

The phenomena discussed in the present paper go well beyond the more
familiar situation of a Mott insulator without orbital degeneracy,
or when the orbital degeneracy is lifted by strong Jahn-Teller (JT)
distortions as for example in the high $T_c$ cuprate superconductors.
In a Mott insulator the optical conductivity is purely incoherent, and
the optical response is found at energies which exceed the optical gap.
When orbital degrees of freedom are absent, the optical gap is
determined by the intraorbital Coulomb interaction element $U$. Naively,
one might expect that the high-energy charge excitations at energy
$\sim U$, which contribute to the optical intensities, are unrelated to
the low-energy magnetic phenomena. However, both energy scales are
intimately related as the superexchange follows from the same charge
excitations which are detected by the optical spectroscopy. The
prominent example of this behavior is the nondegenerate Hubbard model,
where the virtual high-energy excitations determine the superexchange
\cite{And59} energy $J$ --- it decides, together with spin correlations,
about the spectral weight of the upper Hubbard band at
half-filling.\cite{Bae86,Esk94} When temperature increases to an energy
scale $\sim J$, the spin correlations are modified and the total
spectral weight in the optical spectroscopy follows these changes.
\cite{Aic02}

The superexchange models for transition metal perovskites with partly
filled degenerate orbitals have a more complex structure than for
nondegenerate orbitals and allow both for antiferromagnetic (AF) and for
ferromagnetic (FM) superexchange.\cite{Kug82,Cyr75}
These different contributions to the
superexchange result from the multiplet structure of excited transition
metal ions which depends on the Hund's exchange $J_H$ and generates a
competition between high-spin and low-spin excitations. The exchange
interactions are then intrinsically frustrated {\it even on a cubic
lattice\/}, which enhances quantum effects both for $e_g$,
\cite{Fei97,Kha98,Ole00} and for $t_{2g}$ systems.\cite{Kha00,Kha01}
This frustration is partly removed in anisotropic AF phases, which
break the cubic symmetry and effectively may lead to dimensionality
changes, such as in $A$-type AF phase realized in LaMnO$_3$, or in
$C$-type AF phase in LaVO$_3$.

While rather advanced many-body treatment of the quantum physics
characteristic for spin-orbital models is required in general, we want
to present here certain simple principles which help to understand
the heart of the problem and give simple guidelines for interpreting
experiments and finding relevant physical parameters of
the spin-orbital models of {\it undoped\/} cubic insulators. We will
argue that such an approach based upon classical OO is well
justified in many known cases, as quantum phenomena are often quenched
by the Jahn-Teller (JT) coupling between orbitals and the lattice
distortions, which are present below structural phase transitions and
induce OO both in spin-disordered and in spin-ordered phases.
\cite{Kan59} However, we will also discuss the prominent
example of LaVO$_3$, where assuming perfect OO or attempts to
decouple spin and orbital fluctuations,\cite{Mot03} fail in a
spectacular way and give no more than a qualitative
insight into certain limiting cases. Significant corrections due to
quantum phenomena that go beyond such simplified approaches are then
necessary for a more quantitative understanding.

In the correlated insulators with partly occupied degenerate orbitals
not only the structure of the superexchange is complex, but also
the optical spectra exhibit strong anisotropy and temperature
dependence near the magnetic transitions, as found in LaMnO$_3$,
\cite{Tob01,Kov04} the cubic vanadates LaVO$_3$ and YVO$_3$,
\cite{Miy02,Tsv04} and in the ruthenates.\cite{Lee02} In all these
systems several excitations contribute to the excitation
spectra, so one may ask how the spectral weight redistributes between
individual subbands originating from these excitations.
The spectral weight distribution is in general anisotropic
already when OO sets in and breaks the cubic symmetry, but even more so
when $A$-type or $C$-type AF spin order occurs below the N\'eel
temperature.

The effective spin-orbital models of transition metal oxides with partly
filled degenerate orbitals depend in a characteristic way upon those
aspects of the electronic structure which decide whether a given
strongly correlated system can be classified as a Mott insulator or as
a charge transfer (CT) insulator. As suggested in the original
classification of Zaanen, Sawatzky and Allen,\cite{Zaa85} the energy of
the $d-p$ CT excitation $\Delta$ has to be compared with the Coulomb
interaction $U$ --- if $U<\Delta$, the first excitation is at a
transition metal ion and the system is a Mott insulator, otherwise it is
a CT insulator. Both are strongly correlated insulators, yet in one
limit the dominant virtual excitations are of $d-d$ type, whereas in the
other limit they are of $p-d$ type. One may consider this issue
more precisely by analyzing the full multiplet structure, and comparing
the lowest excitation energy (to a high-spin configuration)
{\it at a transition metal ion\/}, $\varepsilon_{\rm HS}=U-3J_H$, with
that of the lowest CT excitation (of energy $\Delta$)
{\it between a transition metal ion and a ligand ion\/}.
\cite{notede} Thus we argue that one can regard a given perovskite as a
{\it charge transfer insulator\/} if $\varepsilon_{\rm HS}>\Delta$, and
as a {\it Mott-Hubbard insulator\/} if $\varepsilon_{\rm HS}<\Delta$.
By analyzing these parameters it has been suggested that the late
transition metal oxides may be classified as CT insulators.\cite{Ima98}
In this case important new contributions to the superexchange arise,
\cite{Goode,Zaa88,Mos04} called below $\Delta$ (charge transfer) terms,
as we shall discuss for two $e_g$ systems: KCuF$_3$ and LaMnO$_3$.

A central aim of this paper is to provide relatively simple expressions
for the magnetic exchange constants and for the optical spectral weights
that can be used by experimentalists to analyze and compare their spin
wave data with optical data. While the full spin-orbital models are
rather complex, they are nevertheless controlled by only very few
physical parameters:
  (i) the superexchange constant $J$,
 (ii) the normalized Hund's exchange $\eta$, and
(iii) the charge transfer parameter $R$.
There are two distinct ways to determine these effective parameters:
either
 (i) from the original multiband Hubbard model, or
(ii) from experimental spin wave and/or optical data.
Here the second approach is of particular interest because the
simultaneous analysis
of magnetism and optics provides a subtle test of the underlying model.

The paper is organized as follows. In Sec. II we introduce the generic
structure of the low-energy effective Hamiltonian in a correlated
insulator with orbital degeneracy, and discuss its connection with the
optical excitations at high-energy.
This general formulation provides the important subdivision of a given
spin-orbital model which is necessary to obtain the partial spectral
weights for individual multiplet transitions. In the remaining part of
the paper we concentrate on some selected cubic perovskites and
demonstrate that this general formulation allows one to
arrive at a consistent interpretation of the magnetic and optical
experiments in these correlated insulators using the superexchange
interactions (Secs. III-VI), and to deduce the parameters relevant for
the theoretical model from the experimental data, whereever available.
We start in Sec. III with the simplest spin-orbital model for $e_g$
holes in KCuF$_3$, and demonstrate that this system is in the CT regime
which changes the commonly used picture of superexchange in this system
in a qualitative way. Next we present and analyze the spin-orbital model
with $e_g$ orbital degrees of freedom for the undoped manganite
LaMnO$_3$ in Sec. IV. Here we show that in this case much smaller
contributions arise from the CT processes, and the system is already in
the Mott-Hubbard regime of parameters, which explains the earlier
success of a simplified effective model based entirely on $d-d$
excitations and sufficient for a semiquantitative understanding.
This justifies our approach to the early transition metal
perovskites with $t_{2g}$ degrees of freedom, titanates in Sec. V and
vanadates in Sec. VI, which we treat as Mott-Hubbard insulators.
For all these systems we analyze the magnetic exchange interactions and
the optical spectral weights, depending on the nature of the spin
correlations in the ground state. The paper is concluded in Sec. VII,
where we provide a coherent view on the magnetic and the optical
phenomena and summarize the experimental constraints on the model
parameters.

\section{General formalism}
\label{sec:general}

We consider here effective models with hopping elements between
transition metal ions,
\begin{equation}
\label{Ht}
H_{0}=
   \sum_{i\alpha\sigma}\varepsilon_{i\alpha}n_{i\alpha\sigma}
+\sum_{ij,\alpha\neq\beta,\sigma}t_{i\alpha,j\beta}
a_{i\alpha\sigma}^{\dagger}a_{j\beta\sigma}^{}.
\end{equation}
Here $\varepsilon_{i\alpha}$ are orbital energies, and
$t_{i\alpha,j\beta}$ are effective hopping elements via ligand orbitals
--- they depend on the type of considered orbitals as discussed in
Refs. \onlinecite{And78} and \onlinecite{Zaa93}. The energy scale for
the hopping is set by the largest hopping element $t$: the $(dd\sigma)$
element in case of $e_g$ systems, and the $(dd\pi)$ element when only
$\pi$ bonds are considered in systems with degenerate and partly filled
$t_{2g}$ orbitals. For noninteracting electrons the Hamiltonian $H_{0}$
would lead to tight-binding bands, but in a Mott insulator the large
Coulomb interaction $U$ suppresses charge excitations in the regime of
$U\gg t$, and the hopping elements can only contribute via virtual
excitations, leading to the superexchange.

The superexchange in the $3d$ {\it cubic\/} systems with orbital
degeneracy is described by spin-orbital models, where both degrees of
freedom are coupled and the orbital state (ordered or fluctuating)
determines the spin structure and excitations, and vice versa. The
numerical and analytical structure of these models represents a
fascinating challenge in the theory, as it is much more complex than
that of pure spin models. The spin-orbital models have been derived
before in several cases, and we refer for these derivations to the
original literature.\cite{Ole00,Fei99,Kha00,Kha01} They describe
in low energy regime the consequences of virtual charge excitations
between two neighboring transition metal ions,
$d_i^md_j^m\rightleftharpoons d_i^{m+1}d_j^{m-1}$,
which involve an increase of energy due to the Coulomb interactions.
Such transitions are mediated by the ligand orbitals between the two
ions and have the same roots as the superexchange in a Mott insulator
with nondegenerate orbitals\cite{And59} at $U\gg t$ --- thus the
resulting superexchange interactions will be called {\it $U$ terms}.
The essential difference which makes it necessary to analyze the
excitation energies in each case separately is caused by the existence
of several different excitations. Their energies have to be determined
first by analyzing the eigenstates of the local Coulomb interactions,
\begin{eqnarray}
\label{Heegen}
H_{int}&=&
   U\sum_{i\alpha}n_{i\alpha  \uparrow}n_{i\alpha\downarrow}
+\sum_{i,\alpha<\beta}\Big(U_{\alpha\beta}-\frac{1}{2}J_{\alpha\beta}\Big)
                    n_{i\alpha}n_{i\beta}                 \nonumber \\
&+& \sum_{i,\alpha<\beta}J_{\alpha\beta}
\Big( d^{\dagger}_{i\alpha\uparrow}d^{\dagger}_{i\alpha\downarrow}
      d^{       }_{i\beta\downarrow}d^{       }_{i\beta\uparrow}
     +d^{\dagger}_{i\beta\uparrow}d^{\dagger}_{i\beta\downarrow}
      d^{       }_{i\alpha\downarrow}d^{       }_{i\alpha\uparrow}\Big)
                                                          \nonumber \\
&-&2\sum_{i,\alpha<\beta}J_{\alpha\beta}
    \textbf{S}_{i\alpha}\cdot\textbf{S}_{i\beta},
\end{eqnarray}
with $\bar{\sigma}=-\sigma$, which in the general case depend on the
three Racah parameters $A$, $B$ and $C$,\cite{Gri71} which may be
derived from somewhat screened atomic values. While the intraorbital
Coulomb element
\begin{equation}
\label{U}
  U=A+4B+3C,                                                   \\
\end{equation}
is identical for all $3d$ orbitals, the interorbital Coulomb and
exchange elements, $U_{\alpha\beta}$ and $J_{\alpha\beta}$, are
anisotropic and depend on the involved pair of orbitals; the values of
$J_{\alpha\beta}$ are given in Table \ref{tab:uij}. The Coulomb
and exchange elements are related to the intraorbital element $U$ by a
relation which guarantees the invariance of interactions in the orbital
space,
\begin{equation}
\label{Uab}
  U=U_{\alpha\beta}+2J_{\alpha\beta}.
\end{equation}

\begin{table}[b!]
\caption{
On-site interorbital exchange elements $J_{\alpha\beta}$ for $3d$
orbitals as functions of the Racah parameters $B$ and $C$ (for more
details see Ref. \onlinecite{Gri71}). }
\vskip .2cm
\begin{ruledtabular}
\begin{tabular}{cccccc}
  orbital      & $xy$ &  $yz$  & $zx$ &$x^2\!-\!y^2$&$3z^2\!-\!r^2$ \cr
\colrule
  $xy$         & $0$  & $3B+C$ & $3B+C$ &    $C$ & $4B+C$  \cr
  $yz$         & $3B+C$ & $0$  & $3B+C$ & $3B+C$ &  $B+C$  \cr
  $zx$         & $3B+C$ & $3B+C$ & $0$  & $3B+C$ &  $B+C$  \cr
 $x^2\!-\!y^2$ &    $C$ & $3B+C$ & $3B+C$ &  $0$ & $4B+C$  \cr
 $3z^2\!-\!r^2$& $4B+C$ &  $B+C$ &  $B+C$ & $4B+C$ &  $0$  \cr
\end{tabular}
\end{ruledtabular}
\label{tab:uij}
\end{table}

\begin{figure}[t!]
\includegraphics[width=7.7cm]{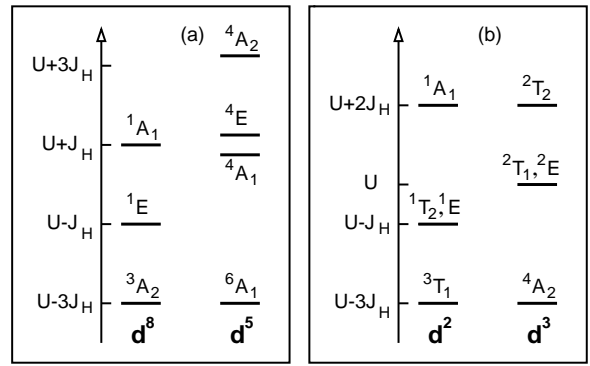}
\caption{
Energies of $d_i^{m}d_j^{m}\rightarrow d_i^{m+1}d_j^{m-1}$ charge
excitations in selected cubic transition metal oxides,
as obtained from Eq. (\ref{Hee}) for:
(a) $e_g$ excitations of Cu$^{3+}$ ($d^8$) and Mn$^{2+}$ ($d^5$) ions
[in the $d^5$ case the spectrum\cite{noteabc} was obtained from Eq.
(\ref{Heegen})];
(b) $t_{2g}$ excitations of Ti$^{2+}$ ($d^2$) and V$^{2+}$ ($d^3$)
ions.
The splittings between different states are due to Hund's exchange
element $J_H$ which refers to a pair of $e_g$ electrons in (a),
and to a pair of $t_{2g}$ electrons in (b), respectively.
}
\label{fig:levels}
\end{figure}

In cases where only the orbitals of one type ($e_g$ or $t_{2g}$) are
partly filled, however, as e.g. in the titanates, vanadates, or copper
fluorides, all relevant exchange elements $J_{\alpha\beta}$ are the same
(see Table \ref{tab:uij}) and one may use a simplified form of onsite
interactions,\cite{Ole83}
\begin{eqnarray}
\label{Hee}
H_{int}^{(0)}&=&
   U\sum_{i\alpha}n_{i\alpha  \uparrow}n_{i\alpha\downarrow}
 +\Big(U-\frac{5}{2}J_H\Big)\sum_{i,\alpha<\beta}n_{i\alpha}n_{i\beta}
                                                          \nonumber \\
&+& J_H\sum_{i,\alpha<\beta}
\Big( d^{\dagger}_{i\alpha\uparrow}d^{\dagger}_{i\alpha\downarrow}
      d^{       }_{i\beta\downarrow}d^{       }_{i\beta\uparrow}
     +d^{\dagger}_{i\beta\uparrow}d^{\dagger}_{i\beta\downarrow}
      d^{       }_{i\alpha\downarrow}d^{       }_{i\alpha\uparrow}\Big)
                                                          \nonumber \\
&-&2J_H\sum_{i,\alpha<\beta}\textbf{S}_{i\alpha}\cdot\textbf{S}_{i\beta}.
\end{eqnarray}
with only two parameters: the
Coulomb element $U$ (\ref{U}) and a Hund's exchange element $J_H$, being
$4B+C$ for $e_g$ and $3B+C$ for $t_{2g}$ systems, respectively. We
emphasize that in the general case when both types of orbitals are
partly filled (as in the manganites) and both thus participate in charge
excitations, the Hamiltonian (\ref{Hee}) is only approximate, and the
full excitation spectra of the transition metal ions\cite{Gri71} which
follow from Eq. (\ref{Heegen}) have to be used instead. A few examples
of spectra for $d_i^md_j^m\rightleftharpoons d_i^{m+1}d_j^{m-1}$ charge
excitations at transition metal ions are shown in Fig. \ref{fig:levels}.
As a universal feature, the high-spin excitation is found at energy
$U-3J_H$ in all cases, provided that $J_H$ is understood as Hund's
exchange for that partly filled manifold ($e_g$ or $t_{2g}$) of
degenerate $d$ orbitals which participate in charge excitations. The
structure of the excited states depends on the partly occupied orbitals
\cite{noteabc} and on the actual valence $m$ --- the distance between
the high-spin and low-spin excitations increases with the number of
electrons for $m\leq 5$ (holes for $m>5$).

At orbital degeneracy the superexchange which connects ions at sites
$i$ and $j$ along the bond $\langle ij\rangle$ involves orbital
operators which depend on the bond direction. Therefore, we introduce
the index $\gamma=a,b,c$ to label the three {\it a priori\/} equivalent
directions in a cubic crystal. In order to analyze the consequences of
each individual charge excitation $n$ that contributes to the
superexchange in a given transition metal compound with degenerate $d$
orbitals, we shall use below a general way of writing the effective
low-energy Hamiltonian as a superposition of such individual terms on
each bond $\langle ij\rangle$,
\begin{equation}
\label{HJ}
{\cal H}_U = \sum_n\sum_{\langle ij\rangle\parallel\gamma}
             H_n^{(\gamma)}(ij),
\end{equation}
with the energy unit [absorbed in individual $H_n^{(\gamma)}(ij)$ terms]
given by the superexchange constant,
\begin{equation}
\label{J}
J = \frac{4t^2}{U}.
\end{equation}
It follows from $d-d$ charge excitations with an effective hopping
element $t$ between transition metal ions, and is the same as that
obtained in a Mott insulator with nondegenerate orbitals in the regime
of $U\gg t$.\cite{And59} While $U$ is the uniquely defined on-site
{\it intraorbital Coulomb element\/} (\ref{U}), increasing upon going
from Ti to Cu along the transition metal series,\cite{vdM88,Miz96} the
definition of the hopping $t$ between two nearest neighbor transition
metal ions depends on the system.\cite{Zaa93} If degenerate $e_g$
orbitals are involved, it is the effective $(dd\sigma)$ hopping element
for a $\sigma$-bond which involves $p_{\sigma}$ orbitals on the
intervening ligand ion (e.g. for the hopping
between two directional $3z^2-r^2$ states along the $c$ axis),
while for the systems with degenerate $t_{2g}$ orbitals it stands for
the effective $(dd\pi)$ hopping element due to $\pi$ bonds which involve
$p_{\pi}$ orbitals on the ligand ion.

In the superexchange Hamiltonian Eq. (\ref{HJ}) the contributions
which originate from all possible virtual excitations
$d_i^md_j^m\rightleftharpoons d_i^{m+1}d_j^{m-1}$ just add up to the
total superexchange interaction, in which the individual terms cannot be
distinguished. Yet each of these excitations involves a different state
in the multiplet structure of at least one of the transition metal ions,
i.e., either in the $d^{m+1}$ or in the $d^{m-1}$ configuration or in
both, depending on the actual process and on the value of $m$.
As pointed out elsewhere,\cite{Kha04} the same charge excitations
contribute to the optical conductivity, and here they appear at distinct
energies, thus revealing the multiplet structure of the excited
transition metal ions. Moreover, they convey a rich and characteristic
{\it temperature dependence\/} to the optical spectrum, determined by
the temperature variation of the spin-spin and orbital-orbital
correlations. We emphasize that it is therefore important to
analyze the various multiplet excitations separately, as they depend on
these correlations in a different way, and will also contribute to a
{\it quite different temperature dependence\/},
as we show in this paper on several examples.

As we will see in more detail below, the generic structure of each such
individual contribution is for a bond $\langle ij \rangle$ given by
\begin{eqnarray}
\label{Hngamma}
H_n^{(\gamma)}(ij)
   &=& ( a_n+b_n \: {\vec S}_i\cdot{\vec S}_j ) \:
         Q_n^{(\gamma)} ( {\vec \tau}_i, {\vec \tau}_j )
   \nonumber \\
   &=& a_n Q_n^{(\gamma)} ( {\vec \tau}_i, {\vec \tau}_j )
  \nonumber \\
  && +b_n \: Q_n^{(\gamma)} ( {\vec \tau}_i, {\vec \tau}_j ) \:
     {\vec S}_i\cdot{\vec S}_j ,
\end{eqnarray}
where the orbital dependence of the superexchange is described by means
of orbital projection operators $Q_n^{(\gamma)}$ which are expressed in
terms of components of orbital pseudospin $T=1/2$ operators at sites $i$
and $j$. The coefficients $a_n$ and $b_n$, which measure the strength of
the purely orbital part and of the spin-and-orbital part of the 
superexchange, respectively, follow from second-order perturbation theory
involving the charge excitation $n$. In the present case of perovskites,
where the bond between two transition metal ions through the ligand ion
(F or O) connecting them is close to linear (180$^{\circ}$),
the coefficients $a_n$ and $b_n$ are of similar magnitude (in contrast to
the situation in layered compounds like LiNiO$_2$ with 90$^{\circ}$ bonds
where the purely orbital interaction is stronger by an order of magnitude
than the spin-and-orbital interaction\cite{Mos02}).

Here we consider systems having cubic symmetry at high temperature. Yet
at low temperature this symmetry is frequently spontaneously broken ---
usually driven by the joint effect of (i) the orbital part of the
superexchange interaction, and (ii) the JT coupling of the same
degenerate (and therefore JT active) $3d$ orbitals to lattice modes.
The result is the simultaneous onset of a macroscopic lattice distortion
and of OO, i.e., a cooperative JT effect. At temperatures well below the
transition temperature $T_s$ of this combined structural and orbital
phase transition, the OO is effectively frozen. The remaining
superexchange interactions between the spins may then be obtained by
replacing the orbital projection operators in Eq. (\ref{HJ}) by their
expectation values,
\begin{equation}
\label{Qav}
Q_n^{(\gamma)} ( {\vec \tau}_i, {\vec \tau}_j ) \longrightarrow
\langle Q_n^{(\gamma)} ( {\vec \tau}_i, {\vec \tau}_j ) \rangle
= Q_n^{(\gamma)} (\langle {\vec \tau}_i \rangle ,
                \langle {\vec \tau}_j \rangle )  .
\end{equation}
Obviously, this leads to anisotropic magnetic interactions,
\begin{equation}
\label{HspinSE}
H_s = J \sum_n\sum_{\langle ij\rangle\parallel\gamma}
 b_n \: \langle Q_n^{(\gamma)} \rangle
 \: {\vec S}_i\cdot{\vec S}_j  ,
\end{equation}
which will in general induce a further magnetic phase transition at
lower temperature. It is noteworthy that in this situation the spin
degrees of freedom get decoupled from the orbital degrees of freedom,
although the purely orbital ($a_n$) and spin-and-orbital ($b_n$)
superexchange terms are of similar strength. Responsible for this
behavior is the JT contribution to the structural phase transition,
which enhances $T_s$ above the value it would have if the transition 
were driven by orbital superexchange alone.

Starting from the microscopic spin-orbital superexchange models, we will
analyze the effective spin models which arise after such a symmetry
breaking at low temperature. Rewritten from Eq. (\ref{HspinSE}), they
are of the generic form
\begin{equation}
\label{Hs}
    H_s =J_c\sum_{\langle ij\rangle_c}   {\vec S}_i\cdot{\vec S}_j
     +J_{ab}\sum_{\langle ij\rangle_{ab}}{\vec S}_i\cdot{\vec S}_j,
\end{equation}
with two different effective magnetic exchange interactions: $J_c$ along
the $c$ axis, and $J_{ab}$ within the $ab$ planes. The latter $J_{ab}$
interactions could in principle still take two different values in case
of inequivalent lattice distortions (caused, e.g., by octahedra tilting
or pressure effects) making the $a$ and $b$ axes inequivalent, but we
intend to limit the present analysis to idealized structures with these
two axes being equivalent. We shall show that the spin-spin correlations
along the $c$ axis and within the $ab$ planes,
\begin{equation}
\label{spins}
   s_c=\langle\vec{S}_i\cdot\vec{S}_j\rangle_{c},   \hskip .7cm
s_{ab}=\langle\vec{S}_i\cdot\vec{S}_j\rangle_{ab},
\end{equation}
next to the orbital correlations, play an important role in the
intensity distribution in optical spectroscopy.

The spectral weight in the optical spectroscopy is determined by the
kinetic energy,\cite{Mal77} and reflects the onset of magnetic order
\cite{Cha00,Mil00} and/or orbital order.\cite{Mac99} As shown by Ahn and
Millis,\cite{Mil00} in the weak coupling regime one can analyze the
total spectral weight in optical absorption using the Hartree-Fock
approximation for the relevant tight-binding Hamiltonian.
In a correlated insulator the electrons are
almost localized and the only kinetic energy which is left is associated
with the same virtual charge excitations that contribute also to the
superexchange. Therefore, we will discuss here the individual kinetic
energy terms $K_n^{(\gamma)}$, which can be determined from the
superexchange (\ref{HJ}) using the Hellman-Feynman theorem,\cite{Bae86}
\begin{equation}
\label{hefa}
K_n^{(\gamma)}=-2\big\langle H_n^{(\gamma)}(ij)\big\rangle.
\end{equation}
For convenience, we define the $K_n^{(\gamma)}$ as positive
quantities. Making use of the generic form of the superexchange
contribution $H_n^{(\gamma)}(ij)$ given by Eq. (\ref{Hngamma}),
and assuming as above that spin and orbital degrees of freedom are
decoupled in the temperature range of interest, we obtain
\begin{equation}
\label{Kngamma}
K_n^{(\gamma)}=-2 J
 \big(a_n+b_n\:\langle{\vec S}_i\cdot{\vec S}_j\rangle_{\gamma}\big) \:
 \big\langle Q_n^{(\gamma)} ({\vec \tau}_i,{\vec \tau}_j)\big\rangle .
\end{equation}
Each term $K_n^{(\gamma)}$ (\ref{hefa}) originates from a given charge
excitation $n$ for a bond $\langle ij\rangle$ along axis $\gamma$.
These terms are directly related to the {\it partial optical sum rule\/}
for individual Hubbard bands, which reads\cite{Kha04}
\begin{equation}
\label{opsa}
\frac{a_0\hbar^2}{e^2}\int_0^{\infty}\sigma_n^{(\gamma)}(\omega)d\omega=
\frac{\pi}{2}K_n^{(\gamma)},
\end{equation}
where $\sigma_n^{(\gamma)}(\omega)$ is the contribution of band $n$ to
the optical conductivity for polarization along the $\gamma$ axis,
$a_0$ is the distance between transition metal ions, and the
tight-binding model with nearest neighbor hopping is implied.
Comparison with Eq. (\ref{Kngamma}) shows that the intensity of each
band is indeed determined by the underlying OO together with the
spin-spin correlation along the direction corresponding to the
polarization.

One has to distinguish the above partial sum rule (\ref{opsa}) from the
sum rule for the total spectral weight in the optical spectroscopy
for polarization along a cubic direction $\gamma$, involving
\begin{equation}
\label{opsatot}
K^{(\gamma)}=-2J\sum_n\big\langle H_n^{(\gamma)}(ij)\big\rangle,
\end{equation}
which stands for the total intensity in the optical excitations (due to
$d-d$ excitations). This quantity is of less interest here as it has
a much weaker temperature dependence and does not allow for a direct
insight into the nature of the electronic structure.
In addition, it might be also more difficult to resolve from experiment.

When the low-energy excitations are of CT type, two holes could also be
created within a $2p$ orbital on a ligand (oxygen or fluorine) ion in
between two transition metal ions, described by
$d_i^mp^6d_j^m\rightleftharpoons d_i^{m+1}p^4d_j^{m+1}$ processes ---
these CT contributions lead to additional superexchange contributions,
called below {\it $\Delta$ terms\/}.
While the latter terms can be safely neglected in Mott-Hubbard systems,
they substantially modify the superexchange in CT insulators, and may
even represent there the dominating contribution.\cite{Zaa88,Mos04} Below
we will analyze them in two situations which involve $e_g$ degrees of
freedom, viz. in the cubic copper fluoride KCuF$_3$ (Sec. III), and in
the cubic manganite LaMnO$_3$ (Sec. IV), and we will show that in
KCuF$_3$ they represent an essential part of the superexchange.

\section{Copper fluoride perovskite: KC\lowercase{u}F$_3$}
\label{sec:d9}

\subsection{Superexchange Hamiltonian}
\label{d9:se}

The simplest spin-orbital models are obtained when transition metal
ions are occupied by either one electron ($m=1$), or by nine electrons
($m=9$); in these cases the Coulomb interactions (\ref{Hee}) contribute
only in the excited state (in the $d^2$ or the $d^8$ configuration)
after a charge excitation
$d_i^md_j^m\rightleftharpoons d_i^{m+1}d_j^{m-1}$
between two neighboring ions. Here we start with the case of a single
hole in the $d$ shell, as realized for the Cu$^{2+}$ ions in KCuF$_3$
with the $d^9$ configuration ($m=9$). Due to the splitting of the $3d$
states in the octahedral field within the CuF$_6$ octahedra, the hole
at each magnetic Cu$^{2+}$ ion occupies one of the $e_g$ orbitals.
The superexchange coupling (\ref{HJ}) is usually analyzed in terms of
$e_g$ holes in this case,\cite{Kug82} and this has become a textbook
example of spin-orbital physics by now.\cite{Faz99,Mae04}

Orbital order occurs in KCuF$_3$ below the structural transition at
$T_s\sim 800$ K. At $T<T_s$ the structure is tetragonal, with longer
Cu$-$Cu distances within the $ab$ planes ($d_{ab}=8.28$ \AA) than along
the $c$ axis ($d_c=7.85$ \AA),\cite{Kad67} which favors strong AF
interactions along the $c$ axis. Below the magnetic transition at
$T_N\simeq 38$ K, long-range magnetic order of $A$-type sets in,
\cite{Sat80,Pao02} and the ordered moment is $\mu_0=0.48\mu_B$.
\cite{Hut69}

The superexchange between the Cu$^{2+}$ ions in KCuF$_3$,
\begin{equation}
\label{som9}
{\cal H}(d^9)= {\cal H}_U(d^9)+{\cal H}_{\Delta}(d^9),
\end{equation}
consists of two terms: the $U$ term ${\cal H}_U$ (\ref{HJ}), and the CT
term ${\cal H}_{\Delta}$. First we introduce the $U$ term
${\cal H}_U(d^9)$ following the general approach of Sec.
\ref{sec:general}. It originates from three different excitations,
leading to an intermediate $d^8$ configuration at an excited Cu$^{3+}$
ion. Using the model Hamiltonian (\ref{Hee}) to describe the Coulomb
interactions between the $e_g$ electrons, one finds an equidistant
excitation spectrum of $^3\!A_2$, $^1E$ ($^1E_{\theta}$ and
$^1E_{\epsilon}$) and $^1\!A_1$ states, with energies:\cite{Gri71,Ole00}
$U-3J_H$, $U-J_H$ and $U+J_H$, as shown in Fig. \ref{fig:levels}(a).
This excitation spectrum is exact, and the element $J_H$ for a pair of
$e_g$ electrons is given by the Racah parameters $B$ and $C$
(see Table I):
\begin{equation}
\label{JHe}
J_H=  4B+ C.
\end{equation}
This definition of $J_H$ will be used for two systems with $e_g$ orbital
degrees of freedom: for the copper fluoride KCuF$_3$ (considered here),
and for the manganite LaMnO$_3$ (in Sec. \ref{sec:d4}).

In what follows, we will parametrize the multiplet structure of the
different transition metal ions by the ratio of the Hund's element
$J_H$ and the intraorbital Coulomb element $U$,
\begin{equation}
\label{eta}
\eta=\frac{J_H}{U}.
\end{equation}
Using Eqs. (\ref{HJ}) and (\ref{eta}), one finds for each bond
$\langle ij\rangle$ along a $\gamma$ axis ($\gamma=a,b,c$) four
contributions:\cite{Ole00}
\begin{eqnarray}
\label{som91}
H_1^{(\gamma)}\!\!&=&-\frac{J}{2}r_1
    \Big(\vec{S}_i\!\cdot\!\vec{S}_j+\frac{3}{4}\Big)
    \Big(\frac{1}{4}-\!\tau_i^{(\gamma)}\tau_j^{(\gamma)}\Big),   \\
\label{som92}
H_2^{(\gamma)}\!\!&=& \!\frac{J}{2}r_2
    \Big(\vec{S}_i\!\cdot\!\vec{S}_j-\frac{1}{4}\Big)
    \Big(\frac{1}{4}-\!\tau_i^{(\gamma)}\tau_j^{(\gamma)}\Big),   \\
\label{som93}
H_3^{(\gamma)}\!\!&=& \!\frac{J}{2}r_3
    \Big(\vec{S}_i\!\cdot\!\vec{S}_j-\frac{1}{4}\Big)\!
    \Big(\frac{1}{2}-\!\tau_i^{(\gamma)}\Big)\!
    \Big(\frac{1}{2}-\!\tau_j^{(\gamma)}\Big),                    \\
\label{som94}
H_4^{(\gamma)}\!\!&=& \!\frac{J}{2}r_4
    \Big(\vec{S}_i\!\cdot\!\vec{S}_j-\frac{1}{4}\Big)\!
    \Big(\frac{1}{2}-\!\tau_i^{(\gamma)}\Big)\!
    \Big(\frac{1}{2}-\!\tau_j^{(\gamma)}\Big),
\end{eqnarray}
with coupled spin and orbital operators. The coefficients,
\begin{equation}
\label{ri}
r_1=\frac{1}{1-3\eta}, \hskip .2cm
r_2=r_3=\frac{1}{1-\eta}, \hskip .2cm
r_4=\frac{1}{1+\eta},
\end{equation}
follow from the above multiplet structure of $d^8$ ions.\cite{notejt}

As explained below, it is straightforward to understand the generic
structure of the superexchange term ${\cal H}_U$, given by Eqs.
(\ref{som91})--(\ref{som94}). Here $\vec{S}_i$ is a spin $S=1/2$
operator, and
\begin{equation}
\label{spinpro}
P_{1}(ij)=\vec{S}_i\!\cdot\!\vec{S}_j+\frac{3}{4}, \hskip .7cm
P_{0}(ij)=\frac{1}{4}-\vec{S}_i\!\cdot\!\vec{S}_j,
\end{equation}
are the spin projection operators on triplet ($S=1$) and singlet
($S=0$) spin states for a bond $\langle ij\rangle$, so one recognizes
the high-spin term (\ref{som91}) and three low-spin terms
(\ref{som92})--(\ref{som94}), respectively. The spin operators in Eqs.
(\ref{som91})--(\ref{som94}) are accompanied by orbital pseudospin
operators $\propto\tau_i^{(\gamma)}$, which select the type of orbitals
{\it occupied by holes\/} at sites $i$ and $j$, and simultaneously
dictate the allowed excited states.

The orbital operators $\tau_i^{(\gamma)}$ depend on the direction of a
considered bond $\langle ij\rangle$, and are given by
\begin{equation}
\label{orbop}
\tau^{(ab)}_i=-\frac{1}{4}\big(\sigma^z_i\mp\sqrt{3}\sigma^x_i\big),
\hskip .7cm
\tau^{(c)}_i=\frac{1}{2}\sigma^z_i,
\end{equation}
where $\sigma^z_i$ and $\sigma^x_i$ are Pauli matrices acting on the
orbital pseudospins and the signs $\pm$ in $\tau^{(ab)}_i$ correspond
to $a$ and $b$ axis, respectively. With the help of $\tau^{(\gamma)}_i$
one defines next the projection operators in the orbital subspace,
\begin{equation}
\label{orbpro}
{\cal Q}_{i\xi}^{(\gamma)}  =\frac{1}{2}+\tau_i^{(\gamma)},
\hskip .7cm
{\cal Q}_{i\zeta}^{(\gamma)}=\frac{1}{2}-\tau_i^{(\gamma)}.
\hskip .7cm
\end{equation}
For a given cubic axis $\gamma$ they project (at site $i$) either on the
planar orbital $|\xi \rangle$ in the plane perpendicular to the $\gamma$
axis, or on the orthogonal directional orbital $|\zeta\rangle$ along
this axis. For instance, in the case where $\gamma$ is the $c$ axis,
they project on the $x^2-y^2$ orbital in the $ab$ plane, and on the
directional $3z^2-r^2$ orbital along the $c$ axis.

Using the projection operators (\ref{orbpro}), the orbital
dependence in Eqs. (\ref{som91})-(\ref{som94}) becomes transparent.
First of all, $(\frac{1}{4}-\tau_i^{(\gamma)}\tau_j^{(\gamma)})$ in Eq.
(\ref{som91}) accompanies the high-spin $^3\!A_2$ excitation
as this state may occur only when a pair of orthogonal orbitals is
occupied at sites $i$ and $j$, described formally by a
superposition of two possibilities, $\frac{1}{2}
({\cal Q}_{i\zeta}{\cal Q}_{j\xi}+{\cal Q}_{i\xi}{\cal Q}_{j\zeta})$.
In contrast, the operator ${\cal Q}_{i\zeta}{\cal Q}_{j\zeta}$
selects two orbitals oriented along the bond $\langle ij\rangle$ for
the high-energy low-spin $^1\!A_1$ excitation, see Eq. (\ref{som94}).
Finally, the second $H_2^{(\gamma)}$ (\ref{som92}) and third
$H_3^{(\gamma)}$ (\ref{som93}) term correspond to the doubly degenerate
low-spin $^1E$ state which consists of two singlet excitations:
(i) an interorbital singlet with two different orbitals occupied
($^1E_{\epsilon}$), and
(ii) a double occupancy within a directional orbital at either site
($^1E_{\theta}$) ---
these two excitations have thus quite different orbital dependences,
identical with those of the $^3\!A_2$ and the $^1\!A_1$ excitation,
respectively.
The sum over all the terms $H_n^{(\gamma)}$, with $n=1,\cdots,4$, gives
the simplest version of the spin-orbital model for the cubic copper
fluoride KCuF$_3$ with {\it degenerate\/} $e_g$ orbitals. Its derivation
and more details on the classical phase diagram can be found in Ref.
\onlinecite{Ole00}.

By considering further the electronic structure of KCuF$_3$, one can
elucidate the role played by the CT part ${\cal H}_{\Delta}$ in the
superexchange Hamiltonian (\ref{som9}). By analogy with the CuO$_2$
planes of the high-temperature superconductors, where the CT processes
give the dominating contribution to the AF superexchange interaction,
\cite{Zaa88,Zaa90} one expects that they are also important for
a cubic copper(II) fluoride and modify the superexchange in KCuF$_3$.
The CT term,
\begin{eqnarray}
\label{som9d}
{\cal H}_{\Delta}(d^9)&=& JR\sum_{\langle ij\rangle\parallel\gamma}
    \Big(\vec{S}_i\!\cdot\!\vec{S}_j-\frac{1}{4}\Big)
    \Big(\frac{1}{2}-\tau_i^{(\gamma)}\Big)
    \Big(\frac{1}{2}-\tau_j^{(\gamma)}\Big),           \nonumber \\
\end{eqnarray}
with the coefficient
\begin{equation}
\label{r}
R =\frac{2U}{2\Delta+U_p}
\end{equation}
resulting from the two-hole charge excitation at a common neighboring
$2p_{\sigma}$ orbital of a fluorine ion in between two copper ions,
in the process $d_i^9p_{\langle ij\rangle}^6d_j^9
\rightleftharpoons d_i^{10}p_{\langle ij\rangle}^4d_j^{10}$.
As a double hole excitation is generated at a single bonding orbital
$2p_{\sigma}$ within each Cu$-$F$-$Cu unit, this term is necessarily
AF. Two holes can move to fluorine from two neighboring Cu ions only if
both of them occupy a directional $e_g$ orbital $|\zeta\rangle$,
oriented along the considered bond (e.g., $3x^2-r^2$ orbitals along the
$a$ axis), being the simplest CT term discussed recently by Mostovoy and
Khomskii.\cite{Mos04} Therefore this process leads to the same orbital
dependence in Eq. (\ref{som9d}) as the low-spin $^1E_{\theta}$ and
$^1\!A_1$ excitations which involve double occupancies of directional
$|\zeta\rangle$ orbitals.

\subsection{Spin exchange constants and optical intensities}
\label{d9:jando}

A characteristic feature of spin-orbital superexchange models with $e_g$
orbital degrees of freedom is the presence of the purely orbital
interactions in Eqs. (\ref{som91})--(\ref{som94}) and (\ref{som9d}),
which favor particular type of occupied orbitals. LDA+U calculations
\cite{Lic95,Med02} have indicated that such purely electronic
interactions would already drive the instability towards the $C$-type OO
($C$-OO) phase, with {\it alternating} orbitals in the $ab$ planes, and
{\it repeated\/} orbitals along the $c$ axis, which induces FM spin
exchange in the $ab$ planes, and strong AF exchange between the planes.
Experimentally, this OO sets in below the structural transition at
$T_s\sim 800$ K,\cite{Pao02} i.e., at much higher temperature than the
characteristic energy scale of the magnetic excitations,\cite{Ten93}
suggesting that the JT effect plays an important role in this
instability. This observation is consistent with the large difference
between $T_s$ and the N\'eel temperature $T_N\sim 38$ K,
\cite{Hut69} the latter being controlled by the magnetic part of the
superexchange, and thus the orbital correlations decouple from the
spin-spin correlations. This motivates one to analyze the dependence of
the magnetic exchange interactions and of the optical spectral weights
on the type of OO stabilized below the structural transition.

\begin{figure}[t!]
\includegraphics[width=7.0cm]{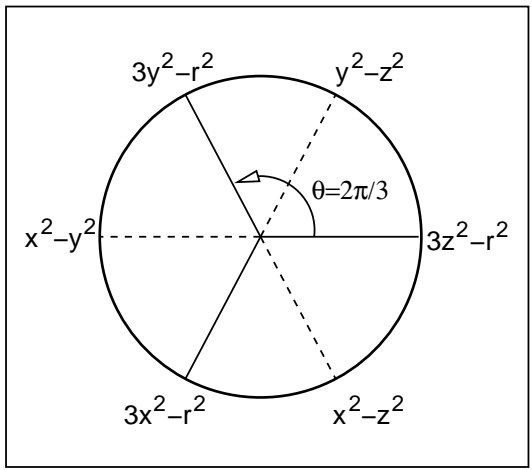}
\caption{
Schematic representation of $e_g$ orbitals as obtained for different
angle $\theta$ in Eq. (\protect\ref{mixing}). Pairs of orthogonal
orbitals, forming a basis in the orbital space, differ by angle
$\theta=\pi$. Directional $e_g$ orbitals $|\zeta\rangle$ along $a$ and
$b$ axes are obtained from the $\{|3z^2-r^2\rangle,|x^2-y^2\rangle\}$ 
basis (\ref{egbasis}) by the transformation (\ref{mixing}) with angle
$\theta=\pm 2\pi/3$. Dashed lines for $\theta=\pm\pi/3$ indicate the
possible OO of $|y^2-z^2\rangle/|x^2-z^2\rangle$ orbitals occupied by
holes in KCuF$_3$, suggested for $T<T_N$ in Ref.
\protect\onlinecite{Kug82}.
}
\label{fig:eg}
\end{figure}

Here we are interested in the low temperature range of $T<500$ K, so we
assume perfect OO given by a classical ansatz for the ground state,
\begin{equation}
\label{oo}
|\Phi_0\rangle=\prod_{i\in A}|\theta_A\rangle_i
               \prod_{j\in B}|\theta_B\rangle_j,
\end{equation}
with the orbital states, $|\theta_A\rangle_i$ and $|\theta_B\rangle_j$,
characterized by opposite angles ($\theta_A=-\theta_B$) and alternating
between two sublattices $A$ and $B$ in
the $ab$ planes. The orbital state at site $i$:
\begin{equation}
\label{mixing}
|\theta\rangle=\cos\Big(\frac{\theta}{2}\Big)|z\rangle
              +\sin\Big(\frac{\theta}{2}\Big)|x\rangle,
\end{equation}
is here parametrized by an angle $\theta$ which defines the
amplitudes of the orbital states
\begin{equation}
\label{egbasis}
|z\rangle\equiv (3z^2-r^2)/\sqrt{6}, \hskip .7cm
|x\rangle\equiv ( x^2-y^2)/\sqrt{2},
\end{equation}
being a local $e_g$ orbital basis at each site. This and other
equivalent orbital bases are shown schematically by pairs of solid and
dashed lines (corresponding to pairs of orbitals
$\{|\theta\rangle,|\theta+\pi\rangle\}$) in Fig. \ref{fig:eg}. The OO
state specified in Eq. (\ref{oo}) is thus defined by:
\begin{eqnarray}
\label{ood9}
|\theta_A\rangle_i&=&\cos\Big(\frac{\theta}{2}\Big)|z\rangle_i
                    +\sin\Big(\frac{\theta}{2}\Big)|x\rangle_i,\nonumber\\
|\theta_B\rangle_j&=&\cos\Big(\frac{\theta}{2}\Big)|z\rangle_j
                    -\sin\Big(\frac{\theta}{2}\Big)|x\rangle_j,
\end{eqnarray}
with $\theta_A=\theta$ and $\theta_B=-\theta$.

The magnetic superexchange constants $J_{ab}$ and $J_{c}$ in the
effective spin model (\ref{Hs}) are obtained by
decoupling spin and orbital variables and next averaging the orbital
operators in the spin-orbital model (\ref{som9}) over the classical
state $|\Phi_0\rangle$ as given by Eq. (\ref{oo}). The relevant averages
are given in Table \ref{tab:eg}, and they lead to compact expressions
for the superexchange constants in Eq. (\ref{Hs}),
\begin{eqnarray}
\label{jc9}
J_c &=&\frac{1}{8}J\Big\{-r_1\sin^2\theta+(r_2+r_3)(1+\cos\theta)
                                                           \nonumber \\
    & &\hskip .5cm +(r_4+2R)(1+\cos\theta)^2\Big\},                  \\
\label{jab9}
J_{ab}&=&\frac{1}{8}J\Big\{-r_1\Big(\frac{3}{4}+\sin^2\theta\Big)
    +(r_2+r_3)\Big(1-\frac{1}{2}\cos\theta\Big)            \nonumber \\
    & &\hskip .5cm +(r_4+2R)\Big(\frac{1}{2}-\cos\theta\Big)^2\Big\},
\end{eqnarray}
which depend on three parameters, viz. $J$ (\ref{J}), $\eta$ (\ref{eta})
and $R$ (\ref{r}), and on the OO (\ref{ood9}) specified by the orbital
angle $\theta$. It is clear that the FM term $\propto r_1$ competes with
all the other AF low-spin terms. Nevertheless, in the $ab$ planes, where
the occupied $e_g$ orbitals alternate, the large FM contribution (when
$\sin^2\theta\sim 1$) still makes the magnetic superexchange $J_{ab}$
weakly FM ($J_{ab}\lesssim 0$), while the stronger AF superexchange
along the $c$ axis ($J_{c}\gg |J_{ab}|$) favors quasi one-dimensional
(1D) spin fluctuations.

\begin{table}[b!]
\caption{
Averages of the orbital projection operators standing in the
spin-orbital interactions in Eqs. (\ref{som91})--(\ref{som94}) for the
$C$-type (or $G$-type) OO of occupied $e_g$ orbitals which alternate in
$ab$ planes, as given by Eqs. (\ref{ood9}). Nonequivalent cubic
directions are labelled by $\gamma=ab,c$.
}
\vskip .3cm
\begin{ruledtabular}
\begin{tabular}{ccc}
        $\gamma$ & $ab$ &  $c$ \cr
\colrule
$\big\langle\big(\frac{1}{2}-\tau_i^{(\gamma)}\big)
            \big(\frac{1}{2}-\tau_j^{(\gamma)}\big)\big\rangle$ &
	$\frac{1}{4}\big(\frac{1}{2}-\cos\theta\big)^2$         &
	$\frac{1}{4}\big(1+\cos\theta\big)^2$                   \cr
$\big\langle\frac{1}{4}-\tau_i^{(\gamma)}\tau_j^{(\gamma)}\big\rangle$ &
        $\frac{1}{4}\big(\frac{3}{4}+\sin^2\theta\big)$         &
	$\frac{1}{4}\sin^2\theta$                               \cr
$\big\langle\big(\frac{1}{2}+\tau_i^{(\gamma)}\big)
            \big(\frac{1}{2}+\tau_j^{(\gamma)}\big)\big\rangle$ &
	$\frac{1}{4}\big(\frac{1}{2}+\cos\theta\big)^2$         &
	$\frac{1}{4}\big(1-\cos\theta\big)^2$                   \cr
\end{tabular}
\end{ruledtabular}
\label{tab:eg}
\end{table}

By considering the superexchange model, one can derive as well the pure
orbital interactions which stabilize the OO. The superexchange
interactions are anisotropic below the structural transition at $T_s$.
In contrast, at sufficiently high temperature $T>T_s$, when also spin
correlations may be neglected, one finds isotropic orbital interactions,
\begin{equation}
\label{kc9}
J_c^{\tau} =J_{ab}^{\rm orb}=\frac{1}{8}J\big(3r_1-r_4-2R\big),
\end{equation}
which multiply $\tau_i^{(\gamma)}\tau_j^{(\gamma)}$ for each bond,
contributing to an orbital instability towards alternating $G$-type OO,
while actually $C$-type OO is observed below $T_s$. It is thus clear
from experiment that this instability cannot be of purely electronic
origin, and that, similarly to what is the case in LaMnO$_3$,
\cite{Mil96} it is supported by the lattice. In fact, although it has
been argued that the OO is caused primarily by the superexchange,
\cite{Lic95} the electronic interactions (\ref{kc9}) predict that
$T_s\sim 0.1J$, and not $T_s\sim J$, as observed. Note also that as
soon as the AF spin correlations develop along the $c$ axis, one finds
anisotropic orbital interactions, $J_{ab}^{\tau}>J_c^{\tau}$,
which amplifies the ongoing symmetry breaking in the tetragonal phase.

The spectral weights of the optical subbands also follow from the
superexchange processes, and are determined from the effective
Hamiltonian (\ref{som9}) by the general relations given by Eqs.
(\ref{hefa}) and (\ref{opsa}). Following the excitation spectrum of
Fig. \ref{fig:levels}, one finds optical absorption at three different
energies (the degeneracy of the $^1E$ state is not removed), so we
label the respective kinetic energy contributions $K_n^{(\gamma)}$ by
$n=1,2,3$. They are determined at low temperature $T\lesssim 500$ K by
rigid $C$-type OO (\ref{ood9}), with the classical averages of the
orbital operators given in Table \ref{tab:eg}. So one finds for
polarization along the $c$ axis
\begin{eqnarray}
\label{wc91}
K_1^{(c)}&=&\frac{1}{4}Jr_1\Big(\frac{3}{4}+s_c\Big)\sin^2\theta,   \\
\label{wc92}
K_2^{(c)}&=&\frac{1}{4}J(r_2+r_3)\Big(\frac{1}{4}-s_c\Big)(1+\cos\theta),
                                                                    \\
\label{wc93}
K_3^{(c)}&=&\frac{1}{4}Jr_4\Big(\frac{1}{4}-s_c\Big)(1+\cos\theta)^2,
\end{eqnarray}
and for polarization in the $ab$ plane
\begin{eqnarray}
\label{wab91}
\!\!K_1^{(ab)}&\!=\!&\frac{1}{4}Jr_1\Big(\frac{3}{4}+s_{ab})
              \Big(\frac{3}{4}+\sin^2\theta\Big),                    \\
\label{wab92}
\!\!K_2^{(ab)}&\!=\!&\frac{1}{4}J(r_2\!+\!r_3)
  \Big(\frac{1}{4}-\!s_{ab}\Big)\Big(1\!-\frac{1}{2}\cos\theta\Big), \\
\label{wab93}
\!\!K_3^{(ab)}&\!=\!&\frac{1}{4}Jr_4\Big(\frac{1}{4}-s_{ab}\Big)
              \Big(\frac{1}{2}-\cos\theta\Big)^2.
\end{eqnarray}
Similar to the exchange constants $J_{ab}$ and $J_c$, the kinetic
energies depend on the multiplet structure described by two parameters,
viz. $J$ (\ref{J}), $\eta$ (\ref{eta}), and
on the OO (\ref{ood9}) specified by the angle $\theta$.
Note that they depend on these parameters also
indirectly, since the spin-spin correlations are governed by $J_{ab}$
and $J_c$ as well. We analyze this dependence in Sec. \ref{d9:kcufopt}.

\subsection{Magnetic interactions in KCuF$_3$}
\label{d9:kcufmag}

In order to apply the above classical theory to KCuF$_3$ we need to
determine the microscopic parameters which decide about the
superexchange constants, given by Eqs. (\ref{jc9}) and (\ref{jab9}). In
principle, if the optical data would also be available, with this
experimental input one would be able to fix the values of the relevant
parameters $J$ and $\eta$, and the orbital angle $\theta$. Having only
magnetic measurements, we give here an example of another approach which
starts from the microscopic parameters for the local Coulomb interaction
and Hund's element suggested by the electronic structure calculations
performed within the LDA+U method:\cite{Lic95,Med02} $U=7.5$, $J_H=0.9$
eV --- they lead to $\eta=0.12$. Note that these parameters are somewhat
smaller than the values $U=8.96$ and $J_H=1.19$ eV deduced for Cu$^{2+}$
ions in the CuO$_2$ planes of the high-temperature superconductors by
Grant and McMahan using the fixed charge method,\cite{Gra92} but we
believe that they reflect better the partly screened interactions within
CuF$_6$ units. We are not aware of any estimation of the remaining
microscopic parameters until now, but taking into account the expected
contraction of the $2p$ wavefunctions by going from O$^{2-}$ to F$^{-}$
ions, we argue that $t_{pd}$ is reduced, while $\Delta$ and $U_p$ could
be similar to their respective values for CuO$_2$ planes.\cite{Gra92}
Therefore, it is reasonable to adopt: $t_{pd}=1.0$, $\Delta=4.0$, and
$U_p=4.5$ eV. Note that although the values of $t_{pd}=1.0$, $\Delta$
and $U_p$ could not be really estimated, in the present approach they
are not independent parameters; also only a linear combination of
$\Delta$ and $U_p$ enters Eq. (\ref{r}), so a change in the value of
$U_p$ could to some extent compensate a modified value of $\Delta$.
The present parameters lead to $J\simeq 33.3$ meV and $R=1.2$.

\begin{figure}[t!]
\includegraphics[width=8.0cm]{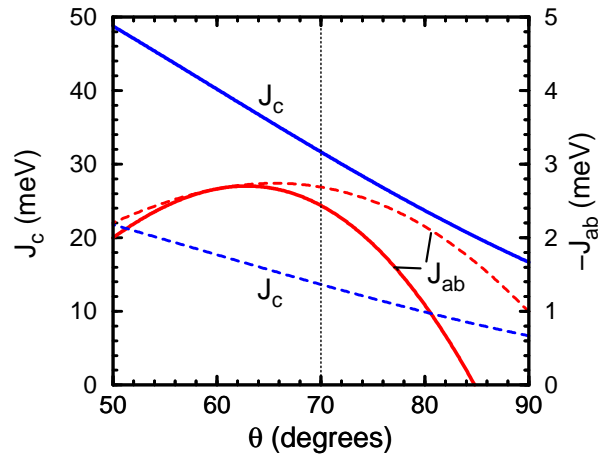}
\caption{(Color online)
Exchange interactions (\ref{jc9}): $J_c$ (\ref{jc9}) and $J_{ab}$
(\ref{jab9}) for the copper fluoride model as functions of the orbital
angle $\theta$ which describes the OO [see Eq. (\protect\ref{oo})].
The dashed lines show the $U$ term alone, while the solid lines include
the CT contributions as well. The OO induced by the JT distortions
($\theta\simeq 70^{\circ}$) is indicated by dotted line.
Parameters: $J\simeq 33.3$ meV, $\eta=0.12$, $R=1.2$.
}
\label{fig:j9}
\end{figure}

Consider now the OO of the occupied orbitals (by holes) in KCuF$_3$.
Recent resonant x-ray scattering experiments suggest that both
sublattices are equivalent, with $\theta_A=-\theta_B=\theta$ in Eq.
(\ref{oo}),\cite{Pao02} but the precise shape of the occupied orbitals
in KCuF$_3$ remains unresolved. There are different views concerning the
type of orbitals that participate in the OO state. On the one hand, it
is believed that the orbital angle $\theta$ should be close to the
angle $\theta_{\rm JT}\simeq 70^{\circ}$ ($\sim 0.39\pi$), as given by
$\cos\theta\simeq\frac{1}{3}$, which follows from the local lattice
distortions.\cite{Kad67} On the other hand, the electronic interactions
in the symmetry broken $A$-AF phase below $T_N$ would favor instead
alternating $(y^2-z^2)/(x^2-z^2)$ orbitals,\cite{Kug82}
with $\theta_{\rm SE}=60^{\circ}$. In reality, one expects rather a
certain compromise between the electronic interactions for finite spin
correlations and those induced by the lattice. Thus, in the present
study of the magnetic exchange constants and optical spectral weights
we shall consider a range of possible values of
$60^{\circ}<\theta<90^{\circ}$, focusing in particular on the above
values favored by the above individual terms in the effective
Hamiltonian.

First, we demonstrate that the model Eq. (\ref{som9}) is capable of
describing the experimentally observed exchange constants
$J_c^{\rm exp}\simeq 35$ meV and $J_{ab}^{\rm exp}\simeq -2$ meV.
\cite{Ten93} Remarkably, the value of $|J_{ab}^{\rm exp}|$ is smaller
by more than one order of magnitude than $J_c^{\rm exp}$, being some
challenge for the theoretical model. Consider first the values of $J_c$
and $J_{ab}$ for varying angle $\theta$ which tunes the OO (Fig.
\ref{fig:j9}). When only the $U$ part of the superexchange ${\cal H}_U$
is considered, one finds $J_c\sim 15$ meV and $J_{ab}\sim -2$ meV.
The FM term $\propto r_1$ gives the largest contribution for the $ab$
planes, which follows from the alternation of hole orbitals in the $ab$
planes, being close to the planar orbitals $(x^2-z^2)/(y^2-z^2)$
($\theta=60^{\circ}$) suggested early on by Kugel and Khomskii,
\cite{Kug82} but is partly compensated by the AF terms.
While $J_{ab}$ is only
weakly depending on $\theta$ near this type of OO, $J_c$ decreases
steadily with increasing $\theta$ (Fig. \ref{fig:j9}), as the overlap
between the orbitals occupied by holes
along the $c$ axis decreases when the amplitude of the $|z\rangle$
states is reduced. One finds that using only the $U$ term in the
superexchange, rather extreme parameters, such as $U<4$ eV and
$J_H\sim 0.3$ eV (with the present value of $t$), would have to be
assumed to reproduce the experimental values of $J_c$ and $J_{ab}$.

The CT term (\ref{som9d}) with $R=1.2$ enhances the AF interaction
$J_c$ by a factor larger than two but hardly changes $J_{ab}$ (see Fig.
\ref{fig:j9}). Only then $J_c$ comes close to the experimental value
$J_c^{\rm exp}\simeq 35$ meV.\cite{Ten93} As in the $U$ model (at
$R=0$), the value of $J_c$ decreases with increasing $\theta$, and there
is no serious difficulty to fit the parameters in order to obtain a
reasonable agreement with experiment, once the value of the orbital
angle $\theta$  would be known. As an illustrative example we show the
results obtained with the present parameters in Fig. \ref{fig:j9} ---
one finds $J_c\simeq 32$ meV for the OO which agrees with the lattice
distortions ($\theta_{\rm JT}\simeq 70^{\circ}$), and $J_c\simeq 40$ meV
for the Kugel-Khomskii $(x^2-z^2)/(y^2-z^2)$ orbitals
($\theta_{\rm SE}=60^{\circ}$). These results demonstrate that the CT
superexchange term plays an essential role in KCuF$_3$ and so this
system has to be classified as a {\it charge transfer insulator\/}.

Remarkably, the value of $J_{ab}$ is almost unaffected by the CT term
(Fig. \ref{fig:j9}). This is due to the alternating OO in the $ab$
planes, which makes the value of the orbital projection
$\big\langle\big(\frac{1}{2}-\tau_i^{(\gamma)}\big)
            \big(\frac{1}{2}-\tau_j^{(\gamma)}\big)\big\rangle$
in Eq. (\ref{som9d}) very small indeed in the physical range of $\theta$
(compare Table \ref{tab:eg}). In fact, for the alternating planar
$(x^2-z^2)/(y^2-z^2)$ orbitals one of the operators
$\{\big(\frac{1}{2}-\tau_i^{(\gamma)}\big),
   \big(\frac{1}{2}-\tau_j^{(\gamma)}\big)\}$ equals zero, and the CT
contribution vanishes, so one cannot reduce the value of $|J_{ab}|$ by
increasing the AF CT term that follows from ${\cal H}_{\Delta}$.

\begin{figure}[t!]
\includegraphics[width=7.7cm]{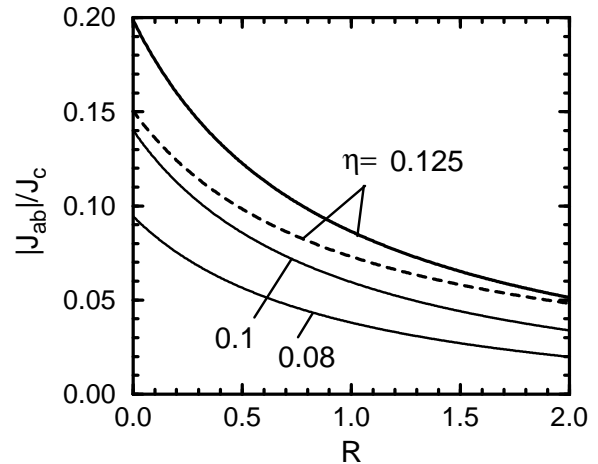}
\caption{
Ratio of the exchange interactions $|J_{ab}|/J_c$ in KCuF$_3$, given by
Eqs. (\ref{jc9}) and (\ref{jab9}), for increasing charge transfer
parameter $R$, obtained for a few values of $\eta$ and for the OO
induced by the JT effect ($\theta\simeq 70^{\circ}$, solid lines).
The dashed line shows $|J_{ab}|/J_c$ obtained for $\eta=0.12$ and for
alternating $(x^2-z^2)/(y^2-z^2)$ orbitals
($\theta=60^{\circ}$).
}
\label{fig:ratio}
\end{figure}

The strong anisotropy of the magnetic exchange interactions in KCuF$_3$
is well illustrated in Fig. \ref{fig:ratio} by the ratio $|J_{ab}|/J_c$,
being close to 0.07 for either the JT OO ($\theta\simeq 70^{\circ}$), or
for the OO suggested by the orbital superexchange at $T<T_N$
($\theta=60^{\circ}$). Note that for $R\sim 1$ the ratio $|J_{ab}|/J_c$
does not depend significantly on $\theta$ in the interesting range
between $\theta=70^{\circ}$ and $\theta=60^{\circ}$.

\subsection{Optical spectral weights for KCuF$_3$}
\label{d9:kcufopt}

Now we turn to the optical spectral weights (\ref{hefa}) and determine
the kinetic energies for the corresponding Hubbard subbands.
As discussed in Sec. \ref{sec:general}, they originate from different
multiplet excitations, and depend on the OO and on the spin-spin
correlations (\ref{spins}). Here we analyze in detail the spectral
weight distribution for polarization along the $c$ axis, where strong
exchange interaction $J_c$ controls the spin-spin correlations $s_c$
(\ref{spins}) which remain finite in a broad temperature regime.

Knowing that the interchain FM exchange coupling $J_{ab}$ is so weak,
we describe the temperature variation of the spin-spin correlations
$s_c=\langle\vec{S}_{i}\!\cdot\!\vec{S}_{i+1}\rangle_{c}$ employing the
Jordan-Wigner fermion representation\cite{Mat85} for a 1D spin chain.
One finds for perfect OO at temperature $T<T_s$
\begin{equation}
\label{scd9}
s_c= -\kappa (1+\kappa),
\end{equation}
where
\begin{eqnarray}
\label{kappad9}
\kappa&=&\frac{1}{N}\,\sum_k\;
         |\cos k|\,\tanh\Big(\frac{\varepsilon_k}{2 k_B T}\Big), \\
\label{epsid9}
\varepsilon_k&=&J_c(1+2\kappa)|\cos k|.
\end{eqnarray}
Here $\varepsilon_k$ is the 1D dispersion of pseudofermions. The
exchange interaction $J_c$ is constant as long as the orbitals remain
frozen, and sets the energy scale for the temperature variation of
$s_c$. Eqs. (\ref{kappad9}) and (\ref{epsid9}) were solved
self-consistently to obtain $s_c$ (\ref{scd9}) as a function of
temperature. In the limit $T\to 0$ one finds $\kappa=1/\pi$, and
$s_c=-(1+\pi)/\pi^2\simeq 0.42$. This value represents an excellent
analytic approximation to the exact result,
\begin{equation}
\label{bethe}
s_c^{\rm ex}=-(\log 2-\textstyle{\frac{1}{4}})\simeq -0.4431,
\end{equation}
obtained for the 1D AF Heisenberg chain from the Bethe ansatz.
\cite{Mat85}

\begin{figure}[t!]
\includegraphics[width=7.5cm]{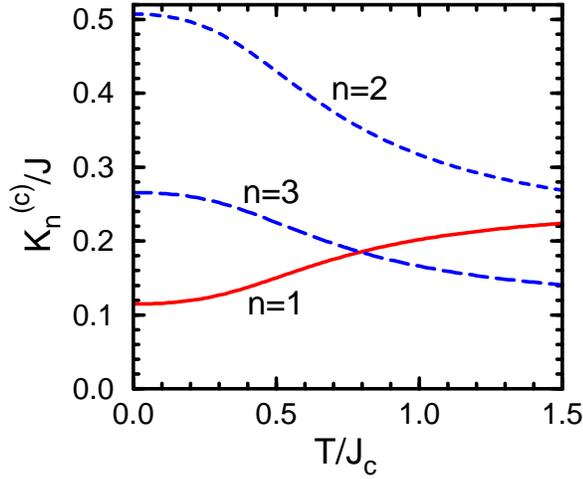}
\vskip -2mm
\caption{(Color online)
Kinetic energy terms per bond (\ref{hefa}), as obtained for the $c$
axis with the OO (\protect\ref{oo}) suggested by the JT distortions
($\theta\simeq 70^{\circ}$) in KCuF$_3$ (in units of $J$): high-spin
$K_1^{(c)}$ (solid line) and low-spin $K_{2,3}^{(c)}$ (dashed lines),
for increasing temperature $T/J_c$. Parameters as in Fig. \ref{fig:j9}.
}
\label{fig:sw9}
\end{figure}

\begin{table}[b!]
\caption{
Kinetic energies of the different Hubbard subbands ($K_n^{(\gamma)}$),
and total kinetic energies ($K^{(\gamma)}$) (in meV), as obtained
for KCuF$_3$ for two representative orbital states:
$\theta_{\rm JT}\simeq 70^{\circ}$ and $\theta_{\rm SE}=60^{\circ}$.
Parameters: $J\simeq 33.3$ meV, $\eta=0.12$.
}
\vskip .2cm
\begin{ruledtabular}
\begin{tabular}{cccccccc}
        & \multicolumn{3}{c}{$\theta_{\rm JT}\simeq 70^{\circ}$} & &
          \multicolumn{3}{c}{$\theta_{\rm SE}=60^{\circ}$}   \cr
     & $T=0$ & $40$ K & $300$ K & & $T=0$ & $40$ K & $300$ K \cr
\colrule
$K_1^{(c)}$  &  3.8 &  3.9 &  6.1 & &  3.2 &  3.3 &  5.2     \cr
$K_2^{(c)}$  & 16.9 & 16.8 & 11.9 & & 19.0 & 18.9 & 13.4     \cr
$K_3^{(c)}$  &  8.9 &  8.8 &  6.2 & & 11.2 & 11.1 &  7.9     \cr
$K^{(c)}$    & 29.6 & 29.5 & 24.2 & & 33.4 & 33.3 & 26.5     \cr
\colrule
$K_1^{(ab)}$ & 21.3 & 16.0 & 16.0 & & 19.5 & 14.6 & 14.6     \cr
$K_2^{(ab)}$ &  0.0 &  3.9 &  3.9 & &  0.0 &  3.6 &  3.6     \cr
$K_3^{(ab)}$ &  0.0 &  0.1 &  0.1 & &  0.0 &  0.0 &  0.0     \cr
$K^{(ab)}$   & 21.3 & 20.0 & 20.0 & & 19.5 & 18.2 & 18.2     \cr
\end{tabular}
\end{ruledtabular}
\label{tab:kcuf3}
\end{table}

The general theory presented in Sec. \ref{sec:general} makes a clear
prediction concerning the temperature dependence of the spectral weights
in optical absorption. First of all, a large anisotropy between the
polarization along the AF $c$ axis and the polarization in the (weakly
FM) $ab$ plane is expected when the AF (FM) spin-spin correlations along
the $c$ axis (within the $ab$ planes) develop. Indeed, using the
self-consistent solution of Eqs. (\ref{kappad9}) and (\ref{epsid9}), one
finds that the kinetic energy $K_1^{(c)}$ (which determines the spectral
weight of the high-spin excitation at energy $U-3J_H$) is rather low
(Fig. \ref{fig:sw9}). In contrast, the low-spin excitations $K_2^{(c)}$
and $K_3^{(c)}$ contribute with large spectral weights in the low
temperature regime, reflecting the AF correlations along the $c$ axis.

When the temperature increases, the spin-spin correlations $s_c$
gradually weaken and the kinetic energy redistributes --- both low-spin
terms $K_2^{(c)}$ and $K_3^{(c)}$ decrease, while the high-spin term
$K_1^{(c)}$ increases  as more high-spin excitations are then allowed
(Fig. \ref{fig:sw9}). For the considered OO given by
$\theta_{JT}\simeq 70^{\circ}$, the changes of the contributions at the
two lower energies which correspond to $n=1$ and $n=2$ are particularly
large between zero and room temperature (up to $k_BT/J_c\simeq 0.8$),
with the increase (decrease) of $K_1^{(c)}$ [$K_2^{(c)}$] by $\sim 60$
($\sim 33$) percent of the reference value at $T=0$. This leads to
rather similar values of all three contributions at room temperature.
This predicted behavior could be verified by future experiments.

The temperature variation of the spectral weights for $ab$
polarization is more difficult to predict as it involves weak $s_{ab}$
spin correlations which develop in the temperature range $T\sim T_N$,
and grow with increasing order parameter, $s_{ab}\propto
|\langle S^z\rangle|$, below $T_N$.\cite{Sch96} Assuming the classical
value for $s_{ab}=0.25$ at $T\to 0$, one finds that the kinetic energy
would come entirely from the high-spin optical excitations $K_1^{(ab)}$,
while the low-spin excitations would be fully suppressed (Table
\ref{tab:kcuf3}). Above $T_N$ the spin system is controlled by the
dominating AF exchange constant $J_c$, and $s_{ab}\simeq 0$. Even then
the high-spin excitations at low energy dominate and have large spectral
weight as a result of the persisting OO. Some decrease of $K_1^{(ab)}$
accompanied by the increase of $K_1^{(c)}$ with increasing temperature
makes the anisotropy between $K_1^{(c)}$ and $K_1^{(ab)}$ considerably
less pronounced, but this anisotropy of the spectra in the low energy
range remains still close to 3:1 even at room temperature (Table
\ref{tab:kcuf3}). Note also that the highest energy excitation for the
$ab$ polarization vanishes at $\theta_{\rm SE}=60^{\circ}$, and gives
a negligible contribution for the JT angle
$\theta_{\rm JT}\simeq 70^{\circ}$, due to the orbital correlations
within the $ab$ planes.

Finally, we would like to emphasize again that knowing only the exchange
constants $J_{ab}$ and $J_c$ in KCuF$_3$, one is not able to determine
all the microscopic parameters of the CT model. We emphasize that a
better understanding of the properties of KCuF$_3$ can be achieved only
by combining the results of magnetic and optical experiments, after the
latter experiments have been performed.

\section{Cubic manganite: L\lowercase{a}M\lowercase{n}O$_3$}
\label{sec:d4}

\subsection{Superexchange model}
\label{d4:se}

Although $e_g$ and $t_{2g}$ electrons behave quite differently in
LaMnO$_3$ and are frequently treated as two subsystems,
\cite{Dag02,Dag03} the neutron experiments\cite{Mou96} which measure
the spin waves in the $A$-AF phase below $T_N$ leave no doubt that an
adequate description of the magnetic properties requires a magnetic
Hamiltonian of the form given by Eq. (\ref{Hs}), describing
superexchange between total $S=2$ spins of the Mn$^{3+}$ ions. The
high-spin $^5E$ ground state at each Mn$^{3+}$ ion is stabilized by
the large
Hund's exchange $J_H$. The situation is here more complex than either
in KCuF$_3$ (Sec. \ref{sec:d9}) or in the $t_{2g}$ systems discussed
in the following Sections, however, as the superexchange terms between
Mn$^{3+}$ ions originate from various charge excitations
$d_i^4d_j^4\rightleftharpoons d_i^5d_j^3$, made either by $e_g$ or by
$t_{2g}$ electrons, leading to different excited states in the
intermediate $d^5$ configuration on a Mn$^{2+}(d_i^5)$ ion.
Such processes determine the $U$ term ${\cal H}_U(d^4)$ defined by Eq.
(\ref{HJ}), and were analyzed in detail in Ref. \onlinecite{Fei99},
and lead to the structure of ${\cal H}_U(d^4)$ given below. However,
the CT processes, ${\cal H}_{\Delta}(d^4)$, contribute
as well and the complete model for LaMnO$_3$ reads
\begin{equation}
\label{som4}
{\cal H}(d^4)= {\cal H}_U(d^4)+{\cal H}_{\Delta}(d^4).
\end{equation}
The superexchange constant $J$ is here defined again by Eq. (\ref{J}),
using an average hopping element along an effective $(dd\sigma)$ bond,
$t=t_{pd}^2/\bar{\Delta}$, where $\bar{\Delta}$ is an average CT
excitation energy, introduced below in Eq. (\ref{bard}).

First we analyze the structure of the $U$ term for LaMnO$_3$,
${\cal H}_U(d^4)$, due to excitations involving $e_g$ electrons.
The energies of the five possible excited states:\cite{Gri71}
(i) the high-spin $^6\!A_1$ state($S=5/2$), and (ii-v) the low-spin
($S=3/2$) states: $^4\!A_1$, $^4E$ ($^4E_{\epsilon}$,
$^4E_{\theta}$), and $^4\!A_2$, will be parametrized again by the
intraorbital Coulomb element $U$ (\ref{U}), and by Hund's exchange $J_H$
between a pair of $e_g$ electrons, defined in Eq. (\ref{JHe}).
\cite{notejh} The energies of the excited states are given in terms of
the Racah parameters in Ref. \onlinecite{Gri71}; in order to parametrize
this spectrum by $J_H$ we apply an approximate relation $4B\simeq C$
which holds for the spectroscopic values of the Racah parameters for a
Mn$^{2+}$ ($d^5$) ion:\cite{Zaa90,Boc92} $B=0.107$ eV and $C=0.477$ eV.
Here we use these atomic values as an example of the
theory --- using them and Eq. (\ref{JHe}) one finds the excitation
spectrum: $U-3J_H$, $U+3J_H/4$, $U+5J_H/4$, $U+5J_H/4$, and $U+13J_H/4$
[Fig. \ref{fig:levels}(a)]. Unlike $J_H$, the value of $U$ is known with
less accuracy --- hence we shall use it here only as a parameter which
can be deduced {\it a posteriori\/} from the superexchange $J$ which is
able to explain the experimental values for two exchange constants
responsible for the $A$-AF phase observed in LaMnO$_3$ well below the
structural transition (here again $T_N\ll T_s$).

Using the spin algebra (Clebsch-Gordon coefficients), and making a
rotation of the terms derived for a bond $\langle ij\rangle\parallel c$
to the other two cubic axes $a$ and $b$, one finds five contributions to
${\cal H}_U(d^4)$ due to different $(t_{2g}^3e_g^1)_i(t_{2g}^3e_g^1)_j
\rightleftharpoons (t_{2g}^3e_g^2)_i(t_{2g}^3)_j$
excitations by $e_g$ electrons,\cite{Fei99}
\begin{eqnarray}
\label{som41}
\!H_1^{(\gamma)}\!&=&\!-\frac{J}{20}r_1
    \big(\vec{S}_i\!\cdot\!\vec{S}_j+6\big)
    \Big(\frac{1}{4}-\tau_i^{(\gamma)}\tau_j^{(\gamma)}\Big),    \\
\label{som42}
\!H_2^{(\gamma)}\!&=& \!\!\frac{3J}{160}r_2
    \big(\vec{S}_i\!\cdot\!\vec{S}_j-4\big)
    \Big(\frac{1}{4}-\tau_i^{(\gamma)}\tau_j^{(\gamma)}\Big),    \\
\label{som43}
\!H_3^{(\gamma)}\!&=&\!\!\frac{J}{32}r_3
    \big(\vec{S}_i\!\cdot\!\vec{S}_j-4\big)
    \Big(\frac{1}{4}-\tau_i^{(\gamma)}\tau_j^{(\gamma)}\Big),    \\
\label{som44}
\!H_4^{(\gamma)}\!&=&\!\!\frac{J}{32}r_4
    \big(\vec{S}_i\!\cdot\!\vec{S}_j-4\big)\!
    \Big(\frac{1}{2}-\!\tau_i^{(\gamma)}\Big)\!
    \Big(\frac{1}{2}-\!\tau_j^{(\gamma)}\Big),                   \\
\label{som45}
\!H_5^{(\gamma)}\!&=& \!\!\frac{J}{32}r_5
    \big(\vec{S}_i\!\cdot\!\vec{S}_j-4\big)\!
    \Big(\frac{1}{2}-\!\tau_i^{(\gamma)}\Big)\!
    \Big(\frac{1}{2}-\!\tau_j^{(\gamma)}\Big),
\end{eqnarray}
where the coefficients
\begin{eqnarray}
\label{r4}
r_1&=&\frac{1}{1-3\eta}, \hskip 1.5cm r_2=\frac{1}{1+3\eta/4}, \nonumber \\
r_3&=&r_4=\frac{1}{1+5\eta/4}, \hskip .3cm r_5=\frac{1}{1+13\eta/4},
\end{eqnarray}
follow from the above multiplet structure of Mn$^{2+}$ ($d^5$) ions,
and $\eta$ (\ref{eta}) stands for the Hund's exchange. The meaning of
the various terms is straightforward: the first term $H_1^{(\gamma)}$
describes the high-spin excitations to the $^6\!A_1$ state
while the remaining ones, $H_n^{(\gamma)}$ with $n=2,\dots,5$, arise
due to the low-spin excited states $^4\!A_1$, $^4E_{\theta}$,
$^4E_{\epsilon}$ and $^4\!A_2$, respectively. The orbital dependence is
given by the same operators (\ref{orbpro}) as in Sec. \ref{sec:d9}.
Similar to the case of the $^1E$ state for the copper fluoride [see Eqs.
(\ref{som92}) and (\ref{som93})], the doubly degenerate $^4E$ state
contributes here with two terms characterized by a different orbital
dependence in Eqs. (\ref{som43}) and (\ref{som44}).
Note that this degeneracy would be removed by the cooperative JT effect,
i.e., the structural phase transition (and associated OO) driven by the
local JT coupling in combination with the elastic lattice forces. The
resulting small level splitting we neglect here, and so we set
$r_3=r_4$.

The superexchange mediated by $t_{2g}$ electrons results from
$(t_{2g}^3e_g^1)_i(t_{2g}^3e_g^1)_j\rightleftharpoons
 (t_{2g}^4e_g^1)_i(t_{2g}^2e_g^1)_j$
excitations which involve $^4T_1$ and $^4T_2$ configurations at both Mn
ions: Mn$^{2+}$ and Mn$^{4+}$. They give low $S=3/2$ spins of Mn$^{2+}$
ions, and this part of the superexchange is AF. Using the present units
introduced in Eqs. (\ref{U}) and (\ref{JHe}), one finds the excitation
energies (not shown in Fig. \ref{fig:levels}):
$\varepsilon(^4T_1,^4T_2)\simeq U+5J_H/4$,
$\varepsilon(^4T_2,^4T_2)\simeq U+9J_H/4$,
$\varepsilon(^4T_1,^4T_1)\simeq U+11J_H/4$, and
$\varepsilon(^4T_2,^4T_1)\simeq U+15J_H/4$, with the first (second)
label standing for the configuration of the Mn$^{2+}$ (Mn$^{4+}$) ion,
respectively. In the actual derivation each of the excited states,
with one $t_{2g}$ orbital being either doubly occupied or empty, has
to be projected on the respective eigenstates and the spin algebra is
next used to construct the interacting total $S=2$ spin states. This
leads to the final contribution to ${\cal H}_U$ which, in a good
approximation, is orbital independent,\cite{Fei99}
\begin{equation}
\label{som46}
H_6^{(\gamma)} = \frac{1}{8}J\beta r_t
\Big(\vec{S}_i\!\cdot\!\vec{S}_j-4\Big).
\end{equation}
Here $\beta=(t_{\pi}/t)^2$ follows from the difference between the
effective $d-d$ hopping elements along the $\sigma$ and $\pi$ bonds,
and we adopt the Slater-Koster value $\beta=1/9$. The coefficient $r_t$
stands for a superposition of the above $t_{2g}$ excitations involved in
the $t_{2g}$ superexchange,
\begin{equation}
\label{rt}
r_t = \frac{1}{4}\Big(
 \frac{1}{1+ \textstyle{\frac{5}{4}}\eta}
+\frac{1}{1+ \textstyle{\frac{9}{4}}\eta}
+\frac{1}{1+\textstyle{\frac{11}{4}}\eta}
+\frac{1}{1+\textstyle{\frac{15}{4}}\eta}\Big).
\end{equation}
There is no need to distinguish between the different excitation
energies; all of them are significantly higher than the first low-spin
excitation energy for the configuration $^4\!A_1$, which occurs after
an excitation by an $e_g$ electron.

While earlier studies of the superexchange interactions in manganites
were limited to model Hamiltonians containing only the $U$ term,
\cite{Dag02,Ish97,Fei99,Feh04} the importance of the CT processes was
emphasized only recently.\cite{Mes01} For our purposes we derived the CT
term ${\cal H}_{\Delta}(d^4)$ by considering again excitations by either
$\sigma$ or $\pi$ electrons on the bond $\langle ij\rangle$, leading
to two-hole excited states at an intermediate oxygen,
$d^4_i(2p_{\langle ij\rangle})^6d^4_j\rightleftharpoons
 d^5_i(2p_{\langle ij\rangle})^4d^5_j$. Unlike in KCuF$_3$ with a unique
CT excitation, however, in the present case a number of different
excited states occurs with the excitation energies depending on the
electronic configuration of the two intermediate Mn$^{2+}$ ions at sites
$i$ and $j$. One finds that these various
excitations can be parametrized by a single parameter $R$ given by Eq.
(\ref{r}), and the excited states on two neighboring transition metal
ions contribute, as for the $U$ term, both FM and AF terms,
\begin{widetext}
\begin{eqnarray}
\label{HCT4e}
H_{\Delta}(d^4)&=&\frac{1}{16}JR\sum_{\langle ij\rangle\parallel\gamma}
    \Big\{ c_1\Big(\vec{S}_i\cdot\vec{S}_j-4\Big)
    \Big(\frac{1}{2}-\tau_i^{(\gamma)}\Big)
    \Big(\frac{1}{2}-\tau_j^{(\gamma)}\Big)           \nonumber \\
&+&\frac{8}{5}\Big[-c_2\Big(\vec{S}_i\cdot\vec{S}_j+6\Big)
        +c_3\Big(\vec{S}_i\cdot\vec{S}_j-4\Big)\Big]
              \Big(\frac{1}{4}-\tau_i^{(\gamma)}\tau_j^{(\gamma)}\Big)
                                                      \nonumber \\
&+&\frac{8}{5}\Big[-c_4\Big(\vec{S}_i\cdot\vec{S}_j+6\Big)
        +c_5\Big(\vec{S}_i\cdot\vec{S}_j-4\Big)\Big]
    \Big(\frac{1}{2}+\tau_i^{(\gamma)}\Big)
    \Big(\frac{1}{2}+\tau_j^{(\gamma)}\Big)\Big\}     \nonumber \\
&+& \frac{1}{8}JR\beta c_t\sum_{\langle ij\rangle}
    \Big(\vec{S}_i\cdot\vec{S}_j-4\Big),
\end{eqnarray}
where the coefficients $c_n$, with $n=1,\cdots,5$, and $c_t$ are all
determined by $\eta$ and $R$ via $\eta'=\eta R$:
\begin{eqnarray}
\label{cn}
c_1&=&\frac{1}{4}\Big(\frac{1}{1+\textstyle{\frac{17}{4}}\eta'}
                     +\frac{2}{1+\textstyle{\frac{21}{4}}\eta'}
                     +\frac{1}{1+\textstyle{\frac{25}{4}}\eta'}\Big),
      \\
c_2&=&\frac{1}{2}\Big(\frac{1}{1+\textstyle{\frac{17}{8}}\eta'}
                     +\frac{1}{1+\textstyle{\frac{25}{8}}\eta'}\Big),
      \\
c_3&=&\frac{1}{16}\Big(\frac{3}{1+4\eta'}
                      +\frac{5}{1+\textstyle{\frac{17}{4}}\eta'}
                      +\frac{3}{1+5\eta'}
                      +\frac{5}{1+\textstyle{\frac{21}{4}}\eta'}\Big),
      \\
c_4&=&\frac{1}{5}
    +\frac{1}{10}\Big(\frac{3}{1+\textstyle{\frac{15}{8}}\eta'}
                     +\frac{5}{1+\textstyle{\frac{17}{8}}\eta'}\Big),
     \\
c_5&=&\frac{3}{5}+\frac{1}{160}\Big(
                  \frac{9}{1+\textstyle{\frac{15}{4}}\eta'}
                +\frac{30}{1+4\eta'}
                +\frac{25}{1+\textstyle{\frac{17}{4}}\eta'}\Big),  \\
c_t&=&\frac{1}{4}\Big(\frac{1}{1+\textstyle{\frac{17}{4}}\eta'}
                     +\frac{2}{1+\textstyle{\frac{19}{4}}\eta'}
                     +\frac{1}{1+\textstyle{\frac{21}{4}}\eta'}\Big).
\end{eqnarray}
\end{widetext}
The coefficients $c_n$ follow from CT excitations by $e_g$ electrons.
As in the copper fluoride case (Sec. \ref{sec:d9}), the lowest
(high-spin) excitation energy will be labeled by $\Delta$,
\begin{equation}
\label{deltaman}
\Delta(^6\!A_1)=\Delta,
\end{equation}
so the other possible individual (low-spin) excitations at each
transition metal ion have the energies
\begin{eqnarray}
\label{deltas}
\Delta(^4\!A_1)&=&\Delta+\frac{15}{4}J_H,     \nonumber  \\
\Delta(^4E)    &=&\Delta+\frac{17}{4}J_H,     \nonumber  \\
\Delta(^4\!A_2)&=&\Delta+\frac{25}{4}J_H,
\end{eqnarray}
These excitation energies are used here to introduce an average CT
energy $\bar{\Delta}$,
\begin{equation}
\label{bard}
\frac{1}{\bar{\Delta}}=\frac{1}{26}\Big(
\frac{8}{\Delta(^6\!A_1)}+\frac{3}{\Delta(^4\!A_1)}+
\frac{10}{\Delta(^4E)}+\frac{5}{\Delta(^4\!A_2)}\Big),
\end{equation}
which serves to define the effective hopping element
$t=t_{pd}^2/\bar{\Delta}$, and the superexchange energy $J$ (\ref{J})
in a microscopic approach. We emphasize, however, that such microscopic
parameters as $\{t_{pd},\bar{\Delta},U,J_H\}$ will not be needed here,
and only the values of the effective parameters $\{J,\eta,R\}$ will
decide about the exchange constants and the optical spectral weights.

Each coefficient $c_n$ (\ref{cn}), with $n=1,\cdots,5$, stands for an
individual process which contributes with a particular orbital
dependence due to an intermediate state arising in the excitation
process, and accompanies either a FM or an AF spin factor, depending on
whether high-spin or low-spin states are involved. As in the $U$ term
(\ref{som41})--(\ref{som46}), a pair of directional $|\zeta\rangle$
orbitals accompanies low-spin excitations, while either high-spin or
low-spin excited states are allowed when two different orbitals, one
directional and one orthogonal to it (planar orbital), are occupied at
the two Mn$^{3+}$ sites. In contrast to the $U$ term, also
configurations with two planar orbitals $|\xi\rangle$ occupied at sites
$i$ and $j$ contribute to ${\cal H}_{\Delta}(d^4)$ in Eq. (\ref{HCT4e}).
These terms are accompanied by the projection operator
$\big\langle(\frac{1}{2}+\tau_i^{(\gamma)})
            (\frac{1}{2}+\tau_j^{(\gamma)})\big\rangle$
in Eq. (\ref{HCT4e}). Note that in the case of the $U$ term such
configurations did not contribute as the $e_g$ electrons were then blocked
and could not generate any superexchange terms. As the electrons from
an oxygen $2p_{\sigma}$ orbital are excited instead to directional
$|\zeta\rangle$ orbitals at two neighboring Mn$^{3+}$ ions, again both
high-spin ($S=5/2$) and low-spin ($S=3/2$) excitations are here possible,
giving a still richer structure of ${\cal H}_{\Delta}(d^4)$.

The OO in LaMnO$_3$ is robust and sets in below the structural
transition at $T_s\simeq 780$ K.\cite{Mur98} The orbital interactions
present in the superexchange Hamiltonian (\ref{som4}) would induce a
$G$-type OO.\cite{Bri99} The observed classical ground state, which can
again be described by Eq. (\ref{ood9}), corresponds instead to $C$-type
OO, as deduced from the lattice distortions. Note that in contrast to
KCuF$_3$, the occupied orbitals refer now to electrons and thus the
values of the expected orbital angle $\theta$ are $>90^{\circ}$ (which
corresponds to $\cos\theta\lesssim 0$) and so are distinctly different
from the copper fluoride case. The averages of the orbital
operators in the orbital ordered state are given in Table \ref{tab:eg},
including the terms $\big\langle(\frac{1}{2}+\tau_i^{(\gamma)})
                    (\frac{1}{2}+\tau_j^{(\gamma)})\big\rangle$
which contribute now to the CT part of superexchange.
The dependence on the orbital angle $\theta$ suggests that, similar to
KCuF$_3$, these new terms are more significant along the $c$ axis for
the OO expected in the manganites.

\subsection{Spin exchange constants and optical intensities}
\label{d4:jando}

For a better understanding of the effective exchange interactions it is
convenient to introduce first the $t_{2g}$ superexchange constant $J_t$
which stands for the interaction induced by the charge excitations of
$t_{2g}$ electrons. When the CT terms are included, $J_t$ consists
of the two contributions given in Eqs. (\ref{som46}) and (\ref{HCT4e}),
\begin{equation}
\label{Jt4}
J_t = J_{tU}+J_{t\Delta} = \frac{1}{8}J\beta (r_t+R c_t).
\end{equation}
This interaction is frequently called the superexchange between the
core spins. We emphasize that this term is orbital independent and thus
isotropic. The coupling constant $J_t$ has a similar origin as the $e_g$
part of the superexchange $J^{(\gamma)}_e$, which however is
orbital dependent and anisotropic. We emphasize that both $J_t$ and
$J^{(\gamma)}_e$ are relatively small fractions of $J$.

The $e_g$ contributions to the effective exchange constants (\ref{Hs})
in LaMnO$_3$ depend on the orbital state characterized again by Eqs.
(\ref{ood9}) via the averages of the orbital operators,
\begin{eqnarray}
\label{je4}
J^{(\gamma)}_e&=&\frac{1}{16}J\Big\{
   -\frac{4}{5}\Big(r_1-\frac{3}{8}r_2-\frac{5}{8}r_3\Big)
    \Big\langle\frac{1}{4}-\tau_i^{(\gamma)}\tau_j^{(\gamma)}\Big\rangle
                                                          \nonumber \\
 &+&\frac{1}{2}(r_4+r_5)
    \Big\langle\Big(\frac{1}{2}-\tau_i^{(\gamma)}\Big)
               \Big(\frac{1}{2}-\tau_j^{(\gamma)}\Big)\Big\rangle\Big\}
                                                          \nonumber \\
 &+&\frac{1}{16}JR\Big\{c_1
    \Big\langle\Big(\frac{1}{2}-\tau_i^{(\gamma)}\Big)
               \Big(\frac{1}{2}-\tau_j^{(\gamma)}\Big)\Big\rangle
                                                          \nonumber \\
 &+&\frac{8}{5}(c_3-c_2)
    \Big\langle\frac{1}{4}-\tau_i^{(\gamma)}\tau_j^{(\gamma)}\Big\rangle
                                                          \nonumber \\
 &+&\frac{8}{5}(c_5-c_4)
    \Big\langle\Big(\frac{1}{2}+\tau_i^{(\gamma)}\Big)
               \Big(\frac{1}{2}+\tau_j^{(\gamma)}\Big)\Big\rangle\Big\}.
\end{eqnarray}
As the structural transition occurs at relatively high temperature
$T_s=780$ K, at room temperature (and below it) the OO may be considered
to be frozen and specified by an angle $\theta$ [see Eqs. (\ref{ood9})].
The orbital fluctuations are then quenched by the combined effect of
the orbital part of the $e_g$ superexchange in Eqs.
(\ref{som41})-(\ref{som45}) and the orbital interactions induced by
the JT effect,\cite{Fei99} and the spins effectively decouple from the
orbitals, leading to the effective spin model (\ref{Hs}). The magnetic
transition then takes place within this OO state,
and is driven by the magnetic part of the superexchange interactions,
which follow from ${\cal H}_U(d^4)$ and ${\cal H}_{\Delta}(d^4)$.
For the $C$-type OO, as observed in LaMnO$_3$,\cite{Mur98} one finds the
effective exchange constants in Eq. (\ref{Hs}) as a superposition of
$J_t$ and $J^{(\gamma)}_e$ after inserting the averages of the orbital
operators (see Table \ref{tab:eg}) in Eq. (\ref{je4}):
\begin{widetext}
\begin{eqnarray}
\label{jab4}
J_{ab}&=&\frac{1}{16}J\Big\{-\frac{1}{5}\Big(r_1-\frac{3}{8}r_2\Big)
      \Big(\frac{3}{4}+\sin^2\theta\Big)
      +\frac{1}{4}r_3\Big(1-\frac{1}{2}\cos\theta\Big)
      +\frac{1}{8}r_5\Big(\frac{1}{2}-\cos\theta\Big)^2\Big\}
                                                        \nonumber \\
 &+&\frac{1}{64}JR\Big\{c_1\Big(\frac{1}{2}-\cos\theta\Big)^2
      +\frac{8}{5}\big(c_3-c_2\big)\Big(\frac{3}{4}+\sin^2\theta\Big)
      +\frac{8}{5}\big(c_5-c_4\big)
                \Big(\frac{1}{2}+\cos\theta\Big)^2\Big\}
      +J_t,                                                       \\
\label{jc4}
J_c &=&\frac{1}{16}J\Big\{-\frac{1}{5}\Big(r_1-\frac{3}{8}r_2\Big)
       \sin^2\theta+\frac{1}{4}r_3(1+\cos\theta)
      +\frac{1}{8}r_5(1+\cos\theta)^2\Big\}             \nonumber \\
    &+&\frac{1}{64}JR\Big\{c_1(1+\cos\theta)^2
      +\frac{8}{5}\big(c_3-c_2\big)\sin^2\theta
      +\frac{8}{5}\big(c_5-c_4\big)(1-\cos\theta)^2\Big\}
      +J_t.
\end{eqnarray}
\end{widetext}
Considering $\beta=1/9$ to be fixed by the Slater-Koster
parametrization, a complete set of parameters which determines $J_c$
and $J_{ab}$ comprises: $J$ (\ref{J}), $\eta$ (\ref{eta}), $R$
(\ref{r}), and the orbital angle $\theta$ which defines the phase with
OO by Eqs. (\ref{ood9}), referring now to the orbitals occupied by
electrons.

Equations (\ref{jab4}) and (\ref{jc4}) may be further analyzed in two
ways: either
 (i) using an effective model which includes only the $U$
     superexchange term due to $d-d$ transitions, as presented in Ref.
     \onlinecite{Fei99} and discussed in Appendix \ref{d4:eff}
(i.e., taking $R=0$ which implies ${\cal H}_{\Delta}(d^4)\equiv 0$), or
(ii) by considering the complete ${\cal H}(d^4)$ model as given by
     Eq. (\ref{som4}), which includes also the CT contributions to the
     superexchange (with $R>0$).
By a similar analysis in Sec. \ref{sec:d9} we have established that the
CT terms are of essential importance in KCuF$_3$ and should not be
neglected, as otherwise the strong anisotropy of the exchange constants
would remain unexplained. Here the situation is qualitatively different
--- as we show in Appendix \ref{d4:eff}, using somewhat
modified parameters $J$ and $\eta$ one may still reproduce the
experimental values of the exchange constants, deduced from neutron
experiments for LaMnO$_3$,\cite{Mou96} within the effective model at
$R=0$, and even interpret successfully the optical spectra.\cite{Kov04}

It is important to realize that the high-spin $e_g$ electron excitations
play a crucial role in stabilizing the observed $A$-AF spin order, as
only these processes are able to generate FM terms in the superexchange.
They compete with the remaining AF terms, and the $A$-AF phase is stable
only when $J_{ab}<0$ and $J_c>0$. We have verified that the terms which
contribute to $J_{ab}$ and $J_c$ in Eqs. (\ref{jab4}) and (\ref{jc4})
are all of the same order of magnitude as all the coefficients
$\{r_n,r_t,c_n,c_t\}$ are of order one. Hence, the superexchange
energy $J$ (\ref{J}) is much higher than the actual exchange constants
in LaMnO$_3$, i.e., $|J_{ab}|\ll J$ and $J_c\ll J$.

Next we consider the kinetic energies associated with the various optical
excitations which determine the optical spectral weights by Eq.
(\ref{hefa}). Again, as in the previously considered case of KCuF$_3$,
the high-spin subband at low energy is unique and is accompanied by
low-spin subbands at
higher energies. While the energetic separation between the high-spin
and low-spin parts of the spectrum $\sim 4J_H$ is large, one may expect
that the low-spin optical excitations might be difficult to distinguish
experimentally from each other. As we will see below by analyzing the
actual parameters of LaMnO$_3$, the low-spin excitations overlap with
the $d-p$ CT excitations, and so such a separation is indeed
impossible --- thus we analyze here
only the global high-energy spectral weight due to the optical
excitations on the transition metal ions, expressed by the total kinetic
energy $K_{\rm LS}^{(\gamma)}$ for all ($e_g$ and $t_{2g}$) low-spin
terms, and compare it with the high-spin one, given by $K_1^{(\gamma)}$.
Using the manganite model (\ref{som4}) one finds
for polarization in the $ab$ plane
\begin{eqnarray}
\label{wab41}
K_1^{(ab)}&=&\frac{1}{40}Jr_1(6+s_{ab})
             \Big(\frac{3}{4}+\sin^2\theta\Big),                   \\
\label{wab42}
K_{\rm LS}^{(ab)}&=&\frac{1}{8}J \Big[
      \frac{3}{40}r_2\Big(\frac{3}{4}+\sin^2\theta\Big)
     +\frac{1}{4} r_3\Big( 1-\frac{1}{2}\cos\theta\Big)  \nonumber \\
   &+&\frac{1}{8} r_5\Big(  \frac{1}{2}-\cos\theta\Big)^2
     +2\beta r_t\Big](4-s_{ab}),
\end{eqnarray}
and for polarization along the $c$ axis
\begin{eqnarray}
\label{wc41}
K_1^{(c)}&=&\frac{1}{40}Jr_1(6+s_c)\sin^2\theta,                   \\
\label{wc42}
K_{\rm LS}^{(c)}&=&\frac{1}{8}J\Big[\frac{3}{40}r_2\sin^2\theta
     + \frac{1}{4}r_3(1+\cos\theta)                      \nonumber \\
    &+&\frac{1}{8}r_5(1+\cos\theta)^2+2\beta r_t\Big](4-s_c).
\end{eqnarray}
The optical spectral weights given by Eqs. (\ref{wab41})--(\ref{wc42})
are determined by $J$, $\eta$, and the orbital angle $\theta$.
Note that the leading term in the low-spin part comes from the $e_g$
optical excitations, while the $t_{2g}$ excitations contribute only
with a relatively small weight $\propto\beta$.

\subsection{Magnetic interactions in LaMnO$_3$}
\label{d4:resj}

It is a challenge for the present theoretical model to describe both
the magnetic exchange constants\cite{Mou96} and the anisotropic optical
spectral weights\cite{Kov04} using only a few effective parameters
$\{J,\eta,R\}$ and the orbital angle $\theta$. We shall proceed now
somewhat differently than in Sec. \ref{sec:d9}, and analyze the
experimental data using primarily these effective parameters, while
we will discuss afterwards how they relate to the expectations based on
the values of the microscopic parameters found in the literature.

The experimental values of the exchange constants,\cite{Mou96}
$J_{ab}=-1.66$ meV and $J_c=1.16$ meV, impose rather severe constraints
on the microscopic parameters and on the possible OO in LaMnO$_3$. The
AF interaction $J_c$ is quite sensitive to the type of occupied orbitals
(\ref{ood9}), and increases with increasing amplitude of $|z\rangle$
orbitals in the ground state, i.e., with decreasing orbital angle
$\theta$. Simultaneously, the FM interaction $-J_{ab}$ is enhanced.
Already with the effective model (at $R=0$) it is not straightforward
to determine the parameters $J$ and $\eta$, as we discuss in Appendix
\ref{d4:eff}. This model is in fact quite successful, and a reasonable
agreement with experiment could be obtained both for the magnetic
exchange constants and for the optical spectral weights, taking the
experimental excitation spectrum.\cite{Kov04} Here we will investigate
to what extent this effective model gives robust results and whether
the CT processes could play an important role in LaMnO$_3$.

By analyzing the CT terms in the superexchange [compare Eqs.
(\ref{jab4}), (\ref{jc4})] one concludes that these contributions are
predominantly AF. Therefore, it might be expected that a higher value of
$\eta$ than 0.16 used in Appendix \ref{d4:eff} would rather be
consistent with the magnetic experiments. Increasing $\eta$ gives
an increased coefficient $r_1$, so not only the FM term in $J_{\gamma}$
is then enhanced, but also the optical spectral weight $K^{ab}_1$ which
corresponds to the high-spin transition. Simultaneously, the angle
$\theta$ is somewhat increased, but the dependence of the spectral
weight $K^{ab}_1$ on the angle $\theta$ is so weak that the higher value
of $\eta$ dominates and a somewhat lower value of $J$ than 170 meV used
in Appendix \ref{d4:eff} has to be chosen.
Altogether, these considerations led us to selecting $J=150$ meV and
$\eta=0.181$ as representative parameters for which we show below that
a consistent explanation of both magnetic and optical data is possible.

\begin{figure}[t!]
\includegraphics[width=7.5cm]{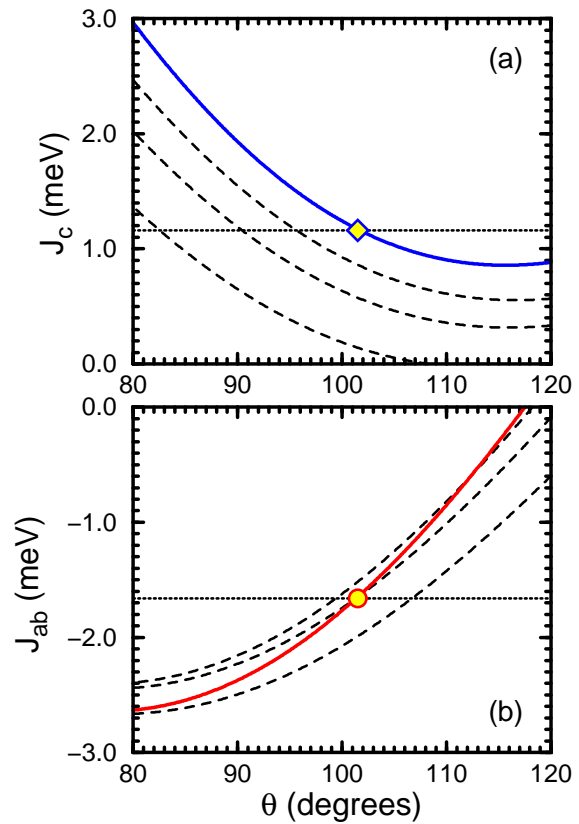}
\caption{(Color online)
Superexchange constants for LaMnO$_3$:
(a) $J_c$ along the $c$ axis, and
(b) $J_{ab}$ in the $ab$ plane,
as functions of orbital angle $\theta$ which defines the $C$-type OO,
see Eqs. (\ref{ood9}).
Dashed lines for increasing $R=0$, 0.15 and 0.30 from bottom to top;
solid lines for $R=0.6$.
Note that the dependence of $J_{ab}$ on $R$ is nonmonotonic.
Horizontal dotted lines indicate the
experimental values of Ref. \onlinecite{Mou96} for LaMnO$_3$;
diamond and circle show the experimental values of $J_c$ and $J_{ab}$
for $\theta=102^{\circ}$. Parameters: $J=150$ meV, $\eta=0.181$.
}
\label{fig:j4r}
\end{figure}

After these two parameter values have been fixed, it is of interest to
investigate the dependence of the effective exchange interactions on the
CT parameter $R$ and on the orbital angle $\theta$. As in the KCuF$_3$
case, one finds a stronger increase of $J_c$ with increasing $R$, while
these terms are weaker and lead to nonmonotonic changes for $J_{ab}$
(Fig. \ref{fig:j4r}). First of all, with the present value of
$\eta=0.181$, at $R=0$ the AF interaction $J_c$ is close to zero for
angles $\theta\sim 100^{\circ}$, while the FM interaction $J_{ab}<-2$
meV is somewhat too strong. With increasing $R$ one finds that $J_c$
increases, while the FM interaction $J_{ab}$ initially becomes weaker
when $R$ increases up to $R=0.3$ and the term $\propto c_1$ dominates
the CT contribution to $J_{ab}$ [see Eq. (\ref{jab4})]. At higher values
of $R$, however, the FM contributions due to $c_3-c_2<0$ and $c_5-c_4<0$
start to dominate, and $J_{ab}$ decreases with increasing $R$,
particularly for small values of $\theta<90^{\circ}$. One finds that the
experimental values of both exchange constants are well reproduced for
$R=0.6$ and at the orbital angle $\theta=102^{\circ}$.

\begin{figure}[t!]
\includegraphics[width=8.0cm]{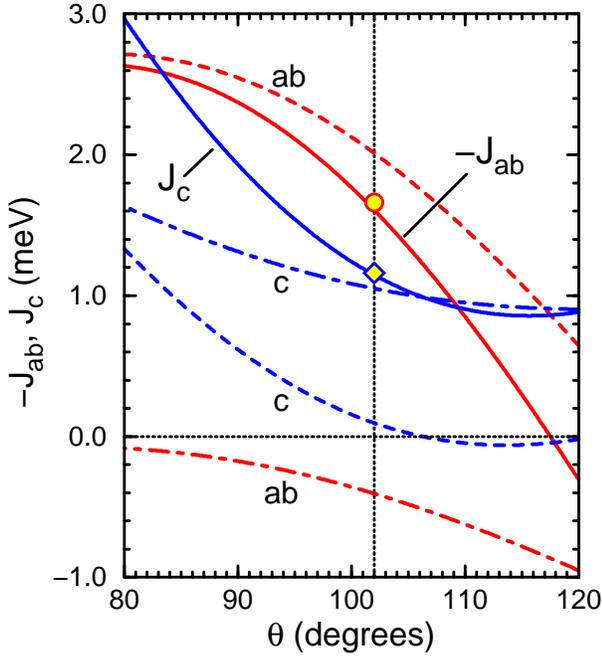}
\caption{(Color online)
Superexchange interactions $-J_{ab}$ (\ref{jab4}) and $J_c$ (\ref{jc4})
as functions of the orbital angle $\theta$, obtained within the
superexchange model (\ref{som4}) for LaMnO$_3$ (solid lines).
Contributions due to the $U$ term and due to the CT term to both
exchange constants ($-J_{ab}$ and $J_c$) are shown by dashed and
dashed-dotted lines, respectively. Isotropic and orbital independent
$t_{2g}$ terms superexchange terms $J_{tU}$ and $J_{t\Delta}$
(\protect\ref{Jt4}) are: 1.45 and 0.83 meV. Experimental values
\cite{Mou96} of $-J_{ab}$ and $J_c$ (indicated by diamond and circle)
are well reproduced at angle $\theta=102^{\circ}$.
Parameters $J=150$ meV, $\eta=0.181$, $R=0.6$.
}
\label{fig:j4}
\end{figure}

Although this fit is not unique, one has to realize that the
experimental constraints imposed on the parameters are indeed rather
severe --- as we show below the present parameters give a very
reasonable and consistent interpretation of the experimental results for
LaMnO$_3$. For the above parameters the FM and AF terms to $J_c$ almost
compensate each other in the $U$ term, and a considerable AF interaction
along the $c$ axis arises mainly due to the CT contributions (Fig.
\ref{fig:j4}). This qualitatively agrees with the situation found in
KCuF$_3$, where the CT term was of crucial importance and helped to
explain the observed large anisotropy between the values of exchange
constants. Also the CT term which contributes to $J_{ab}$ is AF and
increases with increasing angle $\theta$, while the $U$ term is FM but
weakens at the same time. This results in a quite fast dependence of
the FM interaction $J_{ab}$ on $\theta$ (Fig. \ref{fig:j4}).

One thus recognizes a similar dependence of the exchange constants
$-J_{ab}$ and $J_c$ on the orbital angle $\theta$ (Fig. \ref{fig:j4}) as
that found before with the $U$ term alone (see Fig. \ref{fig:j4u} in
Appendix \ref{d4:eff}). The CT terms have mainly two consequences:
(i) a large AF contribution to the interaction along the $c$ axis $J_c$,
and
(ii) an increase of the orbital angle well above $\theta=90^{\circ}$.
These trends are robust in the realistic parameter range. Therefore, one
expects that the occupied $e_g$ orbitals approach the frequently assumed
alternating directional orbitals in the $ab$ planes
$(3x^2-r^2)/(3y^2-r^2)$ with $\theta=120^{\circ}$, but cannot quite
reach them. Indeed, we have verified that orbitals with
$\theta\sim 120^{\circ}$ are excluded by the present calculations,
as then the FM interaction within the $ab$ planes changes sign and
becomes weakly AF. Thus, the mechanism for the observed $A$-AF phase
is lost, and one has to conclude that angles $\theta>\theta_{\rm JT}$
are excluded.

Indeed, we have found that the orbital angle $\theta\simeq 102^{\circ}$
reproduces well the experimental data for both exchange constants (Fig.
\ref{fig:j4}), and is thus somewhat smaller than the angle
$\theta_{\rm JT}=108^{\circ}$ deduced from the lattice distortions in
LaMnO$_3$.\cite{Rod98} This can be seen as a compromise between the
orbital interactions involving the lattice and the purely electronic
superexchange orbital interactions, so it is reasonable to expect
that $\theta<\theta_{\rm JT}$.

Finally, we emphasize that the $e_g$ electron excitations, contributing
both to the $U$ and to the CT processes in the superexchange, are FM
{\it for all cubic directions\/},\cite{noteeg} and only due to a
substantial $t_{2g}$ term which follows from low-spin (AF) excitations,
$J_t=2.28$ meV,
the exchange interaction along the $c$ axis changes its sign and
becomes AF. Alltogether, the present analysis shows that the $t_{2g}$
superexchange plays a decisive role in stabilizing the observed $A$-AF
spin order --- without it already the undoped LaMnO$_3$ would be a
ferromagnet. This peculiar situation follows from the large splitting
between the high-spin and low-spin excitations $\sim 3$ eV in LaMnO$_3$,
which is larger than in any other transition metal compound considered
in this paper, due to the fact that the $d$ shell is half-filled in the
Mn$^{2+}$ excited states.\cite{Ole01} This leads to relatively large FM
contributions, even when the orbitals are not so close to the ideal
alternation of directional and planar states (as found along the $c$
axis), which would maximize the averages of the orbital operators,
$\big\langle\frac{1}{4}-\tau_i^{(\gamma)}\tau_j^{(\gamma)}\big\rangle$,
that control the weight of the high-spin terms, see Table \ref{tab:eg}.
This result is remarkable but again qualitatively the same as found in
the effective model of Appendix \ref{d4:eff}. However, quantitatively
the $t_{2g}$ term is here somewhat stronger, as $J_t$ is now increased
by $\sim 35$\% over its value $J_t=1.7$ meV deduced from the effective
$d-d$ model with $U$ terms only.
As a common feature one finds that $J_t\sim 2$ meV, so we emphasize
that the superexchange promoted by $t_{2g}$ electrons is quite weak and
is characterized by a small value of $J_t\sim 4\times 10^{-3}t$
(or $\sim 0.01t$ for the superexchange between $S=3/2$ core spins
\cite{Dag02,vdB99}) which provides an important constraint for the
realistic models of manganites.

\subsection{Optical spectral weights in LaMnO$_3$}
\label{d4:reso}

Consider now the temperature dependence of the optical spectral weights
(\ref{opsa}). As the orbital dynamics is quenched up to room temperature
$T\sim 300$ K, it suffices to consider the temperature dependence of
the intersite spin-spin correlations (\ref{spins}). We derived these
by employing the so-called Oguchi method\cite{Ogu60} (see Appendix
\ref{app:spincomn}), which is expected to give rather realistic
values of spin correlations functions for the large spins $S=2$ in
LaMnO$_3$ in the entire temperature range. Thereby we
solved exactly the spin-spin correlations $\{s_{ab},s_c\}$ on a single
bond $\langle ij\rangle$ coupled to neighboring spins by mean-field
(MF) terms, proportional to the order parameter $\langle S^z\rangle$.
A realistic estimate of the magnetic transition can be obtained by
reducing the MF result obtained for $S=2$ spins by an empirical factor
$\sim 0.71$.\cite{Fle04} Using the exchange interactions obtained with
the present parameters, one finds $T_N\simeq 143$ K which reasonably
agrees with the experimental value of $T_N^{\rm exp}=136$ K.\cite{Mou96}
The calculations of spin-spin correlations are straightforward, and we
summarize them in Appendix \ref{app:spincomn}. Both correlation
functions $s_{ab}$ and $s_c$ change fast close to $T_N$, reflecting the
temperature dependence of the Brillouin function which deterimes
$\langle S^z\rangle$, and remain finite at $T\gg T_N$.

\begin{figure}[t!]
\includegraphics[width=7.5cm]{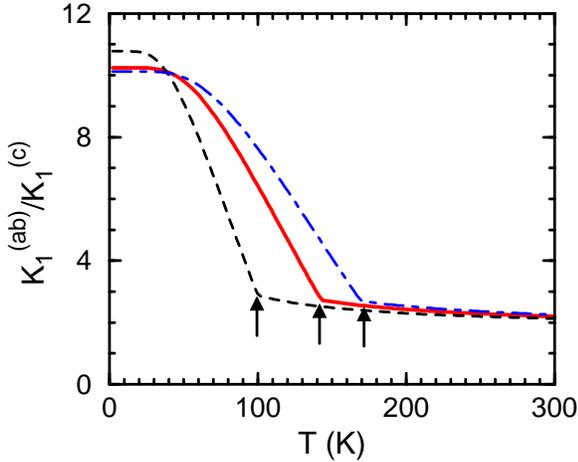}
\caption{(Color online)
Anisotropy of the low-energy spectral weights in LaMnO$_3$, given
by the ratio $K_1^{(ab)}/K_1^{(c)}$, for increasing temperature $T$,
as obtained for:
$\theta=102^{\circ}$ (solid line),
$\theta=108^{\circ}$ (dashed line),
$\theta= 98^{\circ}$ (dashed-dotted line).
The arrows indicate the N\'eel temperature $T_N$ derived from the
exchange constants in each case (see text).
Parameters: $J=150$ meV, $\eta=0.181$.
}
\label{fig:anit}
\end{figure}

The large splitting between the high-spin and low-spin excitations makes
it possible to separate the high-spin excitations from the remaining
ones in the optical spectra, and to observe the temperature
dependence of its intensity for both polarizations. As in the effective
model,\cite{Kov04,Ole04} the present theory predicts that the low-energy
optical intensity exhibits a rather strong anisotropy between the $ab$
and $c$ directions. It is particularly pronounced and close to 10:1 at
low temperatures when the spin correlations $|s_{ab}|$ and $s_c$ are
maximal (Fig. \ref{fig:anit}). Unfortunately, this anisotropy at
$T\to 0$ in only weakly dependent on the orbital angle $\theta$, so it
cannot help to establish the type of OO realized in the ground state of
LaMnO$_3$, and its possible changes with increasing temperature. In
fact, when the parameters are fixed and only the orbital angle $\theta$
is changed, a different value of the N\'eel temperature follows from the
modified exchange constants, being the main reason behind the somewhat
different temperature dependence of the ratio $K_1^{(ab)}/K_1^{(c)}$
(Fig. \ref{fig:anit}).

\begin{figure}[t!]
\includegraphics[width=7.5cm]{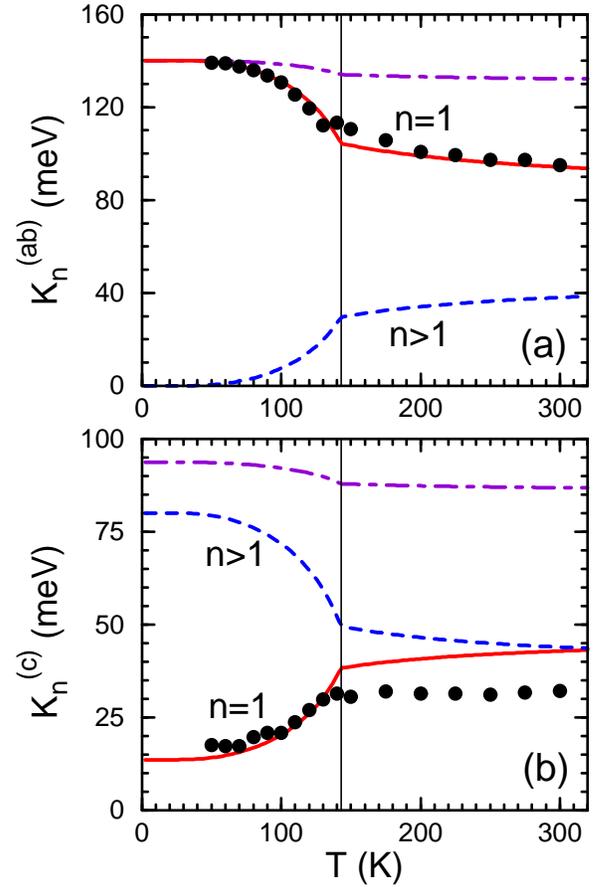}
\caption{(Color online)
Kinetic energies per bond (in meV) as functions of temperature $T$,
obtained for LaMnO$_3$ for the high-spin (solid lines, $n=1$) and for
the low-spin (dashed lines, $n>1$) excitations at the Mn ions for
polarization:
(a) in the $ab$ planes, $K_n^{(ab)}$, and
(b) along the $c$ axis, $K_n^{(c)}$.
The total kinetic energies $K^{(\gamma)}$ are given by the long
dashed-dotted lines, while the vertical dotted lines indicate the value
of $T_N$ derived within the present model (see text).
Filled circles show the experimental intensities \cite{Kov04} at
low energy ($n=1$). Parameters: $J=150$ meV, $\eta\simeq 0.181$,
$R=0.6$, and $\theta\simeq 102^{\circ}$.
}
\label{fig:anis}
\end{figure}

One finds a very satisfactory agreement between the present theory and
the experimental results of Ref. \onlinecite{Kov04}, as shown in Fig.
\ref{fig:anis}. We emphasize, however, that the temperature variation
of the optical spectral weights could also be reproduced within the
effective model at $R=0$,\cite{Kov04,Ole04} showing that the CT terms
lead only to minor quantitative modifications. Note also that at this
stage no fit is made, i.e., the kinetic energies (\ref{hefa}) which
stand for the optical sum rule are calculated using the same parameters
as found above to reproduce the exchange constants in Fig. \ref{fig:j4}.
Therefore, such a good agreement with experiment demonstrates that
indeed the superexchange interactions describe the spectral intensities
in the optical transitions. We note, however, that the anisotropy in
the range of $T>T_N$ is somewhat larger in experiment which might be
due to either some inaccuracy of the Oguchi method or due to the
experimental resolution.

The distribution of the optical intensities $K_n^{(\gamma)}$ and their
changes between the low ($T\simeq 0$) and the high temperature
($T=300$ K) regime are summarized in Table \ref{tab:par}. At temperature
$T=0$ one finds that for the $ab$ polarization the entire spectral
intensity originates from high-spin excitations. This result follows
from the classical value of the spin-spin correlation function
$s_{ab}=-1$ predicted by the Oguchi method. As quantum fluctuations in
the $A$-AF phase are small,\cite{Rac02} the present result is nearly
accurate. When the temperature increases, one finds considerable
transfer of optical spectral weights between low and high energies,
discussed also in Ref. \onlinecite{Kov04}, with almost constant total
intensities $K^{(\gamma)}$ for both $\gamma=ab$ and $c$ (see also Fig.
\ref{fig:anis}). The optical weights obtained for the JT angle
$\theta_{\rm JT}=108^{\circ}$ are similar to those for
$\theta=102^{\circ}$, showing again that the optical spectroscopy is
almost insensitive to small changes of the orbital angle. In contrast,
for $\theta_{\rm JT}$ the exchange constants (and hence also the
estimated value of $T_N$) are too low.

\begin{table}[t!]
\caption{
Exchange constants $J_{\gamma}$ [see Eqs. (\ref{jab4}) and (\ref{jc4})],
kinetic energies for high-spin ($K_1^{(\gamma)}$) and for low-spin
($K_{\rm LS}^{(\gamma)}$) excitations
[see Eqs. (\ref{wab41})--(\ref{wc42})], and total kinetic energies
$K_{\rm tot}^{(\gamma)}$ (all in meV), for a bond within an $ab$ plane
($\gamma=ab$) and along a $c$ axis ($\gamma=c$), as obtained for the OO
given by the orbital angle $\theta\simeq 102^{\circ}$ which reproduces
the experimental exchange constants,\cite{Mou96} and for the angle
suggested by the lattice distortions, $\theta_{\rm JT}$.
Parameters: $J=150$ meV, $\eta\simeq 0.181$, $R=0.6$.}
\vskip .2cm
\begin{ruledtabular}
\begin{tabular}{ccccccc}
&         & \multicolumn{2}{c}{$\theta\simeq 102^{\circ}$}
          & \multicolumn{2}{c}{$\theta_{\rm JT}=108^{\circ}$} & \cr
&       $T$ (K)                   & $0$ & $300$ & $0$ & $300$ & \cr
\colrule
& $J_{ab}$ & \multicolumn{2}{c}{$-1.66$}
           & \multicolumn{2}{c}{$-1.06$} & \cr
& $J_c$    & \multicolumn{2}{c}{$ 1.16$}
           & \multicolumn{2}{c}{$0.95$}  & \cr
\colrule
& $K_1^{(ab)}$         & 140.1 & 94.2  & 135.8 &  88.1  &  \cr
& $K_{\rm LS}^{(ab)}$  & 0.0   & 38.1  & 0.0   &  42.1  &  \cr
& $K_{\rm tot}^{(ab)}$ & 140.1 & 132.3 & 135.8 & 130.2  &  \cr
\colrule
& $K_1^{(c)}$          & 13.7  &  42.9 & 12.6  &  41.2  &  \cr
& $K_{\rm LS}^{(c)}$   & 80.0  &  44.0 & 74.9  &  40.1  &  \cr
& $K_{\rm tot}^{(c)}$  & 93.7  &  86.9 & 87.5  &  81.3  &  \cr
\end{tabular}
\end{ruledtabular}
\label{tab:par}
\end{table}

While the effective parameters $\{J,\eta,R\}$ used in this Section give
a very satisfactory description of both magnetic and optical properties
of LaMnO$_3$, the values of the microscopic parameters, such as the
Coulomb interaction on the Mn ions $U$, and on the oxygen ions $U_p$,
the Hund's exchange $J_H$, the charge transfer parameter $\Delta$, and
the $d-p$ hopping element $t_{pd}$, are not uniquely determined. One
could attempt to fix these parameters using the atomic value of Hund's
exchange $J_H=0.9$ eV. With $\eta=0.181$ it leads to $U\simeq 5.0$ eV,
quite close to other estimates,\cite{Miz96} while the value of $R=0.6$
suggests then that, taking the commonly accepted\cite{Miz96,Boc92} value
of the CT energy $\Delta=5.0$ eV, the oxygen Coulomb element is large,
$U_p\simeq 7$ eV. This value of $U_p$ is larger than usually obtained
$U_p\sim 4$ eV for oxygen ions, as for instance by analyzing the
parameters of the three-band model for CuO$_2$ planes.\cite{Gra92} We
note, however, that the optical data\cite{Kov04} suggest a somewhat
reduced value of Hund's exchange $J_H\sim 0.7$ eV (using the present
units), so it could be that the local exchange interactions are somewhat
screened in reality.

Fortunately, when the optical data are available, also the position of
the low-energy excitation is known, and this may serve as an additional
experimental constraint for the parameters.\cite{Kha04} This excitation
is found at about 2.0 eV,\cite{Kov04} indicating that $U-3J_H\simeq 2.0$
eV. With this constraint one finds, using again $\eta=0.181$, that
$U\simeq 3.7$ eV and $J_H\simeq 0.67$ eV. These parameters give the
low-spin $^4\!A_1$ and $^4E$ excitations close to 4.5 eV, in agreement
with experiment.\cite{Kov04} So the above values of the microscopic
parameters $\{U,J_H\}$ appear to be consistent both with the present
value of $\eta$ and with the spectra. Furthermore, for these empirical
parameters one finds $R=0.6$ either with $\Delta=5.0$ and
$U_p\simeq 2.3$ eV, or with $\Delta=4.2$ and $U_p\simeq 4.0$ eV. These
values, particularly the second set, are perfectly acceptable and in
the usually considereed range.\cite{Zaa90,Boc92} Taking the above value
of $U\simeq 3.7$ and $J=150$ meV, one finds $t=0.37$ eV,
a somewhat lower value than that which
follows from the effective model of Appendix \ref{d4:eff}. Altogether,
these results indicate, contrary to what is frequently assumed,
\cite{Zaa90,vdM88} that the local exchange interactions are somewhat
screened in reality by covalency effects, and that at the same time
the screening of the intraorbital Coulomb interaction $U$ is stronger
than estimated before.\cite{Miz96,Boc92}

\section{Cubic titanites}
\label{sec:d1}

\subsection{Spin-orbital superexchange model}
\label{sec:d1sup}

Perovskite titanates, LaTiO$_3$ and YTiO$_3$, are intriguing examples of
Mott insulators with orbital degrees of freedom due to $t_{2g}$
electrons: in the ground state the Ti$^{3+}$ ions are in the $t_{2g}^1$
configuration. In an ideal perovskite structure the $t_{2g}$ orbitals
are degenerate, but lattice distortions may contribute to the magnetic
ground state\cite{Cwi03,Moc03,Sol04,Pav05} and to the Mott transition
\cite{Pav04} --- here we do not intend to discuss this
controversial issue.

In an ideal cubic system each $t_{2g}$ orbital is perpendicular to a
single cubic axis, for instance the $|xy\rangle$ orbital lies in the
$ab$ plane and is perpendicular to the $c$ axis. It is therefore
convenient to introduce the following short hand notation for the
orbital degree of freedom:\cite{Kha00}
\begin{equation}
|a\rangle \equiv |yz\rangle, \hskip .7cm
|b\rangle \equiv |zx\rangle, \hskip .7cm
|c\rangle \equiv |xy\rangle.
\label{t2g}
\end{equation}
The labels $\gamma=a,b,c$ thus refer to the cubic axes perpendicular to
the planes of the respective orbitals.

The superexchange spin-orbital model (\ref{Hngamma}) in cubic titanates
couples $S=1/2$ spins and the orbital $t_{2g}$ degrees of freedom at
nearest neighbor Ti$^{3+}$ ions. Due to large $U$ the electron densities
satisfy thereby the local constraint at each site $i$,
\begin{equation}
n_{ia}+n_{ib}+n_{ic}=1.
\label{cond1}
\end{equation}
In titanates there is no need to consider CT processes, as these systems
are Mott-Hubbard insulators\cite{Ima98} and no qualitatively new effects
apart from some negligible renormalization of the effective parameters
$\{J,\eta\}$ could arise from CT excitations.
This simplifies our considerations, so we analyze
the superexchange in the leading order of perturbation
theory, given by the contributions which result from virtual excitations
between the neighboring Ti$^{3+}$ ions,
$(t_{2g}^1)_i(t_{2g}^1)_j\rightleftharpoons (t_{2g}^2)_i(t_{2g}^0)_j$.
These charge excitations involve the Coulomb interactions in the $d^2$
configuration of a Ti$^{2+}$ ion, parametrized as before by the
intraorbital Coulomb element $U$, and by Hund's exchange element $J_H$
for a pair of $t_{2g}$ electrons, defined as follows (see Table
\ref{tab:uij}),\cite{Gri71}
\begin{equation}
J_H=3B+C.
\label{JHt}
\end{equation}
The charge excitations lead to one of four different excited states
\cite{Gri71} shown in Fig. \ref{fig:levels}(b):
the high-spin $^3T_1$ state at energy $U-3J_H$, and three low-spin
states --- degenerate $^1T_2$ and $^1E$ states at energy $U-J_H$, and
an $^1\!A_1$ state at energy $U+2J_H$. As
before, the excitation energies are parameterized by $\eta$, defined as
in Eq. (\ref{eta}), and we introduce the coefficients
\begin{equation}
r_1=\frac{1}{1-3\eta}, \hskip .5cm
r_2=\frac{1}{1- \eta}, \hskip .5cm
r_3=\frac{1}{1+2\eta}.
\label{rd1}
\end{equation}
One finds the following compact expressions for the superexchange
${\cal H}_U(d^1)$ as given by Eq. (\ref{HJ}):\cite{Kha00,Kha03}
\begin{eqnarray}
\label{som11}
H_1^{(\gamma)}\!&=&\!\frac{1}{2}Jr_1
    \Big(\vec{S}_i\!\cdot\!\vec{S}_j+\frac{3}{4}\Big)
    \Big(A_{ij}^{(\gamma)}\!-\frac{1}{2}n_{ij}^{(\gamma)}\Big),    \\
\label{som12}
H_2^{(\gamma)}\!&=&\! \frac{1}{2}Jr_2
    \Big(\vec{S}_i\!\cdot\!\vec{S}_j-\frac{1}{4}\Big)
    \Big(A_{ij}^{(\gamma)}\!-\frac{2}{3}B_{ij}^{(\gamma)}\!
    +\frac{1}{2}n_{ij}^{(\gamma)}\Big),                  \nonumber \\
                                                                   \\
\label{som13}
H_3^{(\gamma)}\!&=&\! \frac{1}{3}Jr_3
    \Big(\vec{S}_i\!\cdot\!\vec{S}_j-\frac{1}{4}\Big)B_{ij}^{(\gamma)},
\end{eqnarray}
where
\begin{eqnarray}
\label{som1abn}
A_{ij}^{(\gamma)}&=&
2\Big(\vec{\tau}_i\cdot\vec{\tau}_j+\frac{1}{4}n_in_j\Big)^{(\gamma)},
                                                 \\
B_{ij}^{(\gamma)}&=&
2\Big(\vec{\tau}_i\otimes\vec{\tau}_j+\frac{1}{4}n_in_j\Big)^{(\gamma)},
                                                 \\
n_{ij}^{(\gamma)}&=&n_{i}^{(\gamma)}+n_{j}^{(\gamma)}.
\end{eqnarray}
As in Secs. III and IV the orbital (pseudospin) operators
$\{A_{ij}^{(\gamma)},B_{ij}^{(\gamma)},n_{ij}^{(\gamma)}\}$ depend on
the direction of the $\langle ij\rangle$ bond. Their form follows from
the simple observation\cite{Kha00,Har03} that only two $t_{2g}$
orbitals (flavors) are active along each cubic axis, e.g. if $\gamma=c$,
the active orbitals are $a$ and $b$ [see Eq. (\ref{t2g})], and they give
two components of the pseudospin $T=1/2$ operator $\vec{\tau}_i$. Here
the operators $\{A_{ij}^{(\gamma)},B_{ij}^{(\gamma)}\}$ describe the
interactions between the active orbitals along the particular bond,
which include the quantum effects due to their fluctuations, and take
either the form of a scalar product $\vec{\tau}_i\cdot\vec{\tau}_j$,
or lead to a similar expression,
\begin{equation}
\label{tauxtau}
\vec{\tau}_i\otimes\vec{\tau}_j=
\tau_i^x\tau_j^x-\tau_i^y\tau_j^y+\tau_i^z\tau_i^z,
\end{equation}
which involves double excitations due to $\tau_i^+\tau_j^+$ and
$\tau_i^-\tau_j^-$ terms (as in the $U<0$ Hubbard model). The
interactions along axis $\gamma$ are tuned by the number of
electrons occupying active orbitals, $n_i^{(\gamma)}=1-n_{i\gamma}$,
which is fixed by the number of electrons in the inactive orbital
$n_{i\gamma}$, because of the constraint (\ref{cond1}).

\subsection{Spin exchange constants and optical intensities}
\label{sec:d1jandk}

The exchange constant for a bond $\langle ij\rangle$ along axis $\gamma$
is obtained from Eqs. (\ref{som11})--(\ref{som13}) by averaging over the
orbital states at both sites $i$ and $j$,
\begin{eqnarray}
\label{jd1}
J_{\gamma}&=&
 \frac{1}{2}J(r_1+r_2)\Big\langle A_{ij}^{(\gamma)}\Big\rangle
 -\frac{1}{3}J(r_2-r_3)\Big\langle B_{ij}^{(\gamma)}\Big\rangle
 \nonumber \\
&-&\frac{1}{4}J(r_1-r_2)\Big\langle n_{i}^{(\gamma)}
                                   +n_{j}^{(\gamma)}\Big\rangle.
\end{eqnarray}
The cubic titanates are known to have particularly pronounced quantum
spin-orbital fluctuations, and their proper treatment requires a rather
sophisticated approach.\cite{Kha00,Kha03} Here we shall ignore
this complex quantum problem, and shall illustrate the general theory by
extracting the magnetic exchange constants from Eq. (\ref{jd1}), and the
optical spectral weights (\ref{opsa}), using an ansatz for the OO in the
ground state, in analogy to the approach employed in Secs. \ref{sec:d9}
and \ref{sec:d4} for the more classical $e_g$ systems.

In general a classical orbital state in the titanates with
GdFeO$_3$-type lattice structure can be parametrized as follows,
\begin{eqnarray}
\label{and1}
|\psi_1\rangle&=&\alpha|a\rangle + \beta|b\rangle +\gamma|c\rangle,
                                               \nonumber \\
|\psi_2\rangle&=& \beta|a\rangle +\alpha|b\rangle +\gamma|c\rangle,
                                               \nonumber \\
|\psi_3\rangle&=&\alpha|a\rangle + \beta|b\rangle -\gamma|c\rangle,
                                               \nonumber \\
|\psi_4\rangle&=& \beta|a\rangle +\alpha|b\rangle -\gamma|c\rangle,
\end{eqnarray}
with real coefficients and normalized ($\alpha^2+\beta^2+\gamma^2=1$)
wavefunctions at each site. The occupied orbitals $|\psi_i\rangle$ refer
to four sublattices ($i=1,\cdots,4$), with Ti$_1(000)$, Ti$_2(100)$,
Ti$_3(001)$, Ti$_4(101)$ positions.\cite{Miz96,Saw97}
The minus signs in $|\psi_3\rangle$ and $|\psi_4\rangle$ reflect a
mirror symmetry present in the GdFeO$_3$ structure.
Note that this state resembles $G$-type OO, and is thus different from
the $C$-type OO encountered for $e_g$ orbitals [due to the change of
sign for the $|c\rangle$ orbitals in Eqs. (\ref{and1}) along the $c$
axis].

Using the ansatz (\ref{and1}) one finds after a straightforward
calculation the exchange constants,
\begin{eqnarray}
\label{jd1c}
J_c   &=&\frac{1}{2}J\Big[(r_1+r_2r_3)(1-\gamma^2)^2     \nonumber \\
      & &\hskip .5cm -(r_1-r_2)(1-\gamma^2)\Big],                  \\
\label{jd1ab}
J_{ab}&=&\frac{1}{4}J\Big[2(r_1+r_2r_3)(\gamma^2+\alpha\beta)^2
                                                         \nonumber \\
      & &\hskip .5cm -(r_1-r_2)(1+\gamma^2)\Big].
\end{eqnarray}
They are determined once again by:
(i) the superexchange parameter $J$ given by Eq. (\ref{J}), with $t$
standing now for the effective $(dd\pi)$ hopping element,
(ii) Hund's exchange element $\eta$ (\ref{eta}),\cite{noterr} and
(iii) the orbital state (\ref{and1}), specified in the present case by
the coefficients $\{\alpha,\beta,\gamma\}$.

Following the general theory of Sec. \ref{sec:general}, the optical
excitations corresponding to the high-spin Hubbard band at energy
$U-3J_H$ $(n=1)$, and to the two low-spin Hubbard bands at $U-J_H$
$(n=2)$ and $U+2J_H$ $(n=3)$ of Fig. \ref{fig:levels}(b), have total
intensities given by the respective kinetic energies $K_n^{(\gamma)}$,
see Eq. (\ref{opsa}). Using the classical wavefunctions given by Eqs.
(\ref{and1}), one finds the following optical spectral weights:
\newline
--- for polarization along the $c$ axis,
\begin{eqnarray}
\label{wc11}
K_1^{(c)}&=&Jr_1\gamma^2(1-\gamma^2)\Big(\frac{3}{4}+s_c\Big),      \\
\label{wc12}
K_2^{(c)}&=&\frac{4}{3}Jr_2(1-\gamma^2)\Big(1-\frac{1}{4}\gamma^2\Big)
            \Big(\frac{1}{4}-s_c\Big),                              \\
\label{wc13}
K_3^{(c)}&=&\frac{2}{3}Jr_3(1-\gamma^2)^2\Big(\frac{1}{4}-s_c\Big),
\end{eqnarray}
--- for polarization in the $ab$ plane,
\begin{eqnarray}
\label{wab11}
\hskip -.5cm K_1^{(ab)}\!\!&=&\!\!Jr_1\Big[\frac{1}{2}(1\!+\!\gamma^2)
   -(\gamma^2\!+\!\alpha\beta)^2\Big]\Big(\frac{3}{4}+\!s_{ab}\Big), \\
\label{wab12}
\hskip -.5cm K_2^{(ab)}\!\!&=&\!\!Jr_2\Big[\frac{1}{2}(1\!+\!\gamma^2)
                    +\frac{1}{3}(\gamma^2\!+\!\alpha\beta)^2\Big]
             \Big(\frac{1}{4}-\!s_{ab}\Big),                         \\
\label{wab13}
\hskip -.5cm K_3^{(ab)}\!\!&=&\!\!\frac{2}{3}Jr_3(\gamma^2+\alpha\beta)^2
     \Big(\frac{1}{4}-s_{ab}\Big).
\end{eqnarray}
The kinetic energies $K_n^{(\gamma)}$ depend on the same parameters
$\{J,\eta\}$ as the exchange constants (\ref{jd1c}) and (\ref{jd1ab}),
on the orbital state (\ref{and1}) via the coefficients
$\{\alpha,\beta,\gamma\}$, and on the spin-spin correlations (\ref{spins}).

\subsection{Application to LaTiO$_3$ and YTiO$_3$}

\begin{figure}[t!]
\includegraphics[width=7.5cm]{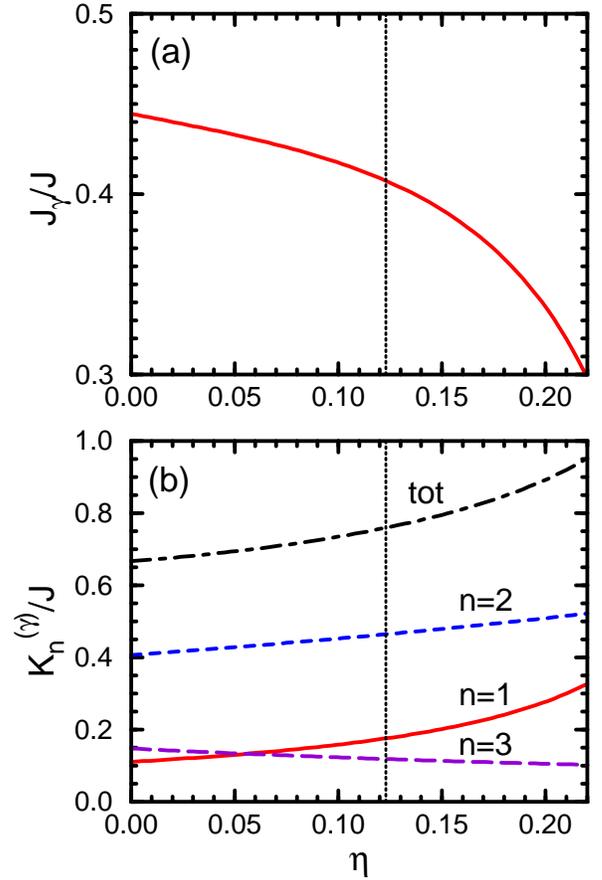}
\caption{(Color online)
Magnetic and optical properties of LaTiO$_3$, as obtained for the
classical wavefunctions (\ref{and1}) with isotropic orbital amplitudes
(\ref{tiso}), for increasing Hund's exchange $\eta=J_H/U$:
(a) exchange interactions $J_c=J_{ab}$, and
(b) kinetic energy terms $K_n^{(\gamma)}$ at $T=0$: high-spin ($n=1$,
solid line), and low-spin intermediate-energy ($n=2$, dashed line) and
high-energy excitation ($n=3$, long-dashed line).
The total kinetic energy $K^{(\gamma)}$ is shown by the dashed-dotted
line.
}
\label{fig:jd1eta}
\end{figure}

It is now straightforward to investigate the dependence of the magnetic
and the optical properties in the four-sublattice classical state
(\ref{and1}) on the effective parameters $\{J,\eta\}$ and on the type of
OO given by the coefficients $\{\alpha,\beta,\gamma\}$. {\it A priori\/},
the average electron density in the $|c\rangle$ orbitals,
$n_c=\langle n_{ic}\rangle$, is different from the densities in the other
two orbitals ($n_a$ and $n_b$), and the cubic symmetry of the expectation
values $J_{\gamma}$ and $K_n^{\gamma}$ is explicitly broken by the OO
given by Eq. (\ref{and1}), unless all the orbital amplitudes are equal,
\begin{equation}
\label{tiso}
\alpha=\beta=\gamma=\frac{1}{\sqrt{3}},
\end{equation}
as argued in Refs. \onlinecite{Moc03,Cwi03}. First we consider this
isotropic state and evaluate the exchange constants $J_{\gamma}$ for
increasing Hund's parameter $\eta$. One finds then AF superexchange
$J_{\gamma}\sim 0.4J$ which decreases with increasing $\eta$ [Fig.
\ref{fig:jd1eta}(a)], but the interactions remain AF in the physically
relevant range of $\eta<0.28$. Note that this classical consideration
seriously overestimates the actual exchange interaction as one expects
instead $J_{\gamma}\sim 0.16J$ when quantum effects are included.
\cite{Kha00} Nevertheless, having no information about the optical
excitations, we give here an example of estimating the exchange
constants and optical spectral weights using the spectroscopic values
\cite{Miz96} for Hund's exchange $J_H\simeq 0.59$, and for Coulomb
intraorbital interaction of $U=4.8$ eV, which give $\eta=0.123$.
Assuming now a hopping parameter $t=0.2$ eV, one finds $J\simeq 33$ meV
which gives $J_{\gamma}\simeq 14$ meV. While one might expect that the
effective $U$ is somewhat reduced as in the case of LaMnO$_3$, and the
value of $\eta$ would then be larger, the present crude estimate is
accidentally quite close to the experimental value of
$J_{\rm exp}=15.5$ meV.\cite{Kei00}

In the AF state, realized in LaTiO$_3$, all three Hubbard subbands
contribute in the optical spectroscopy, and taking the N\'eel state
with $s_{ab}=s_c=-0.25$ one finds the highest spectral weight at $T=0$
for $n=2$ [Fig. \ref{fig:jd1eta}(b)]. For the realistic value
$\eta\sim 0.123$ the spectral weight of the lowest-energy (high-spin)
Hubbard subband $K^{(\gamma)}_1$ is relatively weak, and is similar to
that of the $n=3$ low-spin excitation at the highest energy [Fig.
\ref{fig:jd1eta}(b)]. Note, however, that in early optical experiments
the intensity was found to be practically independent of energy
$\omega$,\cite{Ari93} so different excitations might be difficult to
separate from each other.

\begin{figure}[t!]
\includegraphics[width=7.5cm]{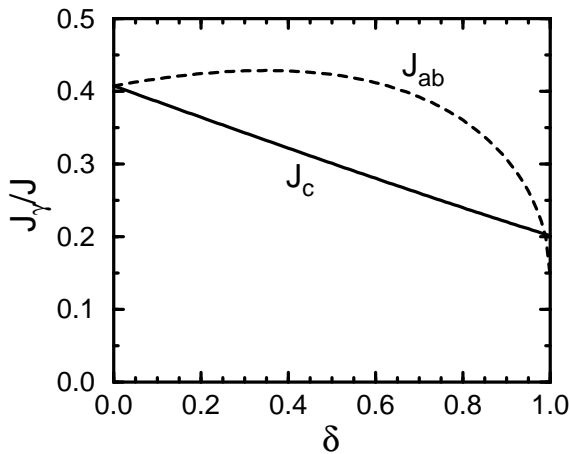}
\caption{
Exchange interactions $J_c$ and $J_{ab}$, Eqs. (\ref{jd1})
(solid and dashed line) for the titanite model for different types
of orbital order given by Eqs. (\ref{and1}). The coefficients in the
wavefunctions (\ref{and1}) vary between the isotropic state
$\alpha=\beta=\gamma=1/\sqrt{3}$ and the state with linear combinations
of $|a\rangle/|b\rangle$ and $|c\rangle$ orbitals, with the coefficients
given by: $\alpha=\gamma=1/\sqrt{3}+\delta(1/\sqrt{2}-1/\sqrt{3})$ and
$\beta=(1-\alpha^2-\gamma^2)^{1/2}$.
Parameters: $\eta=0.123$.
}
\label{fig:jd1}
\end{figure}

Let us verify now whether the wavefunctions as given by Eqs. (\ref{and1})
could also lead to isotropic AF states for other choices of orbital
amplitudes than in Eq. (\ref{tiso}). Thus, we considered equal
amplitudes $\alpha$ and $\gamma$ in the states parametrized by $\delta$,
\begin{equation}
\label{alga}
\alpha=\gamma=(1-\delta)\frac{1}{\sqrt{3}}
                +\delta \frac{1}{\sqrt{2}},
\end{equation}
with $0<\delta<1$. The normalization condition gives
$\beta=(1-\alpha^2-\gamma^2)^{1/2}$ which vanishes at $\delta=1$. As
expected, the exchange interactions are anisotropic for small $\delta>0$
(Fig. \ref{fig:jd1}). The increasing occupancy of the $c$ orbitals with
increasing $\delta$ results in a somewhat enhanced exchange interaction
$J_{ab}$ near the isotropic state (for $\delta<0.5$), while the
interaction $J_c$ decreases almost linearly.\cite{noteli}
At larger orbital anisotropy the interaction $J_{ab}$ decreases as well.
It is remarkable that the AF interactions become again isotropic close
to $\alpha=\gamma=1/\sqrt{2}$ which gives a large charge density
$n_c\simeq 0.5$ in the $|c\rangle$ orbitals, but this feature is not
generic and will likely be modified by quantum effects. Note also that
the state at $\beta=0$ in not a FM state, contrary to some earlier
suggestions.\cite{Miz96}

Finally we remark that the temperature dependence of the optical
spectral weights can again be studied in a similar way as for
LaMnO$_3$, by solving the spin-spin correlation functions at increasing
temperature using the method of Appendix \ref{app:spincomn} for $S=1/2$
spins. Having no experimental data neither for LaTiO$_3$ nor for
YTiO$_3$, we shall limit ourselves to making predictions concerning the
overall change of the spectral weights between low temperature $T\sim 0$
where the spin-spin correlations are maximal, and room temperature
$T\gg T_N$ (or $T\gg T_C$), where the spin-spin correlations may be
neglected. We take rigid classical wave functions with OO (\ref{and1})
and discuss first AF states. Apart from the isotropic AF phase
(\ref{tiso}), called state $A$, we consider in Table \ref{tab:d1} also
somewhat modified orbital amplitudes, as proposed by Itoh
{\it et al.\/}:\cite{Ito99}
\begin{equation}
\label{itoh}
\alpha=0.690, \hskip .7cm \beta=0.452, \hskip .7cm \gamma=0.565,
\end{equation}
called here state $B$. One finds that the exchange interactions $J_c$
and $J_{ab}$ are then rather anisotropic (see Table \ref{tab:d1}), so
we believe that this state can be experimentally excluded
on the basis of the neutron data.\cite{Kei00}

\begin{table}[b!]
\caption{
Exchange constants $J_{\gamma}$, kinetic energies for Hubbard subbands
$K_n^{(\gamma)}$ and total kinetic energies $K^{(\gamma)}$ (all in
units of $J$) for two orbital ordered states suggested for LaTiO$_3$:
state $A$ (\ref{tiso}), and
state $B$ (\ref{itoh}) reported in Ref. \onlinecite{Ito99},
and for the OO state suggested for YTiO$_3$ in Ref. \onlinecite{Kha03}.
Parameter: $\eta=0.123$.}
\vskip .2cm
\begin{ruledtabular}
\begin{tabular}{cccccccc}
&  & \multicolumn{4}{c}{LaTiO$_3$} & \multicolumn{2}{c}{YTiO$_3$}   \cr
&  & \multicolumn{2}{c}{state $A$ (\ref{tiso})}   &
     \multicolumn{2}{c}{state $B$ (\ref{itoh})}   &
     \multicolumn{2}{c}{state $F$ (\ref{fmd1})}  \cr
&  & $T=0$ & $T\gg T_N$ & $T=0$ & $T\gg T_N$ & $T=0$ & $T\gg T_C$   \cr
\colrule
& $J_c$    & \multicolumn{2}{c}{0.407} & \multicolumn{2}{c}{0.428}  &
                                         \multicolumn{2}{c}{-0.148} \cr
& $J_{ab}$ & \multicolumn{2}{c}{0.407} & \multicolumn{2}{c}{0.351}  &
                                         \multicolumn{2}{c}{-0.148} \cr
\colrule
& $K_1^{(c)}$  & 0.176 & 0.264 & 0.172 & 0.258 & 1.057 & 0.792  \cr
& $K_2^{(c)}$  & 0.465 & 0.232 & 0.476 & 0.238 & 0.000 & 0.190  \cr
& $K_3^{(c)}$  & 0.118 & 0.059 & 0.124 & 0.062 & 0.000 & 0.000  \cr
& $K^{(c)}$    & 0.759 & 0.555 & 0.772 & 0.558 & 1.057 & 0.982  \cr
\colrule
& $K_1^{(ab)}$ & 0.176 & 0.264 & 0.207 & 0.311 & 1.057 & 0.792  \cr
& $K_2^{(ab)}$ & 0.465 & 0.232 & 0.451 & 0.226 & 0.000 & 0.190  \cr
& $K_3^{(ab)}$ & 0.118 & 0.059 & 0.107 & 0.053 & 0.000 & 0.000  \cr
& $K^{(ab)}$   & 0.759 & 0.555 & 0.765 & 0.590 & 1.057 & 0.982  \cr
\end{tabular}
\end{ruledtabular}
\label{tab:d1}
\end{table}

Due to the AF order in the ground state for the wavefunctions $A$ and
$B$, the lowest energy high-spin excitations are suppressed at low
temperature, but we predict that their weight should increase by about
$50$\% above the magnetic transition (Table \ref{tab:d1}). An even
stronger temperature dependence is predicted for the low-spin part of
the spectra with its intensity decreasing by a factor 2 when the
magnetic order is lost. Thus, the spectral weight is shifted from the
high-energy to the low-energy part of the optical spectrum with
increasing temperature, and the total spectral weight decreases, both
features being typical of AF bonds. An experimental confirmation of this
trend would also allow one to identify the energy splitting between the
high-spin and low-spin parts of the upper Hubbard band, and to determine
the value of Hund's exchange $J_H$ from experiment. Although the
anisotropy in spectral weights found in the state (\ref{itoh}) proposed
by Itoh {\it et al.}\cite{Ito99} might be too weak to be seen
experimentally, the predicted anisotropy in the exchange interactions
is certainly detectable.

It is quite remarkable that an isotropic FM ground state of YTiO$_3$ can
also be described classically by the OO state with four sublattices, but
the phase factors in the wavefunctions have to be then selected
differently:\cite{Kha03}
\begin{eqnarray}
\label{fmd1}
|\phi_1\rangle&=&\frac{1}{\sqrt{3}}
                 (  |a\rangle + |b\rangle + |c\rangle),  \nonumber \\
|\phi_2\rangle&=&\frac{1}{\sqrt{3}}
                 (- |a\rangle - |b\rangle + |c\rangle),  \nonumber \\
|\phi_3\rangle&=&\frac{1}{\sqrt{3}}
                 (- |a\rangle + |b\rangle - |c\rangle),  \nonumber \\
|\phi_4\rangle&=&\frac{1}{\sqrt{3}}
                 (  |a\rangle - |b\rangle - |c\rangle).
\end{eqnarray}
We call this state the orbital $F$ state. Neglecting again orbital
(quantum) fluctuations, one finds
$\langle A_{ij}^{(\gamma)}\rangle=\langle B_{ij}^{(\gamma)}\rangle=0$ in
this state, while $n_{i}^{(\gamma)}=\frac{4}{3}$. At finite $\eta$ this
leads indeed to an isotropic FM state with exchange constants,
\begin{equation}
\label{jf1}
J_c^{\rm FM}=J_{ab}^{\rm FM}=-\frac{2}{3}J\eta r_1r_2,
\end{equation}
in lowest order $\propto\eta$,\cite{notefet} so this type of OO leads
to a markedly different exchange constant from that given by Eq.
(\ref{jd1c}). Assuming again the same parameters as above one finds
$J^{\rm FM}_{\gamma}\simeq -5$ meV, which is somewhat higher than the
experimental value $J^{\rm exp}_{\gamma}\simeq -3$ meV.\cite{Ulr02}
However, in view of the simplicity of the present considerations this
agreement can be regarded as satisfactory.

Consider now the spectral weights in the optical spectroscopy. One finds
rather simple expressions for the (isotropic) optical spectral weights
for the $F$ state:
\begin{eqnarray}
\label{wf11}
K_1^{(\gamma)}&=&\frac{2}{3}Jr_1\Big(\frac{3}{4}+s_\gamma\Big),     \\
\label{wf12}
K_2^{(\gamma)}&=&\frac{2}{3}Jr_2\Big(\frac{1}{4}-s_\gamma\Big),     \\
\label{wf13}
K_3^{(\gamma)}&=&0.
\end{eqnarray}
Thus, the optical spectrum of YTiO$_3$ would be quite different from
that of LaTiO$_3$ (see Table \ref{tab:d1}). For the FM state at $T=0$
(with $s_\gamma=\frac{1}{4}$) only the high-spin spectral weight
$K_1^{(\gamma)}=\frac{2}{3}Jr_1$ contributes, and for the present
parameters $K_1^{(\gamma)}\simeq J$. Thus, future optical experiments
on YTiO$_3$ combined with magnetic experiments could help to determine
the superexchange constant $J$. Furthermore, for the FM $F$ orbital
state (\ref{fmd1}) one finds at high temperature a moderate reduction
of the low-energy spectral weight by $\sim 25$\%, and a corresponding
increase for the higher energy excitation $n=2$ [the third excitation
does not contribute as long as the OO remains the same as in Eqs.
(\ref{fmd1})]. The energy difference between these excitations is $2J_H$
(Fig. \ref{fig:levels}), so they might serve to determine this parameter
from the optical spectroscopy.

The above results show once again that the magnetic interactions and
the optical spectral weights depend in a crucial way on the underlying
orbital state. Once the OO in $F$ state (\ref{fmd1}) has been fixed, the
exchange constants $J_c^{\rm FM}$ (\ref{jf1}) were completely determined
by the superexchange $J$ (\ref{J}), and by Hund's exchange parameter
$\eta$. For the optical spectral weights $K_n^{(\gamma)}$
(\ref{wf11})--(\ref{wf13}) one needs in addition the spin-spin
correlation function $s_{\gamma}$. Most importantly, a strong
temperature dependence of the optical spectra follows here from a
classical orbital picture, and this should be a crucial experimental
test of the validity of the present approach.

\section{Cubic vanadates}
\label{sec:d2}

\subsection{Spin-orbital superexchange model}
\label{sec:d2sup}

The last example of spin-orbital physics we want to give here are the
cubic vanadates, in which the orbital degrees of freedom originate from
the $t_{2g}^2$ configuration of the V$^{3+}$ ions present in the ground
state. Here our goal will be to answer the question to what extent these
systems could be described by decoupling spin and orbital degrees of
freedom and assuming classical states with OO. The OO in the
cubic vanadates is complementary to the spin order ---
the $C$-type AF phase is found for $G$-type OO, and
the $G$-type AF phase is accompanied by $C$-type OO.\cite{Miy03}
Here we shall consider in more detail $G$-type OO, relevant for the case
of LaVO$_3$. We will show in particular that the superexchange 
spin-orbital model\cite{Kha01} allows one to understand the microscopic 
reasons behind the $C$-AF phase observed in LaVO$_3$, and predicts that 
$|J_c|\sim J_{ab}$, as actually observed.\cite{Kei05}
The comparable size of FM and AF exchange constants $J_c$ and $J_{ab}$,
respectively, is unexpected when considering the
Goodenough-Kanamori-Anderson rules,\cite{Goode} which would suggest
that $|J_c|$ is by a factor $J_H/U$ smaller.

In the case of the vanadates the superexchange between the $S=1$ spins 
of V$^{3+}$ ions in a perovskite lattice results from virtual charge 
excitations
$(t_{2g}^2)_i(t_{2g}^2)_j\rightleftharpoons (t_{2g}^3)_i(t_{2g}^1)_j$.
These charge excitations involve the Coulomb interactions in the $d^3$
configuration of a V$^{2+}$ ion, parametrized for a pair of $t_{2g}$
electrons, as for the titanates, by the intraorbital Coulomb element
$U$, and by Hund's exchange element $J_H$ [see Eq. (\ref{JHt}) and 
Table \ref{tab:uij}].\cite{Gri71} The excitation spectrum which
leads to the superexchange model includes three states:\cite{Kha01}
  (i) a high-spin state $^4\!A_2$ at energy $U-3J_H$,
 (ii) two degenerate low-spin states $^2T_1$ and $^2E$ at energy $U$,
      and
(iii) a $^2T_2$ low-spin state at energy $U+2J_H$
(Fig. \ref{fig:levels}). We parametrize it by two coefficients:
$r_1=1/(1-3\eta)$ and $r_3=1/(1+2\eta)$.
A general Hamiltonian was already given in Ref. \onlinecite{Kha01};
here we shall analyze it assuming that the $xy$ ($|c\rangle$) orbitals
are singly occupied at each V$^{3+}$ ion, as concluded from
experiment\cite{Mah92,Bla01} and from electronic structure
calculations.\cite{Saw96} Therefore, the electron densities in the
remaining two orbitals satisfy at each site $i$ the local constraint,
\begin{equation}
n_{ia}+n_{ib}=1.
\label{cond2}
\end{equation}
One finds then that the superexchange ${\cal H}_U(d^2)$ for a bond
${\langle ij\rangle}$ along the $c$ axis consists of
\begin{eqnarray}
\label{H1cd2}
H_1^{(c)}&=&-\frac{J}{3}r_1 (2+\vec S_i\!\cdot\!\vec S_j)
\Big(\frac{1}{4}-\vec \tau_i\cdot\vec \tau_j\Big),                  \\
\label{H2cd2}
H_2^{(c)}&=&-\frac{J}{12} (1-\vec S_i\!\cdot\!\vec S_j)   \nonumber \\
&&\times\Big(\frac{7}{4}-\tau_i^z\tau_j^z-\tau_i^x\tau_j^x
+5\tau_i^y \tau_j^y\Big),                                           \\
\label{H3cd2}
H_3^{(c)}&=&-\frac{J}{4}r_3 (1-\vec S_i\!\cdot\!\vec S_j) \nonumber \\
&&\times\Big(\frac{1}{4}+\tau_i^z\tau_j^z+\tau_i^x\tau_j^x
-\tau_i^y \tau_j^y\Big),
\end{eqnarray}
and for a bond in the $ab$ plane,
\begin{eqnarray}
\label{H1ad2}
H_1^{(ab)}&=&-\frac{1}{6}Jr_1\Big(\vec S_i\!\cdot\!\vec S_j+2\Big)
\Big(\textstyle{\frac{1}{4}}-\tau_i^z\tau_j^z\Big).                 \\
\label{H2ad2}
H_2^{(ab)}&=&-\frac{1}{8}J\Big(1-\vec S_i\!\cdot\!\vec S_j\Big)
                                                          \nonumber \\
&&\times\Big(\frac{19}{12}\mp \frac{1}{2}\tau_i^z
\mp \frac{1}{2}\tau_j^z-\frac{1}{3}\tau_i^z\tau_j^z\Big),           \\
\label{H3ad2}
H_3^{(ab)}&=&-\frac{1}{8}Jr_3\Big(1-\vec S_i\!\cdot\!\vec S_j\Big)
                                                          \nonumber \\
&&\times\Big(\frac{5}{4}\mp \frac{1}{2}\tau_i^z
\mp \frac{1}{2}\tau_j^z+\tau_i^z\tau_j^z\Big),
\end{eqnarray}
where the operators $\vec\tau_i$ describe orbital pseudospins $T=1/2$
defined (for each direction $\gamma$) by the orbital doublet
$\{|yz\rangle,|xz\rangle\}\equiv\{|a\rangle,|b\rangle\}$. At each site
$i$ there is precisely one electron in these two orbitals, and both of
them are active along the $c$ axis.

It has already been shown before\cite{Kha04} that the quantum 
fluctuations play a decisive role in LaVO$_3$, and the observed
\cite{Miy02} anisotropy and temperature dependence of the high-spin 
excitations in optical spectroscopy is reproduced when the theory 
includes them. In the next Sec. \ref{sec:d2oo} we will present now the 
limitations of a simplified approach which is widely accepted
\cite{Bla01} and assumes that the OO in LaVO$_3$ is quite rigid already 
at the magnetic transition. Quantum fluctuations lead to important 
corrections which go beyond this picture and modify the temperature 
variation of the optical spectral weights, as we discuss in Sec. 
\ref{sec:d2mot}.

\subsection{Spin exchange constants and optical spectral weights
            for alternating orbital order}
\label{sec:d2oo}

After decoupling of spin and orbital variables,
the effective spin exchange constants $J_c$ and $J_{ab}$ in Eq.
(\ref{Hs}) can be obtained from Eqs. (\ref{H1cd2})--(\ref{H3ad2}) by
averaging over orbital correlations, as derived in Ref.
\onlinecite{Kha04},
\begin{eqnarray}
\label{Jcd2}
J_c&=&
-\frac{1}{2}J\Big[\eta r_1-(r_1-\eta r_1-\eta r_3)       \nonumber \\
&&\times\Big(\frac{1}{4}+\langle\vec\tau_i\!\cdot\!\vec\tau_j\rangle\Big)
 -2\eta r_3\langle \tau_i^y \tau_j^y \rangle\Big],                 \\
\label{Jabd2}
J_{ab}&=&\frac{1}{4}J\Big[1-\eta r_1-\eta r_3            \nonumber \\
&&+(r_1-\eta r_1-\eta r_3)\Big(\frac{1}{4}
+\langle\tau_i^z\tau_j^z\rangle\Big)\Big].
\end{eqnarray}
They are determined by the orbital correlations, which result not only
from the superexchange ${\cal H}_U(d^2)$ but also from the interactions
with the lattice (from the JT term),\cite{Kha04} as discussed in detail
in Sec. \ref{sec:d2mot}. Here we shall consider first classical states
with alternating $|a\rangle$ and $|b\rangle$ orbitals in the $ab$
planes:
$G$-type OO when these orbitals alternate also along the $c$ axis, and
$C$-type OO for repeated either $|a\rangle$ or $|b\rangle$ orbitals
along the $c$ axis. These classical states are frequently assumed as a
consequence of a strong JT term and observed GdFeO$_3$ distortions.
\cite{Bla01} Such classical order would naturally follow from a strong
JT effect, but it is still controversial whether the JT interaction is
actually that strong in the vanadates. Fortunately, there are already
experimental results which can help to resolve this controversy, and we
address this issue in more detail in Sec. \ref{sec:d2mot}.

First we consider the limit of strong JT interaction with rigid OO.
This implies $\langle\vec\tau_i\!\cdot\!\vec\tau_j\rangle\simeq
\langle\tau_i^z\tau_j^z\rangle=-\frac{1}{4}$ along each bond for
$G$-type OO, so one finds from Eqs. (\ref{Jcd2}) and (\ref{Jabd2})
fixed values of the exchange constants:
\begin{eqnarray}
\label{Jccaf}
J_c^{C-{\rm AF}}&=&-\frac{1}{2}J\eta r_1,                 \\
\label{Jabcaf}
J_{ab}^{C-{\rm AF}}&=&\frac{1}{4}J[1-\eta(r_1+r_3)].
\end{eqnarray}
The FM interaction $J_c^{C-{\rm AF}}$
increases in lowest order linearly with Hund's exchange $\eta$, and
the above values of the exchange constants give $C$-AF spin order.

Whether or not such a classical OO is realized in the ground state,
can be investigated by analyzing the consequences of the present
theory for the distribution of spectral weight in the optical
spectroscopy. The optical spectral weights follow by averaging the
individual contributions to the superexchange, see Eqs. (\ref{hefa})
and (\ref{opsa}). One finds for $G$-type OO the optical spectral
weights:
\newline
--- for polarization along the $c$ axis,
\begin{eqnarray}
\label{wc21}
K_1^{(c)}&=&\frac{1}{3}Jr_1\big(s_c+2\big),         \\
\label{wc22}
K_2^{(c)}&=&\frac{1}{3}J   \big(1-s_c\big),         \\
\label{wc23}
K_3^{(c)}&=&0,
\end{eqnarray}
--- for polarization in the $ab$ plane,
\begin{eqnarray}
\label{wa21}
K_1^{(ab)}&=&\frac{1}{6}Jr_1\big(s_{ab}+2\big),     \\
\label{wa22}
K_2^{(ab)}&=&\frac{5}{12}J  \big(1-s_{ab}\big),     \\
\label{wa23}
K_3^{(ab)}&=&\frac{1}{4}Jr_3\big(1-s_{ab}\big).
\end{eqnarray}
As in all other cases, they depend on two model parameters, $J$ and
$\eta$, and on the spin-spin correlations $\{s_c,s_{ab}\}$ (the OO
is already fixed).

Again, as in the case of LaMnO$_3$ (Sec. IV), the analysis of the
optical spectra suggests that the effective parameters are somewhat
different from the atomic values, primarily due to the screening of
both Coulomb $U$ and Hund's exchange $J_H$. We use here the parameters
deduced from the neutron experiments\cite{Ulr03} and from the optical
spectra\cite{Miy02} --- one finds\cite{Kha04} $J=40$ meV and
$\eta=0.13$. These values imply that $J_c\simeq -4.3$ meV and
$J_{ab}\simeq 6.8$ meV, which lead to a MF estimate of the transition
temperature $T_N^{\rm MF}\simeq 270$K. This value has to be still
reduced by an empirical factor\cite{Fle04} (close to 68\% for $S=1$) to
estimate the effect of thermal fluctuations, so one finds $T_N\sim 180$
K in reasonable agreement with the experimental value of 140 K.
\cite{Miy00} We note that the above values of the exchange constants
are in good agreement with the neutron experiments.\cite{Kei05}

\begin{figure}[t!]
\includegraphics[width=7.5cm]{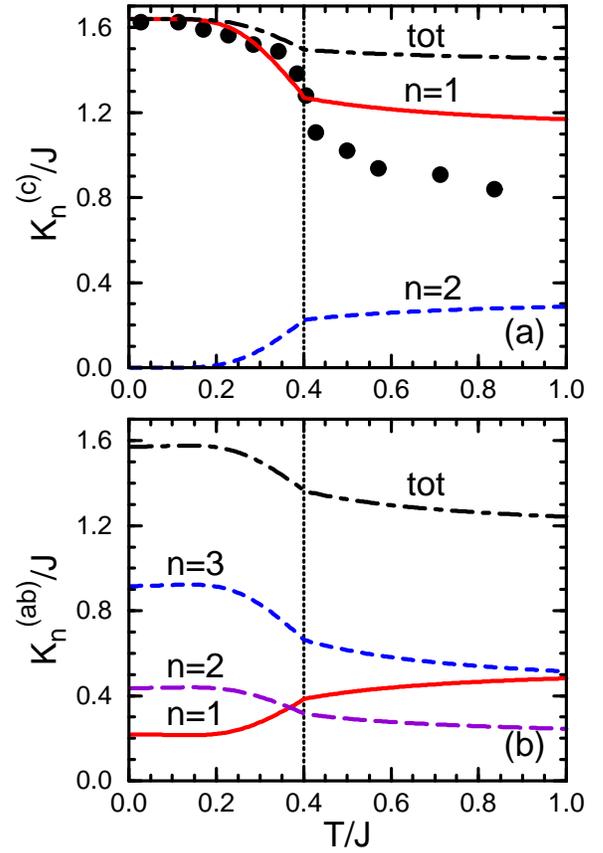}
\caption{
Kinetic energy terms (in units of $J$) per bond (\ref{hefa}),
as obtained for the $C$-AF phase of the cubic vanadates:
high-spin $K_1^{(\gamma)}$ (solid lines) and low-spin
$K_2^{(\gamma)}$ (dashed lines) and $K_3^{(\gamma)}$ (long-dashed lines),
for increasing temperature $T/J$:
(a) along the FM $c$ axis ($\gamma=c$);
(b) within the AF $ab$ plane ($\gamma=ab$).
In part (a) the experimental points from Ref. \onlinecite{Miy03} for the
low-energy spectral weight were reproduced, after proper rescaling to
match the value of $K_1^{(\gamma)}$ at $T\to 0$.
Parameters: $\eta=0.13$.
}
\label{fig:kd2}
\end{figure}

It is instructive to test this classical approach by analyzing its
predictions for the optical spectral weights. We evaluated the spin
correlation functions for bonds in the $ab$ plane and along the $c$
axis using the Oguchi method, and used an order parameter
$\langle S^z\rangle$ at neighboring sites, which acts on the
considered bond by MF terms and vanishes at $T_N=0.4J$, as explained
in Ref. \onlinecite{Kha04}. The correlation $s_c$ for FM bonds along
the $c$ axis can be obtained analytically,\cite{Kha04} while $s_{ab}$
for the AF bonds was determined by a numerical approach described in
Appendix \ref{app:spincov}. As expected, the spectral weight due to
the high-spin excitations dominates for $c$ polarization. However,
when the OO is perfect, the anisotropy between $K_1^{(c)}$ and
$K_1^{(ab)}$ at $T=0$, being close to 8:1 (Fig. \ref{fig:kd2}), is
now less pronounced than in the case of joint spin and orbital
dynamics.\cite{Kha04} At low temperature the low-spin excitations
dominate the optical spectral weight for $ab$ polarization.
One finds that the low-energy spectral weight along the $c$ axis
$K_1^{(c)}$ decreases with increasing temperature. Simultaneously, the
low-energy spectral weight in the $ab$ planes $K_1^{(ab)}$ increases,
and the anisotropy goes down to $\sim 5:2$ at $T=0.85J$ (i.e., at
$T\simeq 300$ K for $T_N=140$ K). It is quite remarkable that the
present classical approach gives qualitatively a very similar
distribution of the spectral weights and their temperature dependence
for the FM and AF bonds in LaVO$_3$ and in LaMnO$_3$ (cf. Figs.
\ref{fig:anis} and \ref{fig:kd2}).

\begin{table}[b!]
\caption{
Exchange constants $J_{\gamma}$ and optical spectral weights
$K_n^{(\gamma)}$ (all in units of $J$) as obtained for:
LaVO$_3$ ($C$-AF phase with $G$-type OO), and for
 YVO$_3$ ($G$-AF phase with $C$-type OO)
at $T=0$ and above the magnetic transition at $T=0.85J$ ($\sim 300$ K).
Parameters: $\eta=0.13$ and $T_N=140$ K.
}
\vskip .2cm
\begin{ruledtabular}
\begin{tabular}{cccccc}
&         & \multicolumn{2}{c}{LaVO$_3$}&
            \multicolumn{2}{c}{YVO$_3$}           \cr
&         & \multicolumn{2}{c}{$C$-AF phase}&
            \multicolumn{2}{c}{$G$-AF phase}      \cr
&  $T$ (K)  &     $0$  &  $300$ &  $0$ &  $300$   \cr
\colrule
& $J_c$     & \multicolumn{2}{c}{-0.107} & \multicolumn{2}{c}{0.224} \cr
& $J_{ab}$  & \multicolumn{2}{c}{0.171}  & \multicolumn{2}{c}{0.171} \cr
\colrule
& $K_1^{(c)}$  & 1.640 & 1.181 & 0.0   & 0.0      \cr
& $K_2^{(c)}$  & 0.0   & 0.280 & 0.586 & 0.494    \cr
& $K_3^{(c)}$  & 0.0   & 0.0   & 0.465 & 0.392    \cr
\colrule
& $K_1^{(ab)}$ & 0.219 & 0.471 & 0.249 & 0.471    \cr
& $K_2^{(ab)}$ & 0.916 & 0.532 & 0.871 & 0.532    \cr
& $K_3^{(ab)}$ & 0.436 & 0.253 & 0.415 & 0.253    \cr
\end{tabular}
\end{ruledtabular}
\label{tab:d2}
\end{table}

A crucial test of the present theory concerns the temperature
dependence of the high-spin spectral weight along the $c$ axis
$K_1^{(c)}$, which according to experiment decreases by about 50\%
between low temperature and $T=300$ K.\cite{Miy02} In the present
theory based upon frozen OO this decrease amounts only to about 27\%
[Fig. \ref{fig:kd2}(a)], and the maximal possible reduction of
$K_1^{(c)}$ reached in the limit of $T\to\infty$ is by 33\%. This result
suggests that the frozen OO scenario in LaVO$_3$ is {\it excluded by
experiment\/}; further arguments supporting this point of view are given
in Sec. \ref{sec:d2mot}. Note also that the actual values of
$K_1^{(c)}$, shown in Refs. \onlinecite{Kha04} and \onlinecite{Ole04},
had to be reduced to match the classical prediction at $T=0$.

A second classical state with $C$-type OO, as proposed for the low
temperature $G$-AF phase of YVO$_3$,\cite{Bla01} gives an AF exchange
constant along the $c$ axis,
\begin{equation}
J_c^{G-{\rm AF}}=\frac{1}{4}J(1-\eta r_3),
\label{jcgaf}
\end{equation}
while the value of $J_{ab}^{G-{\rm AF}}$ is the same as
$J_{ab}^{C-{\rm AF}}$, see Eq. (\ref{Jabcaf}), so in the present
classical approach it does not change when the OO changes from the
$G$-type to $C$-type. In contrast, depending on the type of OO, the
exchange constant along the $c$ axis can be either FM or AF, as
suggested by experiment\cite{Miy03} and confirmed by the data in Table
\ref{tab:d2}. The actual values estimated with the same parameters as
for LaVO$_3$ are: $J_c=9.0$ and $J_{ab}=6.8$ meV. Here one finds that
$J_c>J_{ab}$ as in experiment, but the agreement with the experimental
values,\cite{Ulr03} $J_c^{\rm exp}=3.1$ and $J_{ab}^{\rm exp}=2.6$ meV,
is much poorer. Hence, we conclude that quantum effects beyond the
present classical analysis such as orbital fluctuations,\cite{Kha01}
orbital Peierls effect,\cite{Ulr03,Sir03} and spin-orbit coupling,
\cite{Hor03} play here an important role and have to be included in
a quantitative theory.

Furthermore, while the weights in the $ab$ planes are given by
spin-spin correlations $s_{ab}$, as in Eqs. (\ref{wa21})--(\ref{wa23}),
for the spectral weights in the $C$-type OO one finds along the $c$ axis
\begin{eqnarray}
\label{wc21y}
K_1^{(c)}&=&0,     \\
\label{wc22y}
K_2^{(c)}&=&\frac{1}{4}J   \big(1-s_c\big),     \\
\label{wc23y}
K_3^{(c)}&=&\frac{1}{4}Jr_3\big(1-s_c\big).
\end{eqnarray}
We evaluated these spectral weights (see Table \ref{tab:d2}) for the
standard value of $\eta=0.13$ and for the spin-spin correlation function
$s_c$ determined within the Oguchi method for an AF bond along the $c$
axis (Appendix \ref{app:spincov}), stabilized by the above value of the
AF interaction $J_c$, and taking again $T_N=0.4J$. It is along this axis
where the theory predicts a markedly different behavior of the spectral
weights from that found above for $G$-type OO, see Table \ref{tab:d2}.
In spite of AF bonds in all three directions, the spectral weight
distribution is again anisotropic --- the weights are considerably
higher for the $ab$ planes due to the broken cubic symmetry in orbital
space. In addition, the spectral weights obtained in $ab$ planes at low
temperature differ between the two AF phases, as the MF terms are larger
in $G$-AF phase and make this case somewhat closer to the classical
limit of $s_{ab}=-1$.
Particularly the prediction that $K_1^{(c)}=0$, following from the
classical $C$-type OO, is easy to verify. In fact, the experimental
data are puzzling as one finds instead finite and temperature dependent
spectral weight also for the low-energy regime in the $G$-AF phase of
YVO$_3$,\cite{Miy02,Tsv04} pointing out once again that the present
calculation with frozen OO is oversimplified.

\subsection{ Composite spin-orbital dynamics in LaVO$_3$ }
\label{sec:d2mot}

Finally, we demonstrate that the scenario of strong JT interaction,
quenching the orbital dynamics, cannot apply to LaVO$_3$. We do so by
investigating its consequences for the orbital transition temperature
$T_{\rm o}$ and for the temperature dependence of the optical intensity
$K_1^{(c)}$ of the lowest multiplet transition.
Consider first the transition temperature $T_{\rm o}$ associated with
the phase transtion into the state with OO.
We have already seen in Secs. III and IV that strong orbital-lattice
coupling in a perovskite structure would in fact necessarily decouple
orbital and spin degrees of freedom and lead to $T_{\rm o}\gg T_N$,
contradicting the experiment.\cite{Miy02}

\begin{figure}[t!]
\includegraphics[width=7.7cm]{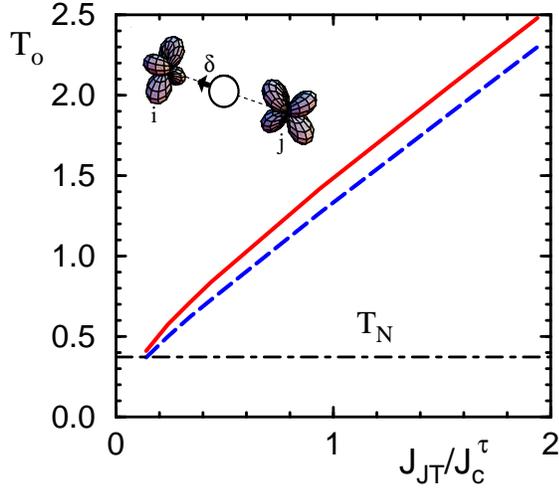}
\caption{(Color online)
Orbital transition temperature $T_{\rm o}$ for increasing JT
interaction $J_{\rm JT}$ in:
orbital-only model [Eq. (\ref{omo}), dashed line], and
spin-orbital model [Eqs. (\ref{omo}) and (\ref{stau}), solid line],
compared with the N\'eel temperature $T_N$ (dashed-dotted line).
Both $T_{\rm o}$ and $T_N$ in units of $J_c^{\tau}$. The inset shows
intersite orbital correlations due to JT distortions in LaVO$_3$.
}
\label{fig:to}
\end{figure}

The JT coupling between the JT-active local lattice modes $Q_i$ and the
pseudospin (orbital) variables $\tau_i^z$ (which refer to the active
$\{|a\rangle,|b\rangle\}$ orbitals along the $c$ axis) may be written as
follows,
\begin{equation}
\label{jtcoup}
{\cal H}_{\rm JT}=g\sum_i Q_i\tau_i^z+\frac{1}{2}\sum_i Q_i^2,
\end{equation}
where $Q_i$ is the appropriate linear combination of coordinates of the
ligand ions next to site $i$. The above local coupling induces local
distortions and an associated energy gain of $E_{\rm JT}=g^2/8$ (the
JT energy) per site, and moreover generates a cooperative JT effect in
the following way. As the oxygens are shared by two neighboring vanadium 
ions in the perovskite structure of LaVO$_3$, the JT distortions $Q_i$ 
at nearest-neighbor sites are not independent from each other. Hence the
electron densities in $|a\rangle/|b\rangle$ orbitals at two vanadium 
ions at sites $i$ and $j$ in the $ab$ plane (see inset in Fig. 
\ref{fig:to}) get coupled to each other, basically because they depend 
on the actual displacement $\delta$ of the shared oxygen ion.
More specifically, this displacement generates interactions between
the orbitals\cite{Kug82,Geh75} and one finds, taking care of the 
orthogonality constraint on the $Q_i$ variables,
\begin{equation}
\label{tautau}
H_{ab}(ij)= J_{\rm JT} \: \tau_i^z\tau_{j}^z,
\end{equation}
with the interaction constant being given by
$J_{\rm JT}= 2 \lambda E_{\rm JT} = \frac{1}{4}\lambda g^2$.
Here the coefficient $\lambda$ is determined by the phonon spectrum,
viz. by all branches in which the local coordinates $Q_i$ participate,
\cite{Geh75} and $\lambda<1$ (see e.g. Refs. \onlinecite{Hal71}
and \onlinecite{Mil96} for how to estimate $\lambda$ for the
perovskite lattice).
The JT interaction (\ref{tautau}) reflects the cooperative nature of
the JT problem. It favors orbital alternation and thus supports the
superexchange orbital interaction $J_{ab}^{\tau}$,\cite{Kha04} and the
orbital model relevant for the $C$-AF phase of LaVO$_3$ is then
\begin{eqnarray}
\label{omo}
{\cal H}_{\tau}=J_c^\tau \sum_{{\langle ij\rangle}\parallel c}
\Big({\vec\tau}_i\cdot{\vec\tau}_{j}-\frac{1}{4}\Big)
+V_{ab} \sum_{{\langle ij\rangle}\parallel {ab}} \tau_i^z\tau_j^z,
\end{eqnarray}
with $V_{ab}=J_{ab}^{\tau}+J_{\rm JT}$. The first term follows from
the spin-orbital model, and $J_c^{\tau}=Jr_1$.

While the superexchange contribution to $V_{ab}$ is small (adopting the
values of Ref. \onlinecite{Mot03}, $J_{ab}^{\tau}\simeq 2$ meV, i.e.,
$J_{ab}^{\tau}\ll J_c^{\tau}\simeq 33$ meV),
one finds that yet $V_{ab}>J_c^{\tau}$ if one assumes $\lambda=1$, i.e.
if one basically identifies the JT interaction with the JT energy, and
further accepts the estimate for $E_{\rm JT}\sim 27$ meV given in Ref.
\onlinecite{Mot03}. Then the Ising term quenches the $J_c^\tau$-driven
orbital dynamics and leads to a cooperative transition at $T_{\rm o}$
which locks the orbitals in all three directions.
However, one should be aware that the total energy decrease produced by
the JT distortion of the lattice, $E'_{\rm JT}$, as obtained in
an {\it ab initio\/} calculation such as in Ref. \onlinecite{Mot03},
actually equals $E'_{\rm JT}= E_{\rm JT}+E_{\rm JT}^{\rm o}$, i.e. it
comprises both the local energy gain $E_{\rm JT}$, which does not
contribute to the ordering,\cite{Geh75} as well as the JT ordering
energy $E_{\rm JT}^{\rm o}=z_{ab}J_{\rm JT}/8=\lambda E_{\rm JT}$.
The JT contribution to $V_{ab}$ is therefore given by
$J_{\rm JT}=2\lambda E'_{\rm JT}/(1+\lambda)$,
and since $\lambda$ is usually appreciably smaller than $1$
one should expect that $J_{\rm JT}$ is definitely smaller than
$E'_{\rm JT}$, so that it is more likely that actually
$V_{ab}\lesssim J_c^{\tau}$.

To gain more insight in the role of the JT interaction we performed 
exact diagonalization of eight-site chains along the $c$ axis, combined 
with a MF treatment of the orbital interactions in the $ab$ plane, to 
determine the orbital transition temperature $T_{\rm o}$. As expected, 
it scales with $J_{\rm JT}$ in the regime of large JT interaction 
$J_{\rm JT}>0.1J_c^{\tau}$ (but still $J_{\rm JT}\lesssim 2 J_c^{\tau}$) 
as shown in Fig. \ref{fig:to}, as was also found in
Ref. \onlinecite{Mot03}. However, we were surprised to see that Motome
{\it et al.\/}\cite{Mot03} have discarded this result considering it
to be ``an artefact of the MF treatment'', and argued that the coupling
between neighboring sites is determined solely by $J_c^{\tau}$.
The latter applies only in the limit of extremely anisotropic coupling,
$V_{ab} \ll J_c^{\tau}$, which is not relevant here. In fact, it is a
classical result \cite{Geh75} that the JT coupling of Eq. (\ref{jtcoup})
also induces intersite interactions between the orbitals
as in Eq. (\ref{tautau}) which may actually
dominate over the superexchange term and determine $T_{\rm o}$ in the
limit of large $g$. As $T_{\rm o}$ is 20-80\% above $T_N$ in other
RVO$_3$ (R= Ce, Pr, Nd, Y, etc.) compounds,\cite{Miy03} it is not
plausible that for R=La it is determined by superexchange alone.
Therefore we argue that as the $ab$ plane correlations are of
Ising-type, the MF result should be in fact a reasonable estimate for
$T_{\rm o}$. Thus, the proximity of the orbital and the magnetic
transitions in LaVO$_3$,\cite{Miy02}
$T_{\rm o}\sim T_N$ ($\simeq 0.4J$), implies that the JT interaction is
small, $J_{JT}\sim 0.1J_c^\tau$, and the JT splitting of the
$xz/yz$-doublet is actually smaller than the superexchange energy scale
itself,
$\Delta_{\rm JT}=z_{ab} J_{\rm JT}/2 = 2 J_{\rm JT} \sim 0.2J_c^\tau$.

Furthermore, as the spin and orbital exchange interactions are
interrelated,\cite{Mot03,Kha04} spin disorder should reduce the
effective orbital exchange $J_c^{\tau}$. Indeed, we have verified that
this follows from the full spin-orbital superexchange model\cite{Kha01}
which contains an extra term,
\begin{equation}
\label{stau}
{\cal H}_{s\tau}\!=\!\frac{1}{2}J_c^\tau\!
\sum_{{\langle ij\rangle}\parallel c}
\big({\vec S}_i\cdot{\vec S}_j-1\big)
\Big[\Big(1-\eta-\eta\frac{r_3}{r_1}\Big)
\Big({\vec\tau}_i\cdot{\vec\tau}_j+\frac{1}{4}\Big)-\eta\Big],
\end{equation}
describing the coupling of spins and orbitals along the $c$ axis.
However, $T_{\rm o}$ obtained from the complete spin-orbital model,
\begin{equation}
\label{Horb}
{\cal H}_U(d^2)\simeq {\cal H}_{\tau}+{\cal H}_{s\tau},
\end{equation}
has about the same value as that found from ${\cal H}_{\tau}$ alone
(see Fig. \ref{fig:to}), and the estimate
$\Delta_{\rm JT}\simeq 0.2J_c^\tau$ remains valid.

\begin{figure}[t!]
\includegraphics[width=8.0cm]{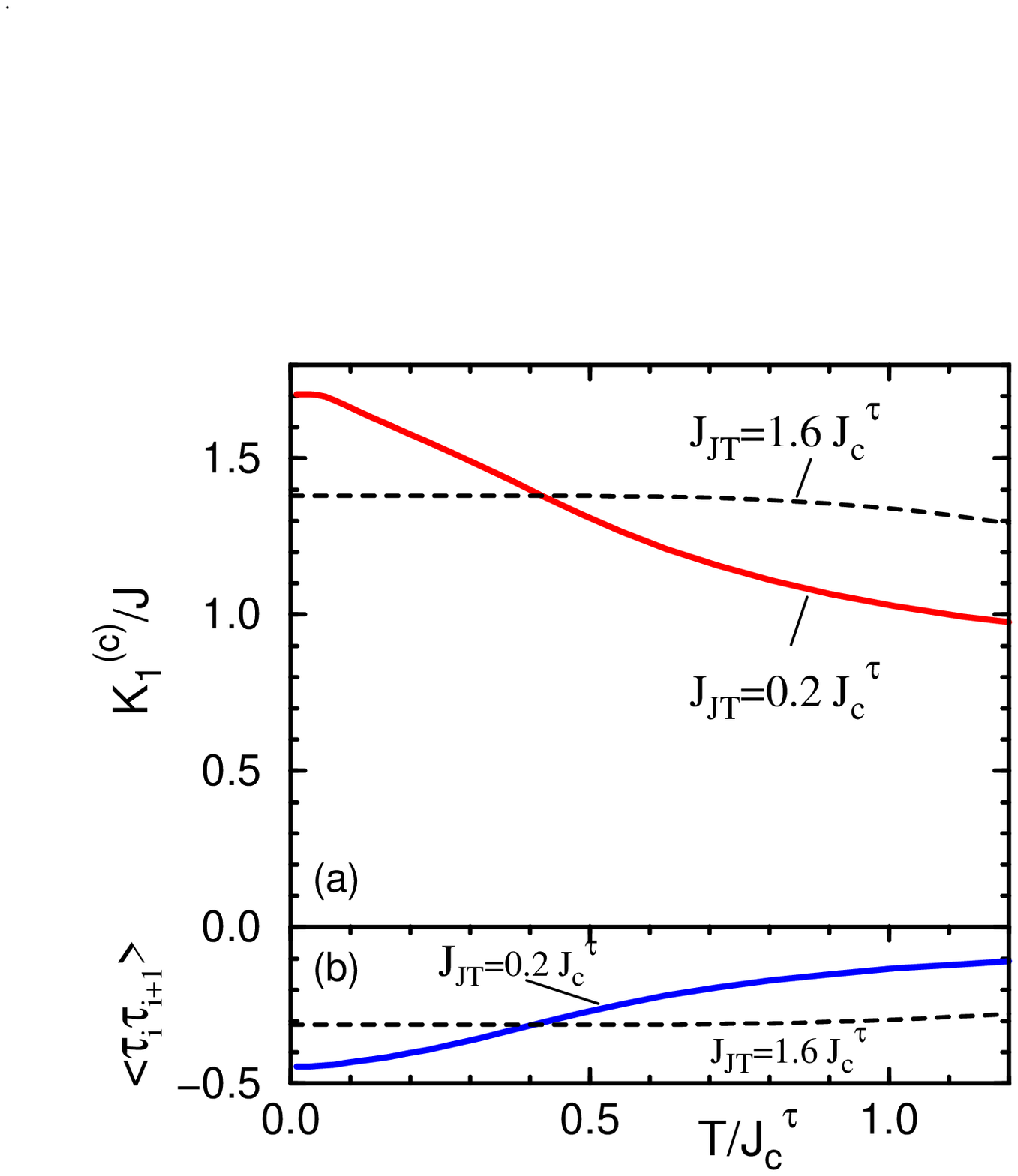}
\caption{(Color online)
Temperature dependence of:
(a) optical spectral weight of the high-spin transition, $K_1^{(c)}$,
in units of $J$ ($J$ is the principal energy scale of the spin-orbital
model), and
(b) orbital correlations
$\langle\vec{\tau}_i\cdot\vec{\tau}_{i+1}\rangle$
along the $c$ axis (bottom), as obtained for
--- the spin-orbital model Eqs. (\ref{Horb}), as given in Ref.
    \onlinecite{Kha01}, with small $J_{JT}=0.2J_c^\tau$ (solid lines),
    and
--- the orbital model Eq. (\ref{omo}) with $J_{JT}=1.6J_c^\tau$,
    as analyzed in Ref. \onlinecite{Mot03} (dashed lines).
}
\label{fig:full}
\end{figure}

Moreover, one finds that for large $J_{\rm JT}$
(e.g., $J_{\rm JT}=2 E'_{\rm JT}=54$~meV, i.e. for $\lambda=1$ and
$E'_{\rm JT}=27$~meV~$\simeq 0.8J_c^\tau$, following
Ref. \onlinecite{Mot03}) the temperature
dependence of the optical intensity $K_1^{(c)}$ derived from
${\cal H}_{\tau}$ is weak below 300 K (Fig. \ref{fig:full}), so adding
orbital correlations in this range of parameters to the spin
correlations of Sec. \ref{sec:d2oo} cannot improve the agreement with
experiment. Hence, this analysis clearly shows that:
(i) a substantial variation of $K_1^{(c)}$ below 300K ($\sim J$, see
Fig. \ref{fig:full}) observed in Ref.~\onlinecite{Miy02} is obtained
only for small $J_{\rm JT}$;
(ii) while neither spin correlations for frozen OO, nor orbital
correlations that follow from ${\cal H}_{\tau}$ alone would suffice,
only the full spin-orbital model (\ref{Horb}) that includes coupled
spin-orbital fluctuations is able to explain a large enhancement of
$K_1^{(c)}$ at low temperature. This was indeed demonstrated in Ref.
\onlinecite{Kha04}, where spin and orbital correlations were treated
self-consistently, and only then the strong temperature dependence of
$K_1^{(c)}$ could be successfully reproduced by the theory.

Summarizing, on comparing the results of Figs. \ref{fig:to} and
\ref{fig:full} with the experimental data,\cite{Miy02} one has to
conclude that the proximity of $T_{\rm o}$ and $T_N$ in LaVO$_3$ implies
that $J_{\rm JT}$ is in fact much smaller than the total energy
associated with the cooperative JT distortion $E'_{\rm JT}\simeq 27$~meV
estimated in Ref. \onlinecite{Mot03}.
The enhanced optical conductivity along the $c$ axis \cite{Miy02} also
suggests that the local splitting $\Delta_{\rm JT}$ is smaller than the
dynamical orbital exchange $J_c^{\tau}$, thus supporting the scenario of
fluctuating orbitals\cite{Kha01} in LaVO$_3$. Therefore, the assumption
of rigid OO, which was so successful for LaMnO$_3$, fails for
LaVO$_3$ and the full quantum spin-orbital many-body problem has to be
treated explicitly.\cite{Kha04}

\section{Summary and conclusions}

The main purpose of this paper was to make the experimental consequences
of the superexchange spin-orbital models for correlated transition
metal oxides with orbital degeneracy more transparent. We formulated
a general approach to the spectral weights in optical spectroscopy
and illustrated it on several examples with different multiplet
structure. While a general feature of all
the superexchange spin-orbital models is a tendency towards enhanced
quantum fluctuations,\cite{Fei97,Kha00} we gave reasons why in many
situations such fluctuations are quenched. One then arrives at
much simpler reduced models, where certain states with OO allow
for a good insight into the mechanisms responsible for the magnetic
interactions and for the optical spectral weights. The common feature
of all these cases is that the knowledge of only a few effective
parameters, the superexchange energy $J$, Hund's exchange $\eta$ and
the charge-transfer parameter $R$, is sufficient to work out the
quantitative predictions of the theory for a given type of orbital
ordered state.
In some of these cases the theoretical models simplify so much
that it is even possible to perform calculations with the help
of a pocket calculator.

The cases of copper fluoride KCuF$_3$ and the manganite LaMnO$_3$
turned out to be simpler, and could be understood with frozen OO and
quenched orbital dynamics below the structural transitions which occur
at much higher temperatures than the N\'eel temperature, $T_s\gg T_N$.
However, we have also seen that particularly in $t_{2g}$ systems, in
the cubic titanates and vanadates, the orbital dynamics may not be
quenched. Therefore, in some cases only the full quantum many-body
problem gives proper answers for the experimental situation.

We came to these conclusions by analyzing in detail the consequences
of decoupling of the spin and orbital degrees in states with rigid
OO and by comparing the predictions of the theory
with the experimental data, wherever available. In the undoped
manganite LaMnO$_3$ we could provide a consistent explanation of the
magnetic and optical experimental data by deducing the values of the
above effective parameters $\{J,\eta,R\}$, and next showing that both
the magnetic exchange constants $J_{ab}$ and $J_c$, and the anisotropy
and the temperature dependence of the low-energy optical spectral
weights can be reproduced by the theory in a satisfactory way. In the
case of the copper
fluoride KCuF$_3$ optical data were not available, but the constraints
in the theory given by the exchange constants are so strong that we
could conclude that the insulating state in this compound has
charge-transfer character. It remains to be verified by future
experiments to what extent the predictions made here concerning the
optical spectral weights and based on the classical picture with ordered
$e_g$ orbitals apply to KCuF$_3$.

Also for LaTiO$_3$ and YTiO$_3$ we investigated the classical states
with OO given by certain wavefunctions which guarantee that the observed
isotropic AF or FM states are realized. As in all other cases, the theory
predicts in such states a rather pronounced temperature dependence and
spectral
weight transfers in the optical spectra near the magnetic transition.
Future experiments will have to establish whether and to what extent
such a scenario relaying on rigid OO could be valid. However, already
without these data we could demonstrate, by looking at the exchange
constants, that there are certain indications that orbital fluctuations
play a role and thus the quantum physics might dominate here over the
thermal fluctuations of the spins alone.

\begin{table}[b!]
\caption{
Values of the effective parameters of spin-orbital models:
superexchange $J$ (in meV), Hund's parameter $\eta$,
the CT parameter $R$, and the microscopic parameters consistent with
these effective parameters: intraorbital Coulomb interaction $U$, Hund's
exchange $J_H$, and the energy of the lowest CT excitation $\Delta$ (all
in eV), deduced from the present analysis of the magnetic and optical
properties of representative transition metal compounds with perovskite
structure. The values of $J_H$ and $U$ in case of LaVO$_3$ and LaMnO$_3$
were obtained from the optical spectra, while the ones for KCuF$_3$ are
the same as in Ref. \onlinecite{Lic95}.}
\vskip .2cm
\begin{ruledtabular}
\begin{tabular}{cccccc}
           &               & LaVO$_3$ & LaMnO$_3$ & KCuF$_3$ \cr
parameters &  orbitals     & $t_{2g}$ &   $e_g$   &  $e_g$   \cr
\colrule
effective  &   $J$         &     40   &    150    &   33     \cr
           &   $\eta$      &    0.13  &   0.18    &   0.12   \cr
           &    $R$        &  $<0.4$  &   0.6     &   1.2    \cr
\colrule
microscopic&      $U$      &   3.8    &    3.8    &   7.5    \cr
           &    $J_H$      &   0.50   &    0.67   &   0.90   \cr
           & $\Delta$      &  $>5.0$  &    3.5    &   4.0    \cr
\end{tabular}
\end{ruledtabular}
\label{tab:uj}
\end{table}

The case of the cubic vanadate LaVO$_3$ really shows that one may
encounter the full complexity of the spin-orbital superexchange model
when the spins and orbitals fluctuate coherently, and these fluctuations
are essential to get a meaningful quantitative description of the
optical data. This case in particular demonstrates the importance of
combining the magnetic and optical data. Whereas the classical analysis
of the spin fluctuations for frozen orbitals seems to suffice to explain
the exchange constants in LaVO$_3$, by looking at the optics one
realizes that the picture of frozen OO induced by large JT coupling is
here misleading, and the complete spin-orbital dynamics has to be
considered instead.

Before concluding this paper, we summarize in Table \ref{tab:uj} the
effective parameters of the spin-orbital models,\cite{notepa} and the
possible values of the microscopic parameters ---
the Coulomb interaction $U$, Hund's exchange
$J_H$, and the charge transfer energy $\Delta$, that are consistent with
these effective parameters. The values of the superexchange constant are
$33<J<40$ meV for KCuF$_3$ and LaVO$_3$, while for LaMnO$_3$ the value
of this parameter is much higher, $J\sim 140$ meV.
This difference reflects a rather high value of the effective hopping
parameter $t$ in the undoped manganite. By considering the optical data
of LaVO$_3$ and LaMnO$_3$ we came to the conclusion that Hund's exchange
$J_H$ is somewhat reduced from the respective atomic values,\cite{Zaa90}
and we give already these reduced parameters in Table \ref{tab:uj},
accompanied by the corresponding values of $U$.

Summarizing, we have illustrated a common approach to the optical and
the magnetic data in Mott insulators with orbital degeneracy, which
provides the basis for a better theoretical understanding of the
experimental constraints on the underlying phenomena and on the model
parameters. It is a unique feature of these systems that the
superexchange interactions, and the spin, orbital, and composite
spin-and-orbital correlations induced by them, are responsible for the
distribution of spectral weight in the optical excitations. We hope that
extending the present analysis of the high-energy excitations in the
upper Hubbard band by an analysis of the low-energy excitations that
occur in doped systems, will allow to develop a quantitative theoretical
approach designed to describe the optical spectra of doped transition
metal insulators.

\acknowledgments
We thank J. van den Brink, N. Kovaleva, B. Keimer,
and C. Ulrich for insightful discussions.
A.~M.~Ole\'s would like to acknowledge support by the Polish Ministry
of Science and Information Technology under Project No.~1 P03B 068 26.

\appendix

\section{ Effective model for L\lowercase{a}M\lowercase{n}O$_3$ }
\label{d4:eff}

Here we present an analysis of the magnetic and optical data
within an effective $d-d$ model for LaMnO$_3$, given by the
${\cal H}_U(d^4)$ term alone, i.e., assuming $R=0$ in Eq. (\ref{HCT4e}).
Unlike in KCuF$_3$, the alternating OO in LaMnO$_3$ refers to the
orbitals {\it occupied by electrons\/}, and is characterized by a single
angle $\theta$ in Eqs. (\ref{ood9}). We performed an analysis of the
exchange constants $J_c$ (\ref{jc4}) and $J_{ab}$ (\ref{jab4}) as
functions of the effective parameters $J$ and $\eta$. Although these
parameters cannot be uniquely determined, we have verified that only
a narrow range of the orbital angle $\theta\sim 90^{\circ}-100^{\circ}$
gives reasonable agreement with experiment. Here we present the results
obtained with $J=170$ meV and $\eta=0.16$.

\begin{figure}[t!]
\includegraphics[width=7.5cm]{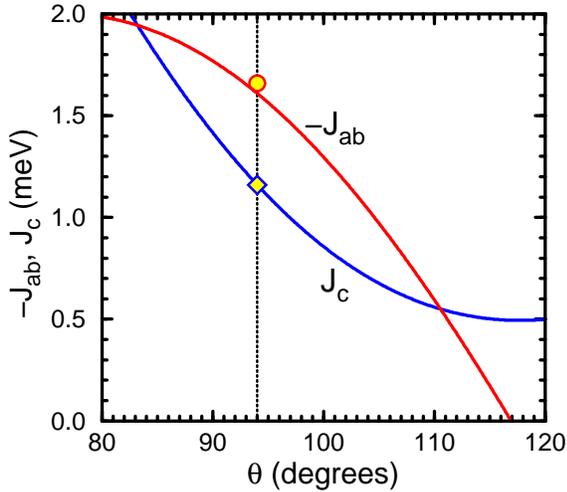}
\caption{(Color online)
Superexchange interactions $-J_{ab}$ (\ref{jab4}) and $J_c$ (\ref{jc4})
as functions of the orbital angle $\theta$ (solid lines), obtained
within the effective superexchange model ${\cal H}_U(d^4)$ ($R=0$).
Experimental values\cite{Mou96} of $-J_{ab}$ and $J_c$ for LaMnO$_3$
(indicated by circle and diamond) are nearly reproduced for the
OO with $\theta=94^{\circ}$.
Parameters: $J=170$ meV, $\eta=0.16$.
}
\label{fig:j4u}
\end{figure}

Both exchange constants exhibit a rather strong dependence on the
orbital angle $\theta$ (Fig. \ref{fig:j4u}). In contrast to the case of
KCuF$_3$, the effective model parameters obtained at $R=0$ suffices to
explain even almost quantitatively the observed exchange constants in
LaMnO$_3$. This result is also consistent\cite{notetg} with the earlier
analysis of Ref. \onlinecite{Fei99}. Due to the strong dependence of the
exchange constants $J_{ab}$ and $J_c$ on the angle $\theta$, one can
exclude the OO of alternating directional $(3x^2-r^2)/(3y^2-r^2)$
orbitals, obtained with $\theta=120^{\circ}$ in Eqs. (\ref{ood9}).

While it is frequently assumed that the $t_{2g}$ superexchange
(\ref{Jt4}) is large, the present analysis shows that a consistent
description of the magnetic properties requires instead a {\it rather
moderate value of\/} $J_t$ in LaMnO$_3$. For the present parameters one
finds $J_t=1.70$ meV,\cite{notes2} and $J_t$ increases by $\sim 35$\%
when the CT terms are included (see Sec. \ref{d4:resj}).
Although it might be argued that $J_t\sim 2$ meV is too large as the
experimental value of $T_N$ in CaMnO$_3$ is only 110 K,
\protect\cite{Wol55} one deals here {\it de facto\/} with different
values of the $U$, $J_H$ and $\Delta$ parameters, namely with those for
Mn$^{2+}$ ions instead of the ones for Mn$^{3+}$ ions, which apply in the
intermediate excited states contributing to the superexchange in
CaMnO$_3$. Yet the differences in these parameter values cannot be large
from the very nature of their physical origin, and so $J_t$ for LaMnO$_3$
cannot differ by more than a factor of 2 from the value of the
$t_{2g}$-induced superexchange in CaMnO$_3$.\protect\cite{Fei99}
This supports our finding that a small ratio $J_t\sim 4\times 10^{-3}t$
corresponds to realistic parameter values for LaMnO$_3$.

\begin{figure}[t!]
\includegraphics[width=7.7cm]{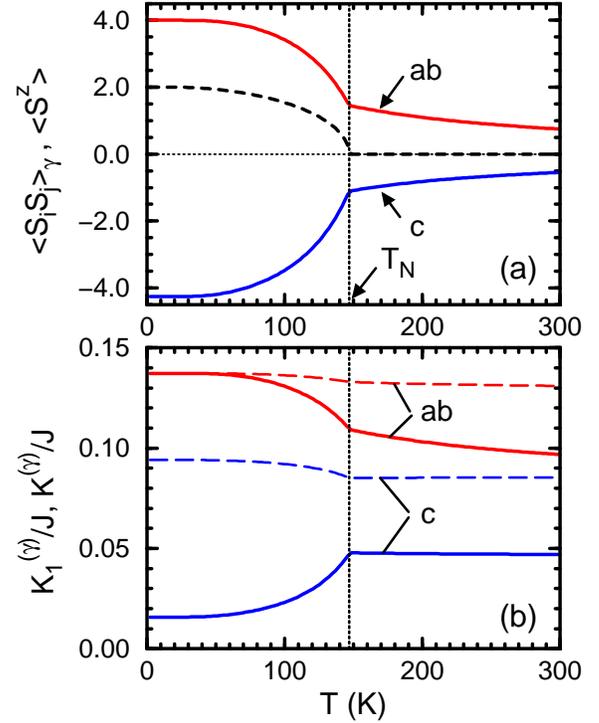}
\caption{(Color online)
Temperature dependence of:
(a) spin-spin correlations
$\langle\vec S_i\cdot\vec S_j\rangle_{\gamma}$ for $\gamma=a,c$,
and the order parameter $\langle S^z\rangle$ (dashed line);
(b) kinetic energies $K_1^{(\gamma)}$ for the high-spin excitations,
standing for the optical intensities (\ref{opsa}) at low energy along
$\gamma=a,c$ axes (solid lines), and total kinetic energies
$K^{(\gamma)}$ (long-dashed lines), as obtained for LaMnO$_3$ within the
effective superexchange model ${\cal H}_U(d^4)$ for the OO given by an
angle $\theta\simeq 94^{\circ}$. Parameters as in Fig. \ref{fig:j4u}.
}
\label{fig:op}
\end{figure}

Next we consider the temperature dependence of the spin-spin
correlations and the optical spectral weights (\ref{opsa}).
As in the full model discussed in Sec. \ref{d4:reso}, one may assume
frozen OO in the relevant temperature range below room temperature,
and derive the temperature dependence of the optical spectral weights
from that of the intersite spin-spin correlations (\ref{spins}), with
the latter determined in a cluster method as explained in Appendix
\ref{app:spincomn}.
Using the exchange interactions obtained with the present parameters at
$\theta=94^{\circ}$ (Fig. \ref{fig:j4u}), one finds the empirical
estimate\cite{Fle04} $T_N\simeq 146$ K, which reasonably agrees with the
experimental value $T_N^{\rm exp}=136$ K.\cite{Mou96} The AF bonds feel
staggered MF terms for $T<T_N$, so the spin-spin correlations $s_c$ had
to be determined by a numerical solution, as explained in Appendix
\ref{app:spincomn}. Both correlation functions change fast close to
$T_N$, reflecting the temperature dependence of the Brillouin function,
and remain finite at $T\gg T_N$ [Fig. \ref{fig:op}(a)].

The effective model also allows one to discuss the qualitative
features of the spectral weight distribution in the optical spectra.
The theory predicts that only {\it high-spin\/} optical excitations are
allowed at $T=0$ for the FM bonds in the $ab$ planes, and one finds for
this polarization a large kinetic energy $K_1^{(ab)}$
[Fig. \ref{fig:op}(b)]. In contrast, the optical excitations for the
AF bonds along the $c$ axis are predominantly of {\it low-spin\/}
character, and thus the kinetic energy $K_1^{(c)}$ is rather small,
resulting in a large anisotropy $K_1^{(ab)} : K_1^{(c)} \sim 10:1$ of
the low-energy optical intensities confirmed by experiment.\cite{Kov04}
When the temperature increases and the spin-spin correlations weaken,
this anisotropy is reduced, but remains pronounced also at $T>T_N$ and
still exceeds $2:1$ at $T=300$ K due to the persisting OO. As shown
elsewhere,\cite{Ole04} also a quantitative analysis of the present
effective model gives a rather satisfactory agreement with the optical
data\cite{Kov04} in the entire temperature range. In contrast, the
total optical
intensities have a much weaker temperature dependence and anisotropy
[Fig. \ref{fig:op}(b)]. Thus, the main features on the experimentally
observed intensity distribution and its temperature variation in the
optical spectra\cite{Kov04} are well reproduced already by the present
effective model.

Finally, we verify whether the used parameters $J=170$ meV and
$\eta=0.16$ could be derived from the microscopic parameters of the CT
model. The value of $J_H=0.90$ eV follows from the spectroscopic values
of the $B$ and $C$ Racah parameters,\cite{Boc92} so $\eta=0.16$ implies
$U\simeq 5.6$ eV. Knowing the value of $J$, this leads to an estimated
effective $d-d$ hopping element $t\simeq 0.49$ eV. Indeed, one finds
that these microscopic parameters are in the expected range. The
effective Coulomb interaction was estimated within the effective $d-d$
model for LaMnO$_3$ as $U\sim 5.5$ eV from spectroscopic data,
\cite{Miz96} so the agreement is close to perfect. Furthermore, taking
the usually accepted values of $t_{pd}=1.5$ eV and $\Delta=5.0$ eV,
following Refs. \onlinecite{Boc92} and \onlinecite{Miz96}, one finds a
very plausible value of the effective hopping parameter
$t=t_{pd}^2/\Delta=0.45$ eV, again quite close to the value derived
above. Note that the experimental magnetic interactions in doped bilayer
manganites were explained with a similar value of $t=0.48$ eV fixed by
experiment.\cite{Ole03}
This favorable comparison emphasizes once again our main conclusion that
the relevant model parameters can be derived by combining the results of
magnetic and optical experiments, whenever available.

\section{ Spin-spin correlations in L\lowercase{a}M\lowercase{n}O$_3$ }
\label{app:spincomn}

Here we describe briefly a simple method which we used to determine the
spin-spin correlation functions $s_{ab}$ and $s_c$
for a pair of interacting spins $S=2$ at nearest neighbor Mn ions in
LaMnO$_3$. The spin-spin correlations were obtained by performing a
statistical average over the exact eigenstates for a single (FM or AF)
bond, found in the presence of MF terms acting at each spin of the bond
and originating from its neighboring spins. This, in fact, is the
simplest cluster MF theory, known in the theory of magnetism as the
Oguchi method.\cite{Ogu60} As long as the MF terms vanish (at $T>T_N$),
one finds for the various eigenstates with degeneracy
$(2S_{\rm tot}+1)$, labeled by the total spin $S_{\rm tot}=0,1,\cdots,4$,
\begin{equation}
\label{ss4ab}
\langle S_{\rm tot}|{\vec S}_i\cdot{\vec S}_j|S_{\rm tot}\rangle
=\frac{1}{2}S_{\rm tot}(S_{\rm tot}+1)-6.
\end{equation}
Depending on $S_{\rm tot}$ the scalar product
$\langle S_{\rm tot}|{\vec S}_i\cdot{\vec S}_j|S_{\rm tot}\rangle$
varies between $-6$ and $4$.

Consider first a FM bond $\langle ij\rangle$ in the $ab$ plane, with the
Hamiltonian given by,
\begin{equation}
\label{H4ab}
H_{ij}^{(ab)}=-|J_{ab}|{\vec S}_i\cdot{\vec S}_j-h_{ab}(S_i^z+S_j^z),
\end{equation}
where for the $A$-AF phase with uniform order parameter in the $ab$
plane $\langle S^z\rangle=\langle S^z_i\rangle$ the identical MF
acting at both spins is
\begin{equation}
\label{h4ab}
h_{ab}=(3|J_{ab}|+2J_c)\langle S^z\rangle.
\end{equation}
At $T>T_N$ the eigenenergies of $H_{ij}^{(ab)}$ follow from Eq.
(\ref{ss4ab}). For $T<T_N$ the order parameter $\langle S^z\rangle$
could in principle be determined self-consistently in the present
cluster approach. However, to gain a qualitative insight into the
temperature dependence of $s_{ab}$ it suffices to use a
self-consistent solution of the MF equation,
\begin{eqnarray}
\label{sz4}
\langle S^z\rangle&=&
\frac{2S+1}{2}\coth\Big(\frac{2S+1}{2}\frac{T_N}{2T}\langle S^z\rangle\Big)
\nonumber \\
&-&\frac{1}{2}\coth\Big(\frac{1}{2}\frac{T_N}{2T}\langle S^z\rangle\Big).
\end{eqnarray}
As the MF value $k_{\rm B}T_N^{\rm MF}=4(2|J_{ab}|+J_c)$ is overestimated,
it is appropriate to use in Eq. (\ref{sz4}) the value of $T_N$ after
an empirical reduction,\cite{Fle04} leading to
$T_N\simeq 0.705T_N^{\rm MF}$.
One finds an analytic solution for $s_{ab}$:
\begin{equation}
\label{s4ab}
s_{ab}=\frac
{4z_4-3z_2-5z_1-6z_0}{z_4+z_3+z_2+z_1+z_0},
\end{equation}
where the terms $z_i$ refer to the subspaces of total spin $S_{\rm tot}$,
\begin{eqnarray}
\label{z4ab}
z_4&=&1+2\cosh x+2\cosh 2x+2\cosh 3x                    \nonumber \\
   &+&2\cosh 4x,                                                  \\
z_3&=&(1+2\cosh x+2\cosh 2x+2\cosh 3x)                  \nonumber \\
   &\times&\exp(-4\beta |J_{ab}|),                                \\
z_2&=&(1+2\cosh x+2\cosh 2x)\exp(-7\beta |J_{ab}|),               \\
z_1&=&(1+2\cosh x)\exp(-9\beta |J_{ab}|),                         \\
z_0&=&\exp(-10\beta |J_{ab}|),
\end{eqnarray}
with $x=\beta h_{ab}$ and $\beta=1/k_{\rm B}T$. Note that the term
$\propto z_3$ is absent in the numerator of Eq. (\ref{s4ab}), because
$\langle S_{\rm tot}=3|{\vec S}_i\cdot{\vec S}_j|S_{\rm tot}=3\rangle=0$
[see Eq. (\ref{ss4ab})].

For an AF bond $\langle ij\rangle$ along the $c$ axis the Hamiltonian is
given by
\begin{equation}
\label{H4c}
H_{ij}^{(c)}=J_c{\vec S}_i\cdot{\vec S}_j-h_c(S_i^z-S_j^z),
\end{equation}
where the molecular field
\begin{equation}
\label{h4c}
h_c=(4|J_{ab}|+J_c)\langle S^z\rangle.
\end{equation}
alternates between the sites $i$ and $j$, so
$\langle S^z_i\rangle=-\langle S^z_j\rangle=\langle S^z\rangle$.
Unlike for the FM bond, the field $h_c$ couples now states which
belong to different values of $S_{\rm tot}$ (but to the same value of
$S_{\rm tot}^z$).  The staggered MF plays no
role for $S_{\rm tot}^z=4$, while for $S_{\rm tot}^z<4$ the eigenstates
have been found by diagonalizing the matrices, with diagonal elements
following from Eq. (\ref{ss4ab}), and offdiagonal ones $\propto h_c$:
\leftline{--- for $S_{\rm tot}^z=3$,}
\begin{equation}
\left( \begin{array}{ccc}
 4J_c   &   -h_c        \\[0.2cm]
 -h_c   &    0
\end{array}
\right),
\label{h4sz3}
\end{equation}
\leftline{--- for $S_{\rm tot}^z=2$,}
\begin{equation}\textstyle{
\left( \begin{array}{ccc}
4J_c  & -2\sqrt{\frac{3}{7}}\;h_c & 0                        \\[0.2cm]
-2\sqrt{\frac{3}{7}}\;h_c &  0   & -4\sqrt{\frac{1}{7}}\;h_c \\[0.2cm]
 0        & -4\sqrt{\frac{1}{7}\;}h_c   & -3J_c
\end{array}
\right),}
\label{h4sz2}
\end{equation}
\leftline{--- for $S_{\rm tot}^z=1$,}
\begin{equation}\textstyle{
\left( \begin{array}{cccc}
4J_c  & -\sqrt{\frac{15}{7}}\;h_c & 0 & 0                 \\[0.2cm]
-\sqrt{\frac{15}{7}}\;h_c& 0 & -\frac{16}{\sqrt{70}}\;h_c & 0 \\[0.2cm]
0 &-\frac{16}{\sqrt{70}}\;h_c& -3J_c& -\sqrt{\frac{21}{5}}\;h_c\\[0.2cm]
0 & 0 & -\sqrt{\frac{21}{5}}\;h_c & -5J_c
\end{array}
\right),}
\label{h4sz1}
\end{equation}
\leftline{--- for $S_{\rm tot}^z=0$,}
\begin{equation}\textstyle{
\left( \begin{array}{ccccc}
4J_c  & -\frac{4}{\sqrt{7}}h_c & 0 & 0 & 0                \\[0.2cm]
-\frac{4}{\sqrt{7}}h_c& -\lambda& -\frac{12}{\sqrt{35}}h_c& 0 & 0 \\[0.2cm]
0 & -\frac{12}{\sqrt{35}}h_c& -3J_c& -2\sqrt{\frac{7}{5}}h_c  & 0 \\[0.2cm]
0 & 0 & -2\sqrt{\frac{7}{5}}h_c& -5J_c& -2\sqrt{2}h_c             \\[0.2cm]
0 & 0 & 0 & -2\sqrt{2}h_c & -6J_c                         \\[0.2cm]
\end{array}
\right).}
\label{h4sz0}
\end{equation}
In this way a complete set of eigenstates $\{|n\rangle\}$ with energies
$\{E_n\}$ for $n=1,2,\cdots,25$ was determined. Finally, the spin-spin
correlation function $s_c$ was found using a standard formula
\begin{equation}
\label{s4c}
s_c=\frac{1}{Z}\sum_n \langle n|{\vec S}_i\cdot{\vec S}_j|n\rangle
\exp(-\beta E_n),
\end{equation}
where $Z=\sum_n \exp(-\beta E_n)$ is the partition function.

\section{Spin-spin correlations in L\lowercase{a}VO$_3$ and in YVO$_3$}
\label{app:spincov}

The short-range spin-spin correlations $s_{\gamma}$ for the cubic
vanadates were determined using the Oguchi method\cite{Ogu60} for a
bond of interacting $S=1$ spins. As in the case of LaMnO$_3$ (see
Appendix \ref{app:spincomn}), we solve exactly a single FM (AF) bond
$\langle ij\rangle$ with interaction $J_c$ ($J_{ab}$), and the MF terms
$\propto\langle S^z\rangle$ originating from neighboring spins and
acting on each spin of the bond.
In the present case the scalar product is given by
\begin{equation}
\label{ss2ab}
\langle S_{\rm tot}|{\vec S}_i\cdot{\vec S}_j|S_{\rm tot}\rangle
=\frac{1}{2}S_{\rm tot}(S_{\rm tot}+1)-2.
\end{equation}

For a FM bond, now along the $c$ axis, one finds an analytic solution.
\cite{Kha04} This problem is analogous to that given by Eq. (\ref{h4ab}).
Using the MF approximation, the order parameter $\langle S^z\rangle$ was
determined from Eq. (\ref{sz4}) with $S=1$, and
$T_N^{\rm MF}=4(2J_{ab}+|J_c|)/3$ was reduced to
$T_N\simeq 0.684T_N^{\rm MF}$ as appropriate for $S=1$ spins.\cite{Fle04}
The final result for $s_c$ reads
\begin{equation}
\label{s2c}
s_c=\frac{z_2-z_1-2z_0}{z_2+z_1+z_0},
\end{equation}
where the terms $z_i$ originate from different subspaces of total spin
$S_{\rm tot}=2,1,0$,
\begin{eqnarray}
\label{z2c}
z_2&=&1+2\cosh x+2\cosh 2x,                                    \\
z_1&=&(1+2\cosh x)\exp(-2\beta |J_c|),                         \\
z_0&=&\exp(-3\beta |J_c|).
\end{eqnarray}
Here
\begin{eqnarray}
\label{h2c}
h_c=(4J_{ab}+|J_c|)\langle S^z\rangle,
\end{eqnarray}
$x=\beta h_c$ and $\beta=1/k_{\rm B}T$; compare with Eq. (\ref{h4c}).

For an AF bond $\langle ij\rangle$ in the $ab$ plane the MF Hamiltonian
is given by
\begin{equation}
\label{H2ab}
H_{ij}^{(ab)}=J_{ab}{\vec S}_i\cdot{\vec S}_j-h_{ab}(S_i^z-S_j^z),
\end{equation}
In analogy to an AF bond in LaMnO$_3$ (see Appendix \ref{app:spincomn}),
the correlation function $s_{ab}$ can be found numerically by
considering the subspaces of $S_{\rm tot}^z$. The molecular field,
\begin{equation}
\label{h2ab}
h_{ab}=(3J_{ab}+2|J_c|)\langle S^z\rangle,
\end{equation}
does not contribute to the highest eigenenergies $E_{8,9}=J_{ab}$ in
the subspace of $S_{\rm tot}^z=2$ [cf. with Eq. (\ref{h4ab})], while
the remaining eigenstates had to be found by diagonalizing the matrices
corresponding to other values of $S_{\rm tot}^z$:
\leftline{--- for $S_{\rm tot}^z=1$,}
\begin{equation}
\left( \begin{array}{cc}
  J_{ab}  &   -h_{ab}        \\[0.2cm]
 -h_{ab}  &   -J_{ab}
\end{array}
\right) ,
\label{h2sz1}
\end{equation}
\leftline{--- for $S_{\rm tot}^z=0$,}
\begin{equation}\textstyle{
\left( \begin{array}{ccc}
J_{ab}              & -\frac{2}{\sqrt{3}}\;h_{ab} & 0
  \\[0.2cm]
-\frac{2}{\sqrt{3}}\;h_{ab} & -J_{ab} & -2\sqrt{\frac{2}{3}}\;h_{ab}
  \\[0.2cm]
       0     & -2\sqrt{\frac{2}{3}}\;h_{ab}   & -2J_{ab}
\end{array}
\right) .}
\label{h2sz0}
\end{equation}
By solving the above eigenproblems, we determined a complete set of
eigenstates $\{|n\rangle\}$, with energies $E_n$, labeled by
$n=1,2,\cdots,9$. Therefore, the spin-spin correlation function $s_{ab}$
for two interactings $S=1$ spins on an AF bond follows in the present
case from an
equation similar to Eq. (\ref{s4c}), with the relevant matrix elements
$\langle n|{\vec S}_i\cdot{\vec S}_j|n\rangle$ now given by
Eq. (\ref{ss2ab}).



\begin{thebibliography}{}
\bibitem{notede} We note, however, that frequently the average energy
                 of a CT type excitation $\bar{\Delta}$ is considered
                 instead and compared with $U$.
\bibitem{noteabc} Fractional contributions due to exchange terms
                  $\propto J_H$ shown in the spectrum for the $d^5$
                  case [see Fig. \ref{fig:levels}(a)] follow from the
                  differences between the exchange elements
                  $J_{\alpha\beta}$ in Hamiltonian $H_{int}$
                  (\ref{Heegen}) (see Table I), and were obtained
                  using the relation $C\simeq 4B$ satisfied
                  approximately by the experimental
                  values for Mn$^{2+}$ ions.\protect\cite{Zaa90}
\bibitem{notejh} Note that a different definition $J_H=2B+C$ was used
                 in Ref. \protect\onlinecite{Fei99} to parametrize the
                 spectrum of Mn$^{2+}$ ions, leading to a seemingly
                 smaller value of $J_H$.
\bibitem{noteoo} Although the OO in LaMnO$_3$ is in principle unknown,
                 we remark that the $C$-type OO with
                 symmetric/antisymmetric combinations of $e_g$ orbitals,
                 $(|z\rangle\pm|x\rangle)/\sqrt{2}$, is favored by a
                 combination of the orbital superexchange and orbital
                 interactions induced by the JT effect at $T>T_N$.
                 \protect\cite{Fei99}
\bibitem{noteeg} We have verified that the amplitude of pairs of two
                 orthogonal orbitals is larger for bonds in the $ab$
                 planes than along the $c$ axis, and thus the FM $e_g$
		 term is here stronger by a factor close to three.
\bibitem{noterr} Equations (\ref{jd1c}) and (\ref{jd1ab}) were simplified
                 using the identity $\frac{1}{3}(r_2+2r_3)=r_2r_3$
                 satisfied by the excitation spectrum of Ti$^{2+}$ ions.
\bibitem{noteli} The nearly linear behavior of $J_c$ is remarkable and
                 one can prove it by performing an expansion in $\delta$.
                 Note also that for $\eta=0$ isotropic exchange
                 interactions $J_{ab}=J_c$ are recovered at $\delta=1$.
\bibitem{notefet} Equation (\ref{jf1}) demonstrates that the FM
                  interaction in lowest order increases linearly with
		  Hund's exchange $\eta$ --- it follows from the
		  identity $\frac{1}{3}(r_1-r_2)=2\eta r_1r_2$.
\bibitem{notepa} Having only the magnetic data\cite{Kei00} for
                 LaTiO$_3$, we could not obtain meaningful values of
                 the effective parameters $J$ and $\eta$ in this case,
                 so this compound was not included in Table \ref{tab:uj}.
\bibitem{notetg} The angle $\theta$ is here different from that given in
                 Ref. \onlinecite{Fei99} as the value of $J_t$ derived
                 there was underestimated by a factor of two.
\bibitem{notes2} Note that this value refers to total spin $S=2$ of
                 Mn$^{3+}$ ions and not to $t_{2g}$ spin $S=3/2$,
                 as frequently defined.
\end{thebibliography}
\end{document}